\documentclass[a4paper,11pt]{article}
\usepackage[utf8]{inputenc}
\usepackage[T1]{fontenc}
\usepackage{jheppub} 
\usepackage[x11names,table]{xcolor}
\usepackage{amsthm, mathtools, slashed}
\usepackage{multirow, rotating, hhline}
\usepackage{subcaption}
\usepackage{tikz, tikz-cd} \usetikzlibrary{positioning,arrows.meta, bending, decorations.pathmorphing}
\usepackage{enumitem}
\usepackage[normalem]{ulem}
\usepackage{multirow}
\usepackage{hyperref}
\usepackage{pifont}

\newlist{todolist}{itemize}{2}
\setlist[todolist]{label=$\square$}

\usepackage{array}
\usepackage{yhmath}
\usepackage{changepage}

\usepackage[export]{adjustbox} 
\newtheorem{theorem}{Theorem}[section]

\theoremstyle{definition}
\newtheorem{definition}[theorem]{Definition}
\newtheorem{proposition}[theorem]{Proposition}
\theoremstyle{remark}

\newcommand{\calA}{\mathcal{A}}
\newcommand{\calB}{\mathcal{B}}

\newcommand{\calD}{\mathcal{D}}

\newcommand{\calF}{\mathcal{F}}
\newcommand{\calG}{\mathcal{G}}
\newcommand{\calH}{\mathcal{H}}
\newcommand{\calI}{\mathcal{I}}
\newcommand{\calJ}{\mathcal{J}}
\newcommand{\calK}{\mathcal{K}}

\newcommand{\calM}{\mathcal{M}}
\newcommand{\calN}{\mathcal{N}}

\newcommand{\calP}{\mathcal{P}}

\newcommand{\calS}{\mathcal{S}}

\newcommand{\calU}{\mathcal{U}}
\newcommand{\calV}{\mathcal{V}}
\newcommand{\calW}{\mathcal{W}}
\newcommand{\calX}{\mathcal{X}}
\newcommand{\calY}{\mathcal{Y}}
\newcommand{\calZ}{\mathcal{Z}}
\newcommand{\e}{\mathrm{e}}
\renewcommand{\i}{\mathrm{i}}
\renewcommand{\d}{\mathrm{d}}

\newcommand{\tr}{\mathrm{tr}}
\newcommand{\Z}{\mathbb{Z}}
\newcommand{\Q}{\mathbb{Q}}
\newcommand{\R}{\mathbb{R}}
\newcommand{\C}{\mathbb{C}}
\newcommand{\Hom}{\mathrm{Hom}}

\newcommand{\tH}{\widetilde{H}}
\newcommand{\tOmega}{\widetilde{\Omega}}

\newcommand{\G}{\mathcal{G}}
\newcommand{\U}{\mathcal{U}}
\renewcommand{\H}{\mathbb{H}}

\usepackage{stackengine}

\def\sc#1{\textcolor{DarkGoldenrod2}{{\bf SC}: #1}}

\usepackage{titlesec}  

\newcounter{subsubsubsection}[subsubsection]

\titleformat{\paragraph}[block]{\normalfont\normalsize\bfseries}{\theparagraph}{1em}{}

\titlespacing*{\paragraph}{0pt}{1.5ex plus 0.5ex minus .2ex}{0.8ex plus .2ex}

\title{Symmetry $\theta$ angles and topological Witten effects}

\author[1]{Shi Chen,}
\emailAdd{chen8743@umn.edu}
\author[1]{Aleksey Cherman,}
\emailAdd{acherman@umn.edu}
\author[1]{Maria Neuzil}
 \emailAdd{neuzi008@umn.edu}
\affiliation[1]{School of Physics and Astronomy, 
University of Minnesota, Minneapolis MN 55455, USA}

\abstract{We introduce a large class of $\theta$ angles in quantum field theory that we call symmetry $\theta$ angles. 
Unlike conventional $\theta$ angles whose definition depends on a choice of a path integral, symmetry $\theta$ angles are intrinsic parameters of a quantum field theory that depend only on its symmetries. 
A frequent consequence of symmetry $\theta$ angles is a phenomenon we call the topological Witten effect, which is a generalization of the standard Witten effect.   
Topological Witten effects modify which charged operators are attached to topological operators as a function of $\theta$. 
Physically, topological Witten effects induce generalized Aharonov-Bohm effects. 
We show that these new $\theta$ angles and Witten effects can appear in many familiar field
theories.}

\begin{document}
\maketitle


\section{Introduction and summary}
\label{sec:introduction}

\begin{figure}[h!tbp]
    \centering
    \includegraphics[width=0.9\textwidth]{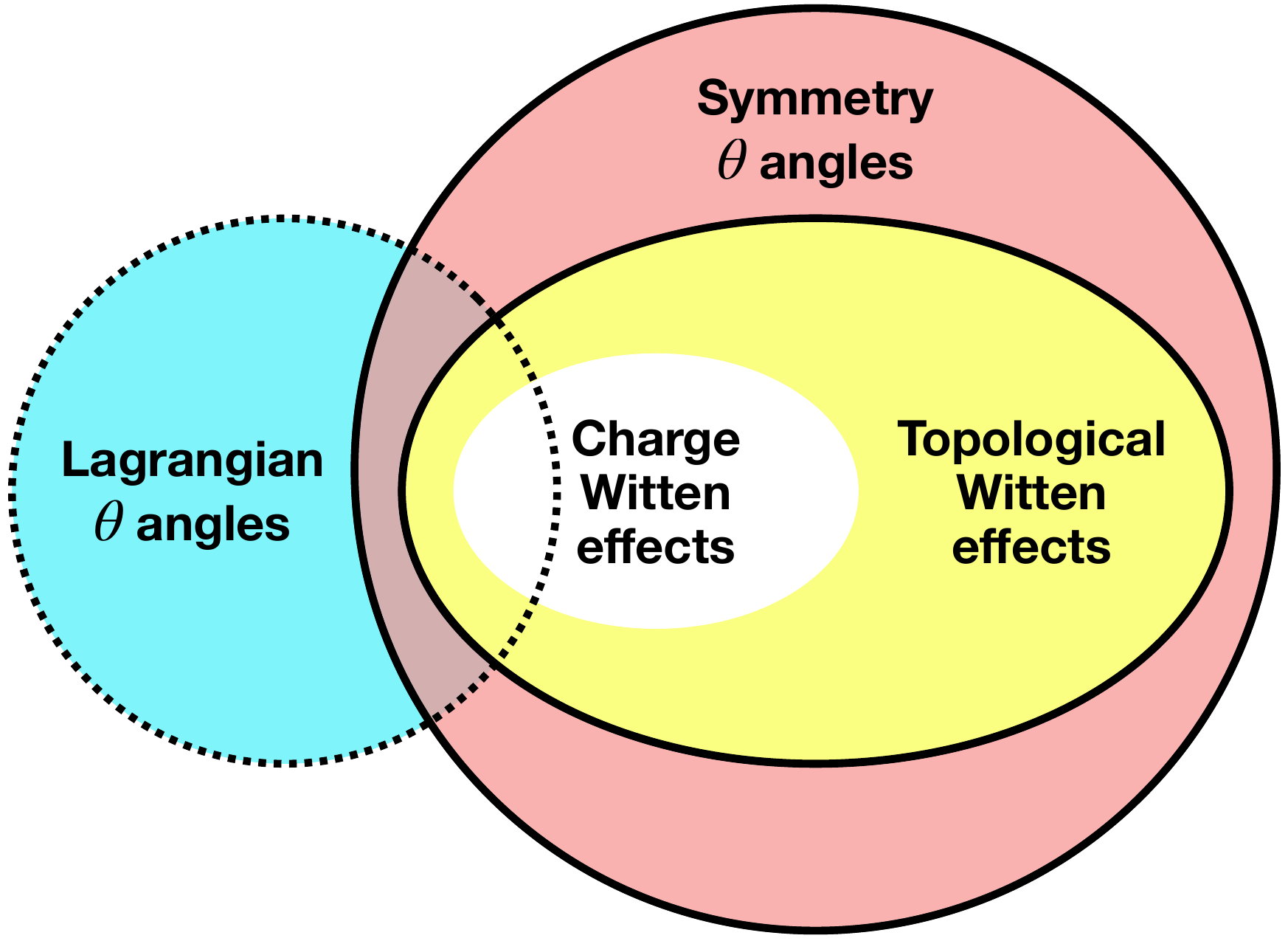}
    \caption{A Venn diagram that illustrates the relationship between Lagrangian
    $\theta$ angles, symmetry $\theta$ angles, and the two types of Witten
    effects. Witten effects can
    only appear as a consequence of symmetry $\theta$ angles, but some symmetry $\theta$
    angles do not induce Witten effects. The charge Witten effect --- which is
    what is normally called the Witten effect --- is always accompanied by the
    topological Witten effect.  But the topological Witten effect can appear
    without any accompanying charge Witten effect.}
    \label{fig:venn_witten}
\end{figure}

Many quantum field theories (QFTs) relevant to both high energy and condensed matter systems possess periodic coupling constants, known as $\theta$ angles, that play a pivotal role in their non-perturbative dynamics. 
In this paper we introduce a large new class of $\theta$ angles, which we call \emph{symmetry
$\theta$ angles}.  
Conventional $\theta$ angles are defined in terms of a particular Lagrangian representation of a QFT, and hence could be termed ``Lagrangian $\theta$ angles.''
In contrast, symmetry $\theta$ angles are Lagrangian-independent.
While in general symmetry $\theta$ angles are unconventional parameters of QFTs that are not visible in the historical approach to the study of $\theta$ angles, in some special cases symmetry $\theta$ angles can be interpreted as Lagrangian $\theta$ angles.  
We illustrate the relationship between Lagrangian and symmetry $\theta$ angles in Fig.~\ref{fig:venn_witten}.  
There are a number of reasons to expect symmetry $\theta$ angles to improve our understanding of non-perturbative aspects of QFT and to have observable consequences in nature. 
For example, many symmetry $\theta$ angles are discrete and hence must be preserved along renormalization-group (RG) flows. 
This means that they might be able to label universality classes of QFTs.


Our definition of symmetry $\theta$ angles is motivated by a reexamination of the well-known Witten effect~\cite{Witten:1979ey}. 
We find that the standard Witten effect of induced electric charge on magnetic monopoles is a special case of a more general --- and more
robust --- phenomenon, which we call the \emph{topological Witten
effect}.  The topological Witten effect changes which operators are
attached to topological symmetry operators.  More technically, it is a
locality-preserving reorganization of the vacuum and twisted sectors of
a global symmetry $G$ with a fixed $G$ charge. We initially discuss the
topological Witten effect and its interplay with the standard ``charge
Witten effect'' in the context of four-dimensional $U(1)$ gauge theory,
and then generalize it to generic QFTs. We find that the topological
Witten effect --- and hence also the standard charge Witten effect ---
shows up in the presence of symmetry $\theta$ angles.  We illustrate the
relationship between these Witten effects and $\theta$ angles in
Fig.~\ref{fig:venn_witten}.  A common physical consequence of the
topological Witten effect is the appearance of Aharonov-Bohm-like
effects for some excitations of the QFT.

We now give a more detailed explanation of the motivation and purpose of this paper using Maxwell quantum field theory, which can be defined on a Euclidean spacetime by a path integral with the action:
\begin{align}
    S[a;\theta] &= \frac{1}{4g^2} \int d^{4}x f_{\mu\nu} f^{\mu\nu} - \frac{i \theta}{32\pi^2} \int d^4{x} \,\epsilon_{\mu\nu\alpha\beta} f^{\mu\nu} f^{\alpha \beta}
    \\
    &=\frac{1}{2g^2}\int \d a\wedge\star \d a\, - \frac{\i\theta}{8\pi^2}\int \d a\wedge\d a \,,
    \label{eq:maxwell_action}   
\end{align}
where in the second line we use standard differential form notation.
One of the results motivating our analysis is that the Maxwell $\theta$ term in
Eq.~\eqref{eq:maxwell_action} can be defined without any reference to a
Lagrangian.  
Specifically, a Maxwell $\theta$ angle can be produced by a procedure known as an ``$S^{\dag}T S$ transformation'' in the literature, see e.g. Ref.~\cite{Choi:2022jqy}.  
At a technical level, the main idea of our paper is to explore the physical meaning of $S^{\dag}T S$ transformations.  We advocate the perspective that these transformations can be used to define a large class of generalized $\theta$ angles --- symmetry $\theta$ angles.  
Symmetry $\theta$ angles go beyond the standard classification of Lagrangian $\theta$ angles in QFT.   Lagrangian $\theta$ angles are classified by bordism groups of the homotopy target space associated with the fields in a path integral~\cite{Freed:2006mx,Seiberg:2010qd,Kapustin:2014gua,Freed:2016rqq,Freed:2017rlk,Cordova:2017vab,Yonekura:2018ufj,Hsin:2018vcg,Hsin:2019fhf,Hsin:2020nts,Chen:2023czk,Chen:2022cyw}.  In contrast, the classification of symmetry $\theta$ angles for a QFT with a non-anomalous symmetry relies on the classification of topological invertible field theories with the dual symmetry.  
We focus on Abelian invertible symmetries in this paper.  

The standard Maxwell $\theta$ term is a particularly simple example of a symmetry
$\theta$ angle.  
The symmetry in question is the $U(1)$ $1$-form magnetic symmetry $U(1)_m^{[1]}$, which acts on 't Hooft line operators~\cite{Gaiotto:2014kfa}.  
The topological Witten effect in 4d $U(1)$ gauge theory is a
$\theta$-dependent, locality-preserving reorganization of line operators of fixed $U(1)_m^{[1]}$ charge. The $\theta$ angle mixes 
$U(1)_m^{[1]}$-vacuum-sector operators (i.e., 't Hooft line
operators), with $U(1)^{[1]}_m$-twisted sector operators (i.e., 't Hooft lines attached to $U(1)_m^{[1]}$
symmetry-generator surface operators).  
The traditional
charge-fractionalization Witten effect of Ref.~\cite{Witten:1979ey}
appears when there is an \emph{additional} $1$-form symmetry with a
mixed 't Hooft anomaly with $U(1)^{[1]}_{m}$.  Pure Maxwell theory of
course has precisely such an extra symmetry, which is the electric $1$-form symmetry
$U(1)^{[1]}_{e}$, which acts on Wilson line operators.


\begin{figure}[!ht]
    \centering
    \includegraphics[width=0.95\textwidth]{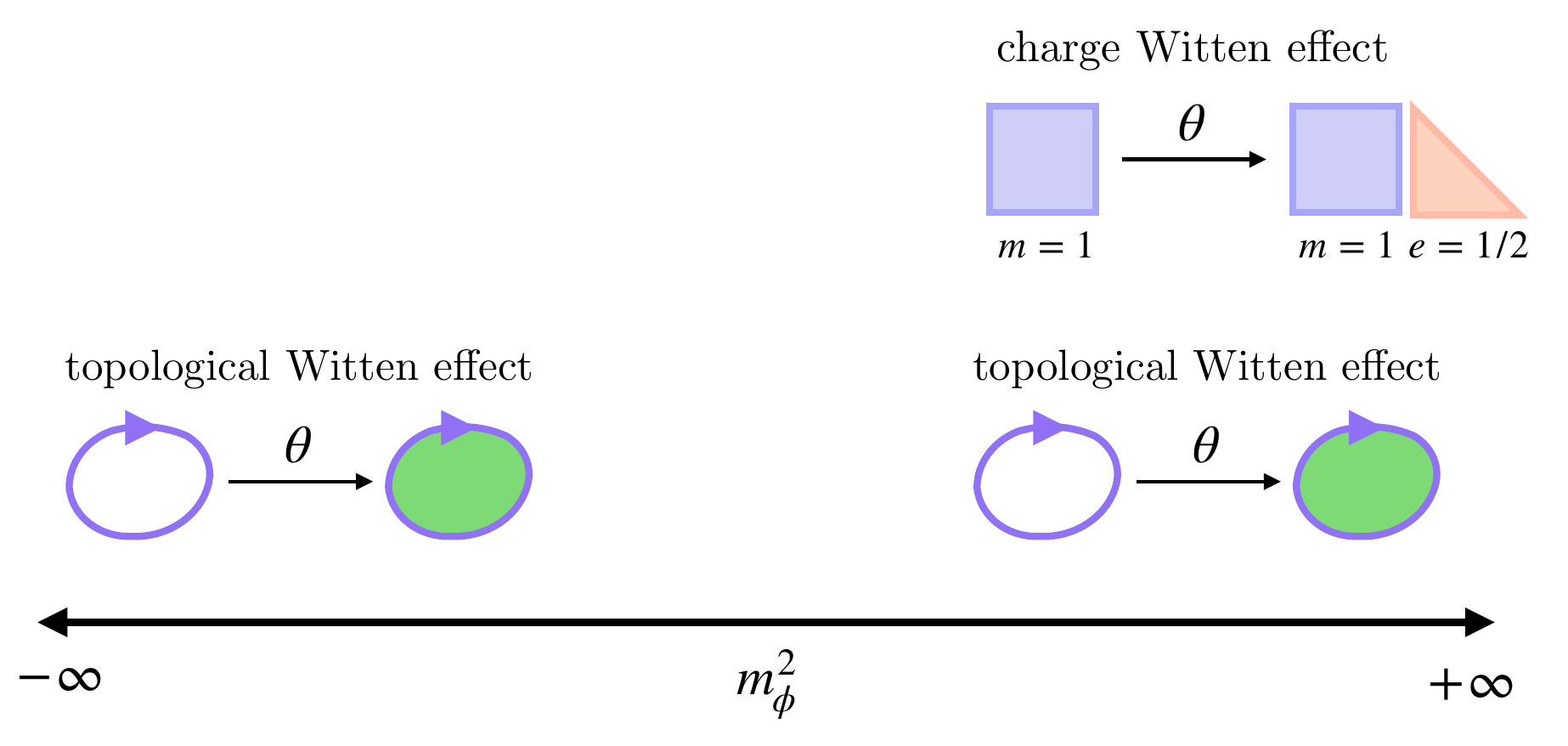}
    \caption{Comparison of Witten effects in a $U(1)$ gauge theory coupled to a
    unit-charge Higgs scalar with mass $m_{\phi}$ at $\theta = \pi$. The charge
    Witten effect (fractional electric charge for magnetic monopoles) vanishes
    in the Higgs phase where $U(1)^{[1]}_{e}$ is completely broken. The
    topological Witten effect persists since $U(1)^{[1]}_{m}$ remains unbroken,
    manifesting as Aharonov-Bohm phases for magnetic excitations. This
    demonstrates that the Maxwell $\theta$ term has physical consequences even
    when electric charge is screened.}
    \label{fig:topological_vs_charge_Witten}
\end{figure}

The $U(1)^{[1]}_{e}$ symmetry can be completely broken by coupling to
unit-charge electric matter. This can produce charge
screening and make $U(1)_e^{[1]}$ quantum numbers meaningless. When
this happens, the standard charge Witten effect gets obliterated. On the
other hand, the topological Witten effect depends solely on
$U(1)^{[1]}_{m}$ symmetry data.  It thus persists even when the
$U(1)^{[1]}_{e}$ symmetry is explicitly broken. Physically, the
topological Witten effect in 4d $U(1)$ gauge theory leads to
Aharonov-Bohm phases for magnetic string excitations moving around 't
Hooft line operators.  We illustrate the behavior of the two Witten
effects in 4d $U(1)$ gauge theory coupled to a unit-charge scalar field
in Fig.~\ref{fig:topological_vs_charge_Witten}. 

\subsection{Organization and highlights}

The organization of this paper is as follows:
\begin{itemize}
    \item We start by explaining the remarks above concerning 4d $U(1)$ theories and their topological and charge Witten effects in Section~\ref{sec:Witten_reappraisal}. This discussion serves as the inspiration for the definition of symmetry $\theta$ angles.
    
    \item We explain how to define symmetry $\theta$ angles for
    finite Abelian symmetries in
    Section~\ref{sec:general_discussion_finite}.  
    It turns out that topological
    Witten effects are always associated with symmetry $\theta$ angles.
    We then highlight some formal comments on our construction in Section~\ref{sec:Theta_vs_others}.
    
    \item We explain how to define symmetry $\theta$ angles for continuous Abelian symmetries in Section
\ref{sec:general_discussion_continuous}.   

    \item  Section~\ref{sec:lagrangian_theta} explains the relationship between conventional Lagrangian $\theta$ angles and symmetry $\theta$ angles, which can overlap in some special cases. The relationships between Lagrangian and symmetry $\theta$ angles and Witten effects are summarized in the Venn diagram in Fig.~\ref{fig:venn_witten}.

    \item We then give some examples of symmetry $\theta$
    angles and Witten effects in QFTs in three and four spacetime dimensions in Section~\ref{sec:examples}, and conclude some
    directions for future research in Section~\ref{sec:outlook}. 
\end{itemize}


We discuss examples of symmetry $\theta$
angles and Witten effects in QFTs in $d=3,4$ spacetime dimensions in Section~\ref{sec:examples}, and then highlight some
interesting directions for future research in Section~\ref{sec:outlook}.
The main aim of this paper is simply to introduce the idea of symmetry $\theta$ angles and highlight the fact that they appear in many familiar QFTs.  We thus give only a brief analysis for each example, leaving more detailed studies for the future.
To whet the reader's appetite for 
digging into this long and technical paper, we end this
introduction by highlighting a few interesting implications of our results:
\begin{itemize}
    \item Four-dimensional Maxwell theory has two symmetry $\theta$ angles; see Section~\ref{sec:Maxwell_revisit}. One of them is the standard $\theta$ angle, which turns out to be a $U(1)_m^{[1]}$ symmetry $\theta$ angle.  The other is a $U(1)_e^{[1]}$ symmetry $\theta$ angle. 
    
    \item The unconventional electric $\theta$ angle can implement a Kennedy-Tasaki transformation in a 4-dimensional charge-$N$ Abelian Higgs model, taking a phase where $(\Z_N)_e^{[1]}$ is spontaneously broken to a $(\Z_N)_e^{[1]}$-SPT phase, see Section~\ref{sec:electric_theta}. 

\item One might naively expect the Maxwell $\theta$ term to be unobservable in the unit-charge Higgs phase of a 4d $U(1)$ gauge theory on $\R^4$.  The reason for this expectation is that there are no $U(1)$ instantons on $\R^4$, and electric charge is ill-defined in the unit-charge Higgs phase, so there cannot be any standard Witten effect.  However, our results in Section~\ref{sec:cheshire_AB_phase} imply that the Maxwell $\theta$ angle leads to observable interference effects for magnetic strings in the Higgs phase, see Fig.~\ref{fig:topological_vs_charge_Witten}. 

\item The exotic --- but potentially measurable --- Cheshire
$\theta$ angle we recently introduced in Ref.~\cite{Chen:2024tsx} (with Gongjun Choi) in
the context of the Standard Model of particle physics coupled to a QCD axion is
actually a symmetry $\theta$ angle associated with a $U(1)^{[0]} \times
U(1)^{[2]}$ symmetry.  This is shown in Section~\ref{sec:cheshire_theta_4d}.  

\item The chiral effective field theory describing the long-distance limit of large $N_c$ QCD with light quarks has a so-far neglected symmetry $\theta$ angle.  It can induce a $U(1)_A$ charge of $\theta/(2\pi)$ for Skyrmions, see Section~\ref{sec:largeN_QCD}.

\item Recognizing that some conventional Lagrangian $\theta$ angles are actually symmetry $\theta$ angles (see e.g.~Section~\ref{sec:lagrangian_theta}) allows us to track them across duality transformations.  We give examples of this in Section~\ref{sec:cheshire_theta_3d}.

\item A generalization of our construction can be used to couple magnetic matter to electric $U(1)$ gauge fields without moving to the magnetic-dual frame.  It can also be used to construct a dual representation of Maxwell-Chern-Simons theory.  These points are discussed in Section~\ref{sec:U(1)_remark}.

\item In this paper we adopt a slightly different perspective on the global structure of 4d gauge theories
than is usually advocated in the literature.
For example, our approach suggests that there are two distinct $SU(2)$ gauge theories, see Section~\ref{sec:4d_YM}. These two $SU(2)$ gauge theories differ by e.g.~whether confining string intersections come with a phase of $-1$ or not.  
Our discussion might also have implications for the global structure of the Standard Model of particle physics, whose symmetries include a $\Z_6^{[1]}$ symmetry; see Section~\ref{sec:outlook}.
\end{itemize}

\section{The Maxwell \texorpdfstring{$\theta$}{theta} angle from a new angle}
\label{sec:Witten_reappraisal}

We start the body of this paper with a careful review of the $\theta$ angle in 4-dimensional $U(1)$ gauge theories, which we shall often abbreviate as
the Maxwell $\theta$ angle. This is a textbook topic, but we shall
discuss it in a somewhat unconventional way. Our approach allows us to
highlight some features that are not normally emphasized in the
literature.  This will let us isolate the topological Witten effect and
see its descendants in later sections.  

After reviewing the classic Witten effect in Section~\ref{sec:Maxwell_Witten}, we explain its reformulation in terms of a surface-attachment picture in Section~\ref{sec:operator_shuffle}.  We then discuss the effect of the Maxwell $\theta$ angle on string excitations in Section~\ref{sec:Maxwell_string_spectrum}, and define the topological and charge Witten effects in Section~\ref{sec:topologicalWitten_vs_chargeWitten}.  We close by explaining some applications of our results to Higgs-phase physics and a non-invertible chiral symmetry in Section~\ref{sec:topologicalWitten_consequences}.

\subsection{The classic Witten effect}
\label{sec:Maxwell_Witten}

Consider a $U(1)$ gauge theory without matter in four spacetime dimensions ---
often referred to simply as \emph{Maxwell theory}. We will work in Euclidean
signature throughout this paper. Maxwell theory with a topological $\theta$
angle can be described by a path integral over a $U(1)$ gauge field $a =
a_{\mu} dx^{\mu}$ defined
on a closed, orientable Euclidean spacetime manifold $X$,
\begin{align}
    \int\calD a\ \exp\left\{ -\,\frac{1}{2g^2}\int_X \d a\wedge\star \d a\,
    +\frac{\i\theta}{8\pi^2}\int_X \d a\wedge\d a\right\}\,.
    \label{eq:maxwell_path_integral}
\end{align}
The quantization condition for the topological charge is coarser on spin manifolds than on generic oriented manifolds:
\begin{equation}
    \int_X \frac{\d a}{2\pi}\wedge\frac{\d a}{2\pi}\ \in\,\begin{cases}
        \Z\,, &\text{on oriented manifolds}\\
        2\Z\,, &\text{on spin manifolds}
    \end{cases}\,.
\end{equation}
Thus, on generic oriented manifolds, the periodicity of $\theta$ is
$4\pi$, see e.g.~Refs.~\cite{Thorngren:2014pza,
Kravec:2014aza,Wang:2018qoy,Ang:2019txy,Kan:2024fuu}. We will focus on
Maxwell theory on spin manifolds, where the periodicity is
$2\pi$.\footnote{A practical motivation for focusing on spin manifolds
is that all the simple manifolds usually considered in particle physics
such as $\R^4, S^4, T^4$ and so on are spinnable.  
The simplest non-spin orientable 
4-manifold is $\mathbb{C}P^2$.} Nevertheless, we emphasize that
Eq.~\eqref{eq:maxwell_path_integral} has no dependence on the spin
structure of $X$, and the Hilbert space
contains bosonic states only.

\subsubsection{On line operators: 't Hooft lines become dyonic}
\label{sec:Maxwell_Witten_on_line}

Maxwell theory possesses two $U(1)$ 1-form symmetries, an
electric symmetry $U(1)^{[1]}_{e}$ and a magnetic symmetry $U(1)^{[1]}_{m}$,
which are both spontaneously broken on $\R^4$.  
The electric symmetry arises from the invariance of the action under the shift 
$a\to a+2\pi\lambda$, where $\lambda$ is any closed 1-form (i.e. $\d\lambda=0$).  
In contrast, the magnetic symmetry is a consequence of the Bianchi identity,
$\d^2 a = 0$. The associated symmetry generators are topological operators
defined on closed 2-manifolds $\Sigma$, and are given by
\begin{subequations}
\begin{align}
    U(1)^{[1]}_{m}: \qquad U_{m}(\alpha; \Sigma) &=
     \exp\left( \i\alpha\int_{\Sigma}\frac{\d a}{2\pi} \right)\,,\;\; 
     \alpha\in\frac{\R}{2\pi\Z}\,,\label{eq:U(1)m-operator}\\
    U(1)^{[1]}_{e}: \qquad U_{e}\,(\beta; \Sigma)\, &= 
    \exp\left[\i\beta\int_{\Sigma}\left( -\frac{\i}{g^2}\star\d a 
    - \frac{\theta}{4\pi^2}\d a \right) \right]\,,\;\; \beta\in\frac{\R}{2\pi\Z}\,.
    \label{eq:U(1)e-operator}
\end{align}
\end{subequations}
The parameter $\alpha$ is $2\pi$-periodic thanks to Dirac quantization,
$\int_{\Sigma} \d a = 2\pi \Z$. The parameter $\beta$ is also $2\pi$-periodic. One
way to see this is to notice that shifts by $\lambda$'s with integral periods
give rise to gauge redundancies, i.e., $\lambda$ parametrizes a $U(1)$ rather than an $\R$ symmetry. 
Alternatively, one may show that $\beta$ is $2\pi$-periodic by invoking electromagnetic duality, which relates the quantity in parentheses in Eq.~\eqref{eq:U(1)e-operator} to a dual $U(1)$ gauge field obeying the Dirac quantization condition. 
The  $U(1)$ nature of the symmetries $U(1)^{[1]}_{m}$ and $U(1)^{[1]}_{e}$ means that 't Hooft and Wilson line
operators have integral charges
\begin{equation}
    m,e\in\Z
\end{equation}
under $U(1)^{[1]}_{m}$ and $U(1)^{[1]}_{e}$.

\begin{figure}[t]
\centering
\begin{subfigure}[t]{0.47\textwidth}
\centering
\includegraphics[width = 0.9 \textwidth]{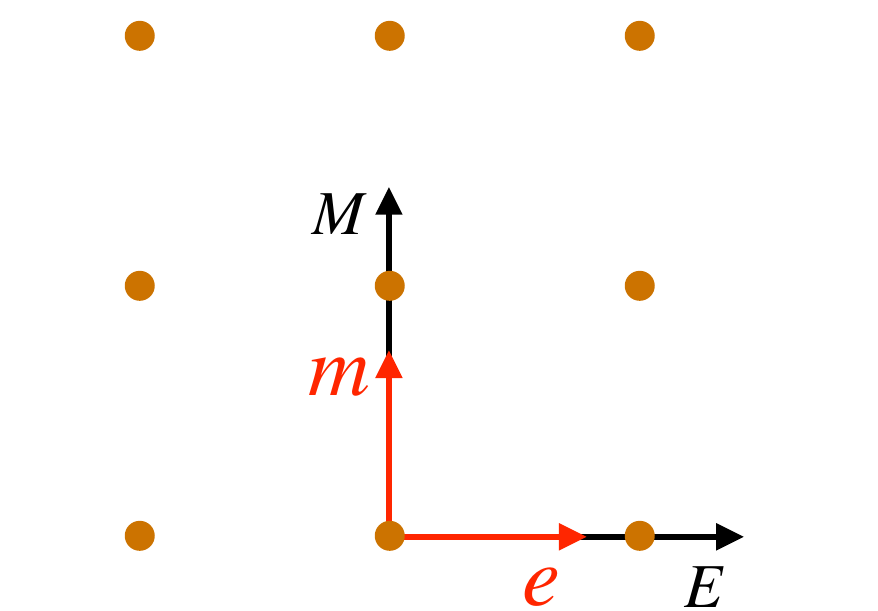}
\caption{$\theta = 0$}
\end{subfigure}
\hfill
\begin{subfigure}[t]{0.47\textwidth}
\centering
\includegraphics[width = 0.88 \textwidth]{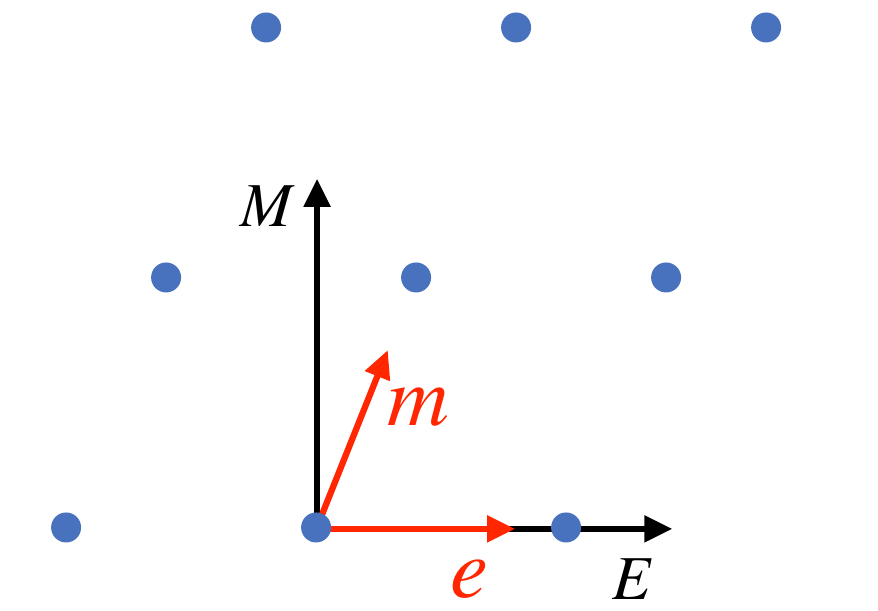}
\caption{$\theta \neq 0$}
\end{subfigure}
\caption{Physical charge lattices of 4d $U(1)$ Maxwell gauge theory.
Left: At $\theta = 0$, the physical and symmetry charges align. Right:
For $\theta \neq 0$, the physical electric charge becomes fractional for
magnetic monopoles. Generic points in both planes represent non-genuine
line operators. The dots on each grid represent genuine line operators
of that theory, which require no surface attachments. The $E,M$ axes
show physical electric and magnetic charges, while the $e,m$ axes
represent the $U(1)_{e,m}^{[1]}$ symmetry charges.}
\label{fig:charge-lattice}
\end{figure}

We next recall that inserting a line operator along the time direction
effectively corresponds to introducing an infinitely heavy probe
particle. The static electromagnetic field surrounding a probe particle
carries a magnetic flux $M$ and electric flux $E$
\begin{equation}
    M=\int\frac{\d a}{2\pi} = m\qquad\text{ and }\qquad 
    E=\int -i\frac{\star\d a}{g^2} = e+\frac{\theta}{2\pi}m\,.
    \label{eq:GaussCharges}
\end{equation}
We refer to $E$ and $M$ as the \textit{physical} charges, since they are calculated from the electric and magnetic versions of Gauss's law.
It is the physical charges that determine the behavior of the correlation functions of a line operator.
Because the $U(1)_{e,m}^{[1]}$ charges of all line operators are always integers, their physical electric charges $E$ are fractional when $\theta \notin 2\pi \Z$.
In particular, 't Hooft lines acquire a $\theta$-induced $E$ and become dyonic.

The phenomenon described above is known as the Witten effect.  
The Witten effect can be visualized by comparing the physical-charge lattices of
line operators at $\theta=0$ and at $\theta\neq 0$, which are shown in
Fig.~\ref{fig:charge-lattice}.
Despite the fractional nature of the physical charges, they always
satisfy the Dirac-Schwinger-Zwanziger quantization condition for any
$\theta$,
\begin{equation}
    M_1 E_2 - M_2 E_1\ \in \mathbb{Z}\,.
\end{equation}
This condition reflects a deep consistency requirement for
the quantum theory: the correlation functions of
\textit{genuine} line operators must remain single-valued.
A non-genuine line operator that allows a non-trivial monodromy must be attached to an appropriate \textit{topological} surface operator which can accommodate a branch cut.
Hence genuine line operators live in the ``vacuum sector'' while non-genuine line operators live in a ``twisted sector,'' which is twisted by its topological surface.
\subsubsection{On stringy excitations: magnetic fluxes become dyonic}
\label{sec:Maxwell_string_spectrum}

While photons --- the familiar excitations of Maxwell theory --- are blind to $\theta$, the theory contains another class of excitations that are sensitive to $\theta$: electromagnetic fluxes.
Fluxes are stringy dynamical objects that carry charges under the 1-form symmetries. Their properties reveal the physical impact of the Witten effect.
However, when the spacetime manifold $X = \R^4$, the $U(1)_{m}^{[1]}$ and $U(1)_{e}^{[1]}$ symmetries are spontaneously broken. 
This makes the fluxes tensionless, which then show up in the form of a spin-1 Goldstone boson, i.e.~the massless photon~\cite{Kovner:1992pu,Gaiotto:2014kfa}.
As a result, the effect of the $\theta$ parameter on flux strings in $\R^4$ in pure Maxwell theory is opaque.

Things become much clearer if we make the string tensions finite.  
This can be accomplished by taking spacetime to be a closed manifold such as
$X\equiv S^2\!\times\!S^1_{L_s}\!\times\!S^1_{L_t}$, where $S^2$ is a sphere
with an arbitrary regular metric, while $S^1_{L_s}$ and $S^1_{L_t}$ are circles
with length $L_s$ and $L_t$, respectively.  
We take $S^1_{L_t}$ as the Euclidean temporal direction.  
Since $X$ is compact, the $U(1)_{ e,m}^{[1]}$ symmetries are not spontaneously
broken on $X$. This particular choice of $X$ also has a very convenient topology
for calculations.
Evaluating the Maxwell path integral on this spacetime $X$, we can obtain a partition function to extract information about the flux strings.

To correctly identify the species of fluxes, we turn on flat background $U(1)^{[1]}_{m}$ and $U(1)^{[1]}_{e}$ gauge fields $A_m$ and $A_e$ given by
\begin{align}\label{eq:temp_holonomies}
    A_e=\frac{\beta_e}{L_s L_t}\, \d x_s\wedge \d x_t\,, \qquad A_m= \frac{\beta_m}{L_s L_t} \, \d x_s\wedge \d x_t\,.
\end{align}
They have zero curvature but non-trivial holonomies on $S^1_{L_s}\!\times\! S^1_{L_t}$,
\begin{equation}
    \int_{S^1_{L_s}\!\times\! S^1_{L_t}} A_e = \beta_e\,,\qquad \int_{S^1_{L_s}\!\times\! S^1_{L_t}} A_m = \beta_m\,.
\end{equation}
Hence a flux stretched on $S^1_{L_s}$ with $U(1)^{[1]}_{e,m}$ charges $e,m$ will get a phase factor $e^{\i\beta_e e+\i\beta_m m}$ in the partition function, so that the parameter $\beta$ acts as a dimensionless imaginary chemical potential. 
The partition function with background fields $A_{m,e}$ is given by
\begin{equation}\label{eq:u1_bckgd_fields}
\begin{gathered}
    \int\calD a\ \exp\left\{ -\,\frac{1}{2g^2}\int_X (\d a\!-\!A_e)\wedge\star(\d a\!-\!A_e)\right. \\
    \left. +\frac{\i\theta}{8\pi^2}\int_X (\d a\!-\!A_e)\wedge(\d a\!-\!A_e)\,+ \frac{\i}{2\pi}\int_X A_m \wedge \d a  \right\} \,.
\end{gathered}
\end{equation}
The last term in the action is not (background) gauge invariant and reflects the
mixed 't Hooft anomaly between $U(1)_{e}^{[1]}$ and $U(1)_{m}^{[1]}$. 

Since the action is quadratic, we can evaluate the partition function analytically. 
First, the path integral decomposes into a sum over topological sectors labeled by the magnetic fluxes of the dynamical gauge field on $S^2\!\times\!S^1_{L_s}\!\times\!S^1_{L_t}$.  
These flux sectors are classified by
\begin{equation}\label{eq:(m,n)}
    (m,n)\in H^2\Bigl( S^2\!\times\!T^2,\,\Z \Bigr) = \Z\times\Z\,.
\end{equation}
To evaluate the path integral in each topological sector, we need to find the saddle point of the action, and then sum over fluctuations around this saddle point.
The saddle-point configurations are independent of $A_{e,m}$, and their curvatures are given by
\begin{align}\label{eq:da_ansatz}
    \d a_{m,n}=\frac{2\pi m}{|S^2|}\,\d\Omega + \frac{2\pi n}{L_s L_t} \,\d x_s\wedge \d x_t \,,
\end{align}
where $\d\Omega$ denotes the volume 2-form on the sphere such that $\int_{S^2}\d\Omega = |S^2|$ gives the area of the 2-sphere.   
There are two classes of fluctuations: zero modes which represent the fact that there are many solutions with the same action, and one-loop fluctuations that have a finite action cost.  
In our case, the zero modes come from the holonomies of $a$, which are not fixed by the equations of motion.
To get the contribution of {\it all} solutions with the flux \eqref{eq:da_ansatz} we have to sum over the holonomies.  
As usual, the remaining one-loop integral is given by a functional determinant.

Performing the computation  described above, after some algebra we obtain the partition function and express it as a (graded) thermal partition function with an inverse temperature $L_t$ associated with the Hilbert space of states on
$S^2\!\times\!S^1_{L_s}$:
\begin{equation}
\label{eq:Maxwell_Z}
    \calZ_{\beta_m,\beta_e}(\theta,X)=\calF(X)\!\sum_{m,e\in\Z}\exp\Bigl\{ - L_t L_s T_{\rm string}(e,m, \theta)  + 
    \i\beta_m m + \i\beta_e e  \Bigr\}\,\,,
\end{equation}
where the $e$-sum is the Poisson resummation of the $n$-sum implied by Eq.~\eqref{eq:(m,n)}.
The appearance of $\exp(\i\beta_m m + \i\beta_e e)$ shows that $m,e$ precisely label $U(1)_{m,e}^{[1]}$ charges.
The function $\calF(X)$ comes from the one-loop and zero-mode contributions together with an overall factor from the Poisson resummation. 
$\calF(X)$ does not depend on $\theta$ or $g$, and is the temperature-$1/L_t$ thermal partition function of a free massless spin-1 boson on the spatial manifold $S^2\!\times\! S^1_{L_s}$ (with the Casimir energy included). 
The function $T_{\rm string}$ is defined as
\begin{align}\label{eq:T_string}
    T_{\rm string}(e,m, \theta) = \frac{1}{2|S^2|}\left[\frac{4\pi^2}{g^2}m^2 + g^2\left(e +\frac{\theta}{2\pi}m\right)^2 \right] \,.
\end{align} 
$T_{\rm string}$ is the tension of a stringy excitation stretched along $S^1_{L_s}$, so that $L_s T_{\rm string}$ gives its energy.
We thus see that the compact transverse area $|S^2|$ confines the electromagnetic fluxes and gives them finite tensions.%
\footnote{ In general, the tension of stringy excitations would depend on the details of the metric on $S^2$. 
The fact that the string tensions in Eq.~\eqref{eq:Maxwell_Z} only care about the area of $S^2$ is a special property of Maxwell theory. }
In particular, we learn from Eq.~\eqref{eq:T_string} that the tensions of string excitations are determined by their {\it physical} charges $M=m$ and $E=e+\frac{\theta}{2\pi}m$ rather than the raw $U(1)_{m,e}^{[1]}$ symmetry charges $m, e$.   
This is precisely a reflection of the Witten effect on the electromagnetic fluxes.

\subsection{Witten effects in generic \texorpdfstring{$U(1)$}{U(1)} gauge theories}
\label{sec:generic_U(1)_gauge_theories}

So far we have focused on pure Maxwell theory, where both $U(1)^{[1]}_e$ and $U(1)^{[1]}_m$ play a role in our description of the Witten effect.
However, we can still write down the Maxwell $\theta$ term in the presence of electric matter fields.
Now the $U(1)^{[1]}_e$ symmetry is explicitly destroyed by the electric matter while  $U(1)^{[1]}_m$ survives.
We now try to formulate the implication of the $\theta$ term using $U(1)^{[1]}_m$ only, without referring to $U(1)^{[1]}_e$ at all.

\subsubsection{On line operators: attaching a magnetic surface}
\label{sec:operator_shuffle}

We first examine the Witten effect on line operators.%
\footnote{As we were finalizing this paper, we learned that the observations in this section were also independently made and very thoroughly explored in Ref.~\cite{GarciaGarcia:2025uub}. }
Let us put closed line operators on the closed manifold $S^4$, which is just an infrared-regulated version of $\R^4$.  
Although the topological charge of $a$ vanishes on $S^4$, 
as we discussed in Section~\ref{sec:Maxwell_Witten_on_line}, the $\theta$ term affects correlation functions of line operators in the Maxwell theory.
Our goals here are to show that this fact is also true in a generic $U(1)$ gauge theory and to find a way to precisely quantify this effect using $U(1)^{[1]}_m$ only.

Let us consider the correlation function $\langle T_m(C)\cdots\rangle_{\theta}$ involving a genuine 't Hooft line operator $T_m(C)$ carrying a $U(1)^{[1]}_{m}$ charge $m$, along with other operators with zero $U(1)_m^{[1]}$ charge represented by the ellipsis.  
We will show that 
\begin{align}
    \Bigl \langle T_m(C)\cdots \Bigr \rangle_{\theta} 
    = \Bigl\langle T_m(C)\exp\left(\frac{\i\theta m}{2\pi}\int_{D}\d a\right)
    \cdots \Bigr \rangle_{0}\,,
    \label{eq:shuffle-operator}
\end{align}
for an arbitrary compact surface $D$ such that $C=\partial D$.
The surface operator defined on $D$ above is precisely a topological operator of $U(1)^{[1]}_m$, and thus Eq.~\eqref{eq:shuffle-operator} contains information about $U(1)^{[1]}_m$ only.
Equation~\eqref{eq:shuffle-operator} is extremely simple to motivate, because a heuristic application of Stokes' theorem would imply
\begin{equation}
    exp\left(\frac{\i\theta m}{2\pi}\int_{D}\d a\right) \ \text{``}\!=\!\text{''}\  \exp\left(\frac{\i\theta m}{2\pi}\int_{C} a\right)\,.
\end{equation}
Hence Eq.~\eqref{eq:shuffle-operator} can be interpreted as saying that magnetic monopoles pick up a fractional electric charge, which is just the classic Witten effect of Ref.~\cite{Witten:1979ey}.
The virtue of Eq.~\eqref{eq:shuffle-operator} is that it holds \emph{regardless} of whether $U(1)_{e}^{[1]}$ charge is a good quantum number, and the presence of the topological surface on the right-hand side has very important physical effects that cannot be ignored. 

To see how Eq.~\eqref{eq:shuffle-operator} arises from a rigorous calculation, let us view $T_m(C)$ as a defect and thicken its support $C$ into a small tubular neighborhood, namely, $C \times B^3_{\epsilon}$, where $B^3_{\epsilon}$ denotes a small open 3-ball.
Removing this tubular neighborhood from $S^4$ leaves a boundary $C\times S^2_{\epsilon}$ in the remaining spacetime, where $S^2_{\epsilon}$ is a small sphere linking $C$.  
The charge $m$ is encoded in the condition $\frac{1}{2\pi} \int_{S^2_{\rm \epsilon}} \d a = m$.  

Because the manifold $S^4 \setminus (C \times B^3_{\epsilon})$ now has a boundary, the Maxwell topological charge is no longer quantized.  
Nonetheless, its topological nature is preserved.
Namely, the $\theta$ term remains invariant under smooth deformations of the gauge field that maintain the prescribed boundary condition on $C\times S^2_{\epsilon}$. 
This means that the Maxwell $\theta$ term associated with a given field configuration can be expressed almost entirely through the boundary data.  
Specifically, choosing an arbitrary surface $D$ with $\partial D = C$, we can rewrite
\begin{align}
    \exp\left\{\frac{\i\theta}{8\pi^2}\int_{S^4 \setminus (B^3_{\epsilon} \times C)}
     \d a\wedge\d a\right\} 
    &=\:\exp\left\{\frac{\i\theta}{4\pi^2}\left(\int_{S^2_{\epsilon}}\d a\right)
    \left(\int_{D}\d a\right)\right\} \nonumber \\
    &= \exp\left(\frac{\i\theta m}{2\pi}\int_{D}\d a\right)\,.
    \label{eq:theta_boundary}
\end{align}
We thus see that the $\theta$ term can be removed from the action at the
cost of inserting the right-hand side of Eq.~\eqref{eq:theta_boundary}
into the correlation function.  Since the partition function on $S^4$ is
insensitive to $\theta$, the correlation functions of line operators at
generic $\theta$ and $\theta=0$ are related as given by
Eq.~\eqref{eq:shuffle-operator}.



\subsubsection{On stringy excitations: shuffling magnetic twists}
\label{sec:genericU1_Witten}

We now examine the Witten effect on flux strings.
Let us start from Maxwell theory on the spacetime $S^2\!\times\!S^1_{L_s}\!\times\!S^1_{L_t}$ as we discussed in Section~\ref{sec:Maxwell_string_spectrum}.
We modify the background gauge field $A_m$ given in Eq.~\eqref{eq:temp_holonomies} into 
\begin{equation}\label{eq:A_m}
    A_m= \frac{\alpha}{|S^2|} d\Omega + \frac{\beta_m}{L_s L_t} \, \d x_s\wedge \d x_t\,, 
\end{equation}
which adds a non-trivial holonomy on $S^2$, 
\begin{equation}
    \int_{S^2} A_m = \alpha\,.
\end{equation}
It thus puts a static background gauge field into  the system, i.e.~activates a $U(1)^{[1]}_{m}$-twist.
Evaluating the Maxwell path integral with this $A_m$, we obtain the same partition function as Eq.~\eqref{eq:Maxwell_Z}, except that the string tension now receives an extra contribution from the twist $\alpha$,
\begin{align}\label{eq:T_string_alpha}
    T_{\rm string}(e,m, \theta; \alpha) = \frac{1}{2|S^2|}\left[\frac{4\pi^2}{g^2}m^2 + g^2\left(e +\frac{\theta}{2\pi}m +\frac{\alpha}{2\pi}\right)^2 \right] \,.
\end{align}
We see that a $U(1)^{[1]}_{m}$ twist can also induce a fractional physical electric charge,
\begin{equation}
    E\to E+\frac{\alpha}{2\pi}.
\end{equation}
This is actually a consequence of the mixed 't Hooft anomaly between
$U(1)^{[1]}_{e}$ and $U(1)^{[1]}_{m}$, which  
requires the sector with $U(1)^{[1]}_{m}$ twist $\e^{\i\alpha}$ to acquire a
fractional physical electric charge $\alpha/{2\pi}$, and vice
versa~\cite{Gaiotto:2014kfa}.

Based on the observation above, we can now describe the $\theta$ effect on flux strings without mentioning anything about $U(1)^{[1]}_e$ at all.
Let $\calH^{\alpha}\left(\theta,\,S^2\!\times\!S^1_{L_s}\right)$ denote the Hamiltonian of the Hilbert space on $\,S^2\!\times\!S^1_{L_s}$ with the $U(1)^{[1]}_m$-twist $\alpha$.
Since $U(1)_m^{[1]}$ is a well-defined symmetry of the quantum theory, states with different charges cannot mix, and therefore the Hilbert space decomposes into a sum of $U(1)_m^{[1]}$-charged subsectors.
The Hamiltonian $\calH^{\alpha}$ also decomposes into the Hamiltonians of these subsectors.
\begin{equation}
    \calH^{\alpha}\left(\theta,\,S^2\!\times\!S^1_{L_s}\right) = 
    \sum_{m\in\Z}\calH_m^{\alpha}\left(\theta,\,S^2\!\times\! S^1_{L_s}\right)\,.
\end{equation}
Comparing the form of Eq.~\eqref{eq:Maxwell_Z} with $\theta = 0$ and $\theta \neq 0$ reveals the isomorphism
\begin{align}
    \calH_m^{\alpha}\left(\theta,\,S^2\times S^1_{L_s}\right) \simeq 
    \calH_m^{\alpha+\theta  m}\left(0,\,S^2\times S^1_{L_s}\right),\qquad \forall\, m\in\Z\,.
\label{eq:hamiltonian_shuffle}
\end{align}
In other words, the charge-$m$ subsector in the $U(1)^{[1]}_{m}$ $\e^{\i\alpha}$-twisted sector at topological angle $\theta$ is equivalent to the charge $m$ $\e^{\i(\alpha+\theta m)}$-twisted sector in the theory with $\theta=0$.  
This Hamiltonian perspective on the Witten effect is the counterpart of the operator shuffling discussed in Section~\ref{sec:operator_shuffle}.


We now want to derive the spectrum shuffle~\eqref{eq:hamiltonian_shuffle} above in an \textit{arbitrary} $U(1)$ gauge theory with electrically-charged matter fields. 
If charged fields are collectively represented by $\Psi$, we can write a formal expression for the path integral of the gauge theory as 
\begin{align}\label{eq:Abelian_path_integral}
\begin{gathered}
    \int\calD a\calD\Psi\ \exp\left\{ -\,\frac{1}{2g^2}\int_X \d a\wedge\star \d a\, - \calS\bigl( \Psi, a \bigr) \right.\\
    \left.+\frac{\i\theta}{8\pi^2}\int_X \d a\wedge\d a \,+ \frac{\i}{2\pi}\int A_m \wedge \d a   \right\}\,,
\end{gathered}
\end{align}
where $\calS\bigl( \Psi, a \bigr)$ is a (gauge-invariant) matter action and
$A_m$ is a background $U(1)^{[1]}_{m}$ gauge field.  
The presence of the electrically-charged $\Psi$ fields means that there is no
longer  
a $U(1)^{[1]}_{e}$ global symmetry at the microscopic level.  

Let us evaluate the partition function again on $X\equiv S^2\!\times\!S^1_{L_s}\!\times\!S^1_{L_t}$ with the same $A_m$ given by Eq.~\eqref{eq:A_m}.
Because $U(1)$ gauge fields admit nontrivial flux configurations,  
the partition function naturally decomposes into a sum over distinct topological
sectors, just as in pure Maxwell theory.  
Each sector is labeled by discrete flux quantum numbers $(m,n)$ given by Eq.~\eqref{eq:(m,n)}.%
\footnote{ If $\Psi$ is a scalar with a nonlinear target space the topological
sectors can be more refined, but this does not alter our argument. }  
Let us denote the contribution to the path integral from the top line of Eq.~\eqref{eq:Abelian_path_integral} in each topological sector by
\begin{equation}\label{eq:Z_m,n}
    \calZ_{m,n}(X)\,.
\end{equation}
In pure Maxwell theory we were able to exactly evaluate $ \calZ_{m,n}(X)$, but
this is no longer possible once charged matter is included. Nevertheless, the
bottom line of Eq.~\eqref{eq:Abelian_path_integral} is purely topological in the
sense that it only depends on the topological numbers $(m,n)$, so we see that
the full partition function $\calZ^{\alpha}_{\beta}(\theta,X)$ depends on
$\calZ_{m,n}(X)$ via 
\begin{equation}\label{eq:sum-Z_m,n}
\begin{split}
    \calZ^{\alpha}_{\beta}(\theta,X)  &=\sum_{m,n\in\Z}\e^{\i\beta m}\e^{\i n(\theta m + \alpha
    )}\,\calZ_{m,n}(X)\,.
\end{split}
\end{equation}
The Hamiltonian interpretation of $\calZ^{\alpha}_{\beta}(\theta,X)$ is 
\begin{equation}
    \calZ^{\alpha}_{\beta}(\theta,X) = \sum_{m\in\Z}\e^{\i\beta m }\,\tr\exp\biggl[ - L_t \calH_m^{\alpha}\left(\theta,\,S^2\!\times\!S^1_{L_s}\right) \biggr]\,,
\end{equation}
where $\calH_m^{\alpha}\left(\theta,\,S^2\!\times\!S^1_{L_s}\right)$ is the
Hamiltonian in the charge-$m$ subsector of the $\e^{\i\alpha}$-twisted sector on
the spatial manifold $S^2\!\times\!S^1_{L_s}$.  Therefore
\begin{equation}
    \tr\exp\biggl[ - L_t \calH_m^{\alpha}\left(\theta,\,S^2\!\times\!S^1_{L_s}\right) \biggr] = \sum_{n\in\Z}\e^{\i n(\theta m + \alpha
    )}\,\calZ_{m,n}(X)\,,\qquad m\in\Z\,.
\end{equation}
This implies the isomorphism
\begin{equation}\label{eq:spectrum_shuffle}
    \calH_m^{\alpha}\left(\theta,\,S^2\!\times\!S^1_{L_s}\right) \simeq \calH_m^{\alpha+\theta m}\left(0,\,S^2\!\times\!S^1_{L_s}\right)\,,\qquad m\in\Z\,.
\end{equation}
Thus, even when electrically-charged matter fields are present,  
the physical effect of $\theta$ is fully captured by a shuffle of $U(1)^{[1]}_{m}$ twists, holding the charges $m$ constant.

\subsection{The topological Witten effect versus the charge Witten effect}
\label{sec:topologicalWitten_vs_chargeWitten}

We are now in a position to summarize what we have learned about the $\theta$
parameter of pure Maxwell theory and to introduce some useful terminology.  As
discussed in Section~\ref{sec:Maxwell_Witten}, one effect of the $\theta$
parameter is to induce a fractional electric charge on magnetic monopoles. In
the literature, this phenomenon is almost universally called the Witten effect,
and indeed we have done so at the beginning of this section. However, from here
onward, we will refer to it as the \emph{charge Witten effect}:
\begin{definition}[Charge Witten effect]
    Line operators that have physical electric and magnetic charges $(E,M)$ at
    $\theta = 0$ pick  up a modified charge $(E+ \frac{\theta}{2\pi}M, M)$ at
    $\theta\neq 0$.
\end{definition}
It is often assumed that the charge Witten effect is the only effect of the
$\theta$ parameter in Maxwell theory. However, as discussed in
Section~\ref{sec:generic_U(1)_gauge_theories}, the $\theta$ parameter also induces
the twisted-sector shuffle effect.  We refer to this as the \emph{topological
Witten effect}:
\begin{definition}[Topological Witten effect]
    Line operators at $\theta\neq 0$ with $U(1)^{[1]}_{m}$ charge $m$ and twist $\alpha$ correspond to line operators at
    $\theta= 0$ with $U(1)^{[1]}_{m}$ charge $m$ and twist $\alpha+\theta m$ .  
\end{definition}

The charge Witten effect and the topological Witten effect always appear
together in pure Maxwell theory. They are, however, fundamentally different. The
definition of the topological Witten effect only requires \emph{one} symmetry,
namely $U(1)^{[1]}_{m}$.  The charge Witten effect requires the topological
Witten effect together with the existence of a second symmetry,
$U(1)^{[1]}_{e}$, with an appropriate mixed 't Hooft anomaly with
$U(1)^{[1]}_{m}$.  This relationship can be summarized as 
\begin{equation}
   \boxed{ 
   \text{charge Witten effect}\quad\Longleftrightarrow\quad \begin{cases}
        \text{topological Witten effect}\\
        \text{extra symmetry with mixed anomaly}
    \end{cases}
    \!\!\!\!\!
    }
\end{equation}
While the additional conditions on the right hold in pure Maxwell theory, they
can fail in more general QFTs.  This makes it interesting to highlight the
practical consequences of the topological Witten effect, and we do so in the
next subsection.

Historically, generalizations of the Witten effect beyond pure Maxwell theory
use the charge Witten effect as their foundation, see
e.g.~Refs.~\cite{Seiberg:1994rs,Seiberg:1994aj} and many others. However,
our discussion implies that the topological Witten effect is more fundamental
than the charge Witten effect, and thus it is a more natural foundation for
generalizations.   In Sections~\ref{sec:general_discussion_finite}, \ref{sec:general_discussion_continuous} and
 \ref{sec:examples} we will discuss generalizations of the topological
Witten effect to a wide set of QFTs, far beyond $U(1)$ gauge theories in four
spacetime dimensions.

\subsection{Physical consequences of the topological Witten effect}
\label{sec:topologicalWitten_consequences}
If a $U(1)$ gauge theory is in a gapless photon phase where $U(1)^{[1]}_m$ spontaneously breaks, then a spontaneously broken $U(1)^{[1]}_{e}$ symmetry arises in the long-distance limit,%
\footnote{See e.g. Ref.~\cite{Cherman:2023xok} for a careful discussion of emergent higher-form symmetries.}
In this situation an emergent charge Witten effect captures most of the physics of the $\theta$ term. 
However, if the $U(1)^{[1]}_m$ symmetry is unbroken, or if there are massless charged fermions in addition to massless photons, the topological Witten effect ends up having very important physical implications.
We discuss these two situations in Sections~\ref{sec:cheshire_AB_phase} and~\ref{sec:non_inv_chiral}, respectively.


\subsubsection{Stringy Aharonov-Bohm effect in the Higgs phase}
\label{sec:cheshire_AB_phase}

The difference between the topological and charge Witten effects is particularly
clear in the electric unit-charge Higgs phase of $U(1)$ gauge theory.  
In the Higgs phase the $U(1)_e^{[1]}$ symmetry gets destroyed, and so
the charge Witten effect disappears as well.  The unit-charge Higgs phase is trivially
gapped.  There is thus no way for the Maxwell
$\theta$ parameter to influence the Higgs-phase ground state.  However, as we advertised in the introduction, we will
now show that $\theta$ affects the interactions of string excitations above the
ground state through the topological Witten effect.

A key fact we will use below is that any 4d QFT with a $1$-form symmetry that is not spontaneously broken has charged
string excitations with finite string tensions, and $U(1)$ gauge theory in its
Higgs phases is no exception. To make things concrete, let us couple the $U(1)$
gauge field to a charge $1$ scalar field $\varphi$, so that the action is 
\begin{equation}
\begin{gathered}
    \int\calD a\calD\varphi\calD\varphi^{\dagger}\ \exp\left\{ -\,\frac{1}{2g^2}\int_{\R^4} \d a\wedge\star \d a\,+\frac{\i\theta}{8\pi^2}\int_{\R^4} \d a\wedge\d a\right.\\
    \left. -\int_{\R^4} D_a\varphi \wedge\star D_a^{\dagger}\varphi^{\dagger}  \,- \lambda \int_{\R^4}\star \left(\varphi \varphi^{\dag} - v^2\right)^2 \right\}\,,
\end{gathered}
\label{eq:abelian_higgs}
\end{equation}
where $D_a \varphi= \d \varphi - \i a \varphi$, $g >0$, $\lambda>0$, and $v^2
\in \R$. When $v^2$ is sufficiently positive, the system is pushed into a
massive Higgs regime where the $U(1)^{[1]}_{m}$ symmetry is not spontaneously
broken.  
Deep in the Higgs phase, the $U(1)^{[1]}_{m}$-charged string excitations can be
viewed as (quantum) solitons.  These solitons are the Abrikosov-Nielsen-Olesen
straight-vortex solutions to the classical equation of
motion~\cite{Nielsen:1973cs}. Unit-charge vortices have a finite transverse size
since the fields decay exponentially away from the vortex worldsheet. For a
unit-charge vortex, the flux $\d a/2\pi$ integrates to $1$ on a plane
perpendicular to the vortex string.  
While it is often helpful to keep this semiclassical description in mind, the
statement that there are string excitations with finite string tensions in the
Higgs phase is much more general, and does not rely on the validity of the
semiclassical approximation.

\begin{figure}[htbp]
    \centering
    \begin{subfigure}[t]{0.48\textwidth}
        \centering
        \includegraphics[width=0.9\textwidth]{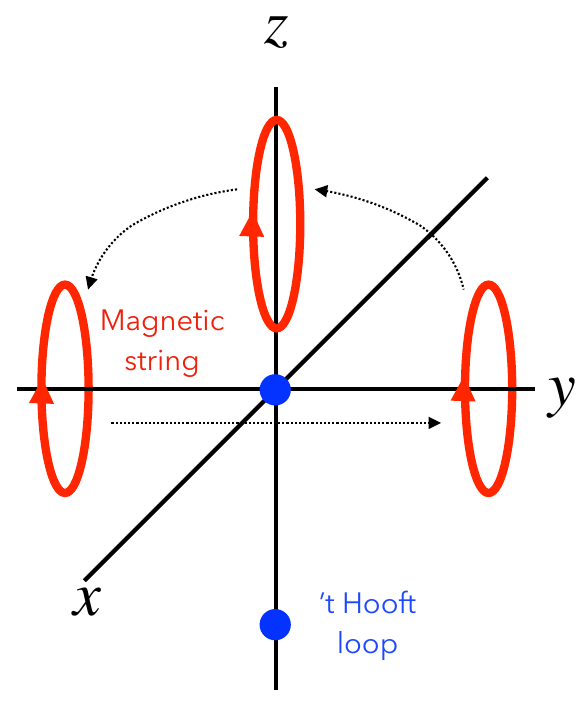}
        \caption{The $\theta \neq 0$ process.}
        \label{fig:AB_theta_QED}
    \end{subfigure}
    \hfill
    \begin{subfigure}[t]{0.48\textwidth}
        \centering
        \includegraphics[width=0.9\textwidth]{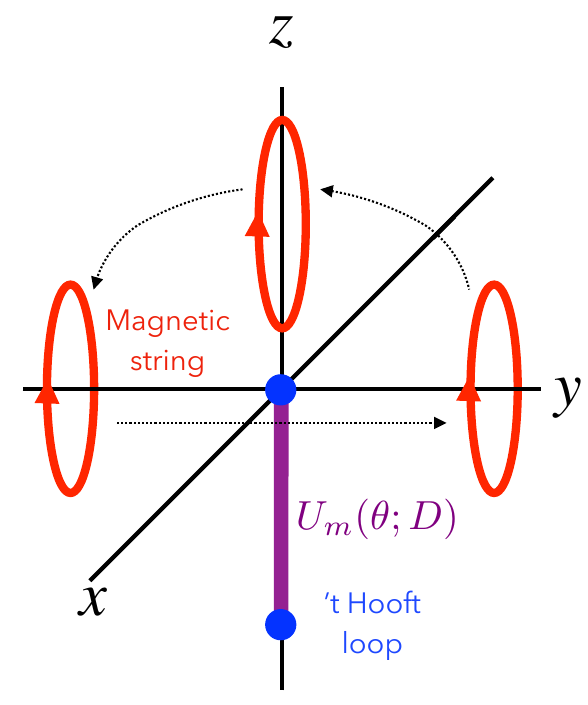}
        \caption{The $\theta = 0$ process.}
        \label{fig:AB_twisted_QED}
    \end{subfigure}
    \caption{The string Aharonov-Bohm effect induced by the $\theta$ parameter in the Higgs phase of $U(1)$ gauge theory. Left: A magnetic string excitation (red loop) encircling a 't Hooft line (blue dots) picks up phase $e^{i\theta}$. Right: The equivalent process at $\theta=0$, where the 't Hooft line is attached to a $U(1)_m^{[1]}$ symmetry surface (purple). The AB phase arises from the intersection of the string worldsheet with this surface.}
    \label{fig:string_AB_effect}
\end{figure}

The $\theta$ parameter induces a generalized Aharonov-Bohm (AB) effect for
magnetic string excitations propagating around 't Hooft operators in the Higgs
phase. We illustrate this string AB effect in Fig.~\ref{fig:string_AB_effect}.
In general, the AB effect can be defined as:
\begin{definition}[Aharonov-Bohm effect]
\label{def:AB_effect}
    In quantum field theory, we say that there is an Aharonov-Bohm effect when a
    $p$-dimensional physical {\it excitation} propagating in the presence of an
    $n$-dimensional {\it operator} picks up a phase shift $e^{i\alpha}$ when the
    worldvolume of the excitation, $C_{p+1}$, has a non-vanishing linking
    number%
    \footnote{Linking number can be defined for two objects with dimension $l$
    and $q$ in $d$-dimensional spacetime if and only if $l+q = d-1$.}
    with the manifold on which the operator is defined, $M_{n}$.  
\end{definition}

The most familiar example of an AB effect in QFT occurs in the charge-$N$ Abelian
Higgs model coupled to a massive, uncondensed unit-charge field.  Here vortex
operators are $2$d surface operators, and they carry a magnetic flux of
$2\pi/N$.  The gauge field holonomy around the vortices is thus $e^{2\pi i /N}$.
As a result, particle  excitations with unit electric charge pick up phases like
$e^{2\pi i /N}$ when they move around vortices.  This clearly fits the
definition of the AB effect we gave above. Note that it is important for the
excitations to be massive for the AB effect to make sense: the excitations need
to have a finite size so that it is physically meaningful to e.g. compare
particle trajectories that pass on different sides of a vortex operator.  It is
also worth emphasizing that this standard AB effect has nothing to do with the
$\theta$ parameter of $U(1)$ gauge theory. 

However, the $\theta$ parameter produces a distinct and less familiar AB effect,
which affects {\it string} excitations rather than particle excitations. Recall
from Eq.~\eqref{eq:shuffle-operator} that correlation functions of 't Hooft loop
operators at finite $\theta$ coincide with certain non-genuine 't Hooft loop
correlation functions in a version of the theory with $\theta$ turned off.
This is a consequence of the topological Witten effect.  The precise relation
between correlators is
\begin{equation}
    \Bigl\langle T_m(C)\cdots\Bigr\rangle_{\theta} \ =\  \Bigl\langle T_m(C)\ U_{m}(\theta m; D)\cdots\Bigr\rangle_{0}\,,
    \label{eq:surface_attachment}
\end{equation}
where $\partial D = C$, and $U_{m}(\theta m; D)$  is a symmetry generator of the
$U(1)^{[1]}_{m}$ symmetry on the open disk $D$.  Symmetry generators act on
charged {\it operators} via phases that depend on linking numbers, but they also
act on the charged {\it excitations} (created by those charged operators) via
phases that depend on the intersection number%
\footnote{The worldvolume of an excitation created by an operator has one more
 dimension than the operator, making the intersection number the natural
 topological quantity in this discussion.   The intersection number of two
 manifolds is well defined when their dimensions sum to the spacetime
 dimension.}
of the worldvolumes of the excitations with the symmetry generator. Now, a
magnetic string worldsheet $\Sigma$ that links with the 't Hooft loop operator
in Eq.~\eqref{eq:surface_attachment} must also have a non-zero intersection
number with the symmetry-generator surface operator in
Eq.~\eqref{eq:surface_attachment}.  This implies that a string excitation with
unit magnetic charge picks up an AB phase $e^{i\theta}$ when moving around a
unit-charge 't Hooft loop operator.   This is illustrated in
Fig.~\ref{fig:string_AB_effect}.

Since 't Hooft line operators act as sources for the magnetic string
excitations, this AB effect can also be interpreted as an interference effect
between two magnetic string excitations with intersecting worldsheets.  For
example, if we have two infinitely-long unit-charge magnetic strings on $\R^4$
that are not parallel, their worldsheets will intersect at a single point in
spacetime.  In such a situation, the  Maxwell $\theta$ parameter will contribute
a phase
\begin{equation}
    \exp\left(\frac{\i\theta}{8\pi^2}\int_{X_1\times X_2}\d a\wedge\d a\right) \sim \exp\left[\pm\frac{\i\theta}{4\pi^2} \left(\int_{X_1}\d a\right) \left(\int_{X_2}\d a\right)\right] \sim \exp\left(\pm\i\theta\right)\,,
\end{equation}
where $X_1$ and $X_2$ are the planes transverse to the two strings. This phase
factor will appear in many dynamical processes such as string scattering, so
that each intersection of string worldsheets produces a factor of
$\exp\left(\pm\i\theta\right)$, with the sign determined by the orientation of
the intersection.  Note that for closed string worldsheets the intersections
always appear in orientation-reversed pairs, so virtual stringy processes are
never accompanied by a nontrivial overall phase factor.

\subsubsection{Massless fermions and non-invertible chiral symmetry}
\label{sec:non_inv_chiral}

We now turn to a different implication of the topological Witten effect,
which is a relation between theories that differ by a shift of the
$\theta$ angle. Something very interesting happens when theories with
different values of $\theta$ are actually isomorphic. Then the
topological Witten effect is a transformation from a theory to itself,
which is a symmetry.

One situation where this occurs is in the presence of massless charged fermions.
For example, let us couple the $U(1)$ gauge field to a massless unit-charge
Dirac fermion $\psi$, i.e.,
\begin{equation}
\begin{gathered}
    \int\calD a\calD\psi\calD\bar{\psi}\ \exp\left\{ -\,\frac{1}{2g^2}\int \d a\wedge\star \d a\,+\frac{\i\theta}{8\pi^2}\int \d a\wedge\d a\,+\int\star\bar{\psi}  (\slash\!\!\!\partial-\i\slash\!\!\!a ) \psi\right\}\,.
\end{gathered}
\label{eq:massless-fermion}
\end{equation}
In the evaluation of this partition function on $X\equiv
S^2 \times S^1_{L_s} \times S^1_{L_t}$ as presented in
Section~\ref{sec:genericU1_Witten}, we have $\calZ_{m,n} = 0$ in
Eq.~\eqref{eq:Z_m,n} as long as $mn\neq 0$. This is due to the
Atiyah-Singer index theorem, 
which implies that these topological sectors come with fermion zero
modes, and thus the functional determinant of the Dirac operator
vanishes. Then Eq.~\eqref{eq:sum-Z_m,n} reduces to
\begin{equation}
\begin{split}
    \calZ^{\alpha}_{\beta}(X)  &=\sum_{\substack{m\in\Z \\ m\neq0}}\e^{\i\beta m}\,\calZ_{m,0}(X) + \sum_{n\in\Z}\e^{\i n\alpha}\,\calZ_{0,n}(X)\,,
\end{split}
\end{equation}
which is independent of $\theta$. Different values of $\theta$ thus yield the
same theory. In particular, as long as $m\neq 0$, the charge-$m$ subsectors of
all $U(1)^{[1]}_m$ twisted sectors are isomorphic, i.e.,
\begin{equation}\label{eq:H=H}
    \calH_m^{\alpha}\left(S^2\!\times\!S^1_{L_s}\right) \,\simeq\, \calH_m^{0}\left(S^2\!\times\!S^1_{L_s}\right)\,,\qquad m\neq0\,
\end{equation}
for any $\alpha\in{\R}/{2\pi\Z}$. In other words, we see a degeneracy between
the stringy sectors (i.e., those with nonzero $U(1)_m^{[1]}$ charges) with
different $U(1)_m^{[1]}$ twists.

In general, degeneracies are supposed to be explained by symmetries. To explain
Eq.~\eqref{eq:H=H}, we need a 0-form symmetry that shifts $\theta$, whose
details are as follows. First, it needs to act on the sector with vanishing
$U(1)_m^{[1]}$ charge as a chiral rotation $\psi\to\psi\e^{\i\theta\gamma^5}$,
since shifts of $\theta$ and chiral rotations are equivalent. At the same time,
it must act on the stringy sectors of the model as
\begin{equation}
    \calH_m^{\alpha}\left(S^2\!\times\!S^1_{L_s}\right) \,\longrightarrow\, \calH_m^{\alpha+\theta m}\left(S^2\!\times\!S^1_{L_s}\right)\,,
    \qquad m\neq 0\,, 
\end{equation}
as implied by the spectrum shuffle~\eqref{eq:spectrum_shuffle}. This implies
that, at the operator level, the symmetry should act on charge-$m$ genuine line
operators by attaching them to the topological surface operators $U_m(\theta ;
D)$ associated with the $U(1)_m^{[1]}$ symmetry.  When spacetime is $\R^4$ or
$S^4$, this can be summarized as a selection rule:
\begin{equation}
    \Bigl\langle T_m(C)\cdots\Bigr\rangle \ =\  \left\langle T_m(C)\exp\left(\frac{\i\theta m}{2\pi}\int_{D}\d a\right)\cdots\right\rangle\,, \qquad \theta \in \frac{\R}{2\pi \Z} \,.
    \label{eq:non-inv-selection-rule}
\end{equation}

The above discussion means that we are not dealing with any conventional
invertible symmetry. An invertible symmetry should give unitary (or
anti-unitary) operators in each twisted sector independently. Symmetries that do
not preserve the locality properties of operators, and hence are able to
intertwine distinct twisted sectors must be non-invertible. A famous example of
this occurs in the 2d critical Ising model, where the non-invertible
Kramers-Wannier symmetry exchanges the spin operator in the vacuum sector and
the defect operator in the twisted sector, see e.g.~Ref.~\cite{Chang:2018iay}.
Therefore, the symmetry explaining Eq.~\eqref{eq:H=H} must be a non-invertible chiral symmetry that acts invertibly on local operators but non-invertibly on line operators.  When $\theta$ takes values
in $2\pi(\Q/\Z)$, topological operators for this chiral symmetry were first constructed in
Refs.~\cite{Choi:2022jqy,Cordova:2022ieu}.


\section{Symmetry \texorpdfstring{$\theta$}{theta} angles: the finite Abelian case}
\label{sec:general_discussion_finite}

In the previous section we saw that the effect of the Maxwell $\theta$
angle in 4-dimensional $U(1)$ gauge theories can be described purely in
terms of data related to its global symmetry $U(1)^{[1]}_{m}$, through
the topological Witten effect.   It is natural to wonder whether we can
somehow \emph{define} the Maxwell $\theta$ angle --- and indeed other
$\theta$ angles --- in terms of global symmetries in the first place. 

In this and the subsequent sections, we show that this is indeed possible by establishing a general theory of symmetry $\theta$ angles. 
It turns out that any non-anomalous Abelian invertible global symmetry%
\footnote{
For simplicity, we do not take the mixing between internal and spacetime symmetry into account throughout this paper.
}
in a general $d$-dimensional QFT can be associated with \textit{symmetry $\theta$ angles}, at least in principle.
This construction also provides a generalization of the topological Witten effect, and we shall find that all possible topological Witten effects must come from symmetry $\theta$ angles. 

Abelian invertible symmetries could be finite symmetries (e.g.~$\Z_k$) or continuous $U(1)$ symmetries. 
In this section, we focus on finite Abelian invertible symmetries, and assume that the symmetry $\G$ is a direct sum of $p$-form symmetries,
\begin{equation}
    \G = G_0^{[0]}\oplus G_1^{[1]}\oplus\ldots\oplus G_{d-2}^{[d-2]}\,,
    \label{eq:bigG}
\end{equation}
such that $G_{0},G_{1},\cdots,G_{d-2}$ are all finite Abelian groups
(some of which could be trivial, of course). 
We first present our construction for $\G$-symmetry $\theta$ angles in Section~\ref{sec:construction}. 
We then derive their hallmark physical consequence, the topological Witten effect, in Section~\ref{sec:topologicalWitten}. 


\subsection{Construction}
\label{sec:construction}

We now describe our definition of symmetry $\theta$ angles.  To do this
we will introduce a class of transformations that systematically create
--- or shift --- symmetry $\theta$ angles.  
These transformations are built out of some well-studied
transformations of QFTs, which are called $S$ and $T$ transformations.%
\footnote{The names of $S$ and $T$ transformations come from the fact that they were first studied in a situation where they obey the relations $S^2 = -1$ and $(ST)^3 = -1$.  In this case they generate a modular group like $SL(2,\Z)$ or $SL(2,\Z_Q)$ depending on whether $T^Q = 1$.  The generators of modular groups are traditionally called $S$ and $T$.  However, we emphasize that this modular-group structure is \emph{not at all universal} in the QFT context.  It only appears when a symmetry is self-dual $\G = \widehat{\G}$.
See Section~\ref{sec:Theta_vs_others} and Appendix~\ref{sec:STSTST=1} for more details.}

We start by reviewing
the definitions of dual symmetries and $S$ and $T$ transformations in Sec.~\ref{sec:S_transformation}
and~Sec.~\ref{sec:T_discrete}. Our construction of symmetry
$\theta$ angles appears in Sec.~\ref{sec:Theta_transformation}.
For a finite Abelian group $G$, $G^{[p]}$ gauge fields must be flat,%
\footnote{This is not necessarily true in lattice field theories. However,
in this paper we focus on continuum theories, where discrete gauge fields are
always flat.}
so they do not contain any local information. 
The only physical information contained in such a gauge field is in its holonomies on non-contractible cycles in spacetime.
The inequivalent gauge fields for a $G^{[p]}$ symmetry on a closed spacetime $X$ are classified by the cohomology group
\begin{equation}
    H^{p+1}(X,G)\,.
\end{equation}
For a general finite Abelian invertible symmetry $\G$, the gauge fields are classified by the direct sum,
\begin{equation}
    \H(X,\G) \equiv H^{1}(X,G_0)\oplus H^2(X,G_1)\oplus \cdots \oplus H^{d-1}(X, G_{d-2})\,.
\end{equation}
For a $\G$-symmetric quantum field theory $\calZ$, we write
\begin{equation}
    \calZ(X,\rho)
\end{equation}
for its partition function on $X$ with a background $\G$ gauge field
$\rho\in\H(X,\G)$.

\subsubsection{Dual symmetries and \texorpdfstring{$S$}{S} transformations}
\label{sec:dual_symmetries}
\label{sec:S_transformation}

Since we assume that the symmetry $\G$ is anomaly-free, we can gauge it by summing over all $\G$ gauge fields.
This operation transforms the original $\G$-symmetric theory $\calZ$ to a new $\widehat{\G}$-symmetric theory $\widehat{\calZ}$. 
Here $\widehat{\G}$ denotes the \textit{dual symmetry} of $\G$, and the transformation from a QFT with symmetry $\G$ to the related QFT with symmetry $\widehat{\G}$ is called an \textit{$S$ transformation}. 
We now briefly review the definition of a dual symmetry and the $S$ transformation.  
Readers for whom these ideas are new should probably also consult other references such as Refs.~\cite{Vafa:1989ih,Gaiotto:2014kfa,Bhardwaj:2017xup,Aharony:2013hda,Tachikawa:2017gyf,Witten:2003ya} for more details.

Consider a finite Abelian group $G$.
Then the dual symmetry of a $G^{[p]}$ symmetry is a $\widehat{G}^{[d-p-2]}$ symmetry, where 
\begin{equation}
    \widehat{G} \equiv \Hom\left( G,\,\tfrac{\R}{2\pi\Z}\right)
\end{equation}
is $G$'s character group.
Characters just mean irreducible unitary representations.
$\widehat{G}$ and $G$ share the same group structure and, in particular, $|\widehat{G}|=|G|$, where $|A|$ stands for the cardinality of $A$.
Between a $G^{[p]}$ gauge field $\rho\in H^{p+1}(X,G)$ and a
$\widehat{G}^{[d-p-2]}$ gauge field $y\in H^{d-p-1}(X,\widehat{G})$, there is a cup product%
\footnote{The cup product is the discrete analog of the wedge product for
differential forms, see e.g.~Ref.~\cite{Hatcher:2001} for a detailed
mathematical discussion, and
Refs.~\cite{Chen:2018nog,Chen:2021ppt,Jacobson:2023cmr} for physics-focused
discussions.  The super-commutativity of the cup products shown in Eq.~\eqref{eq:cup} holds only at the cohomology level. As extensively discussed in the aforementioned references, the relation between $\rho \cup y$ and $y \cup \rho$ is significantly more complicated at the level of cochains. }
\begin{equation}
    \rho\cup y\ =\ (-1)^{(p+1)(d-p-1)}\,y\cup\rho\ 
    \in \,H^d\left(X,\tfrac{\R}{2\pi\Z}\right)\,.
    \label{eq:cup}
\end{equation}
On a closed oriented manifold $X$, through this cup product, Poincar\'e duality identifies a $\widehat{G}^{[d-p-2]}$ gauge field as a character of $H^{p+1}(X,G)$, and vice versa:
\begin{equation}\label{eq:H_pontryagin}
\begin{gathered}
    y\ \mapsto\int_X - \cup y \qquad\Longrightarrow\qquad 
    H^{d-p-1}\left(X,\widehat{G}\right) = 
    \Hom\left( H^{p+1}(X,G),\,\tfrac{\R}{2\pi\Z}\right)\,,\\
    \rho\ \mapsto\int_X - \cup \rho \qquad\Longrightarrow
    \qquad H^{p+1}(X,G) 
    = \Hom\left( H^{d-p-1}\left(X,\widehat{G}\right),\,\tfrac{\R}{2\pi\Z}\right)\,.
\end{gathered}
\end{equation}
This immediately implies that $|H^{p+1}(X,G)|=|H^{d-p-1}(X,\widehat{G})|$.


We can obtain the dual symmetry of a generic non-anomalous finite Abelian invertible symmetry $\G$ in a ``factorized'' way --- that is, term by term in Eq.~\eqref{eq:bigG}. 
Namely, the dual symmetry of $\G=\bigoplus_{\bullet=0}^{d-2}G_{\bullet}^{[\bullet]}$ is $\widehat{\G} = \bigoplus_{\bullet=0}^{d-2}\widehat{G}_{\bullet}^{[\bullet]}$, where
\begin{equation}
    \widehat{G}_{\bullet}\equiv\widehat{G_{d-\bullet-2}}\,.
\end{equation}
The cup products $\rho\cup y$ and $y\cup\rho$ between a $\G$ gauge field
$\rho\in\H(X,\G)$ and a $\widehat{\G}$ gauge field
$y\in\H\bigl(X,\widehat{\G}\bigr)$ are also defined in a factorized way.
Their integrals also induce isomorphisms,
\begin{equation}\label{eq:G(X)_pontryagin}
\begin{gathered}
    y\ \mapsto\int_X -\cup y \qquad\Longrightarrow\qquad \H\bigl(X, \widehat{\G}\bigr) = \Hom\left( \H(X,\G),\,\tfrac{\R}{2\pi\Z}\right)\,,\\
    \rho\ \mapsto\int_X -\cup\rho \qquad\Longrightarrow\qquad \H(X,\G) = \Hom\left( \H\bigl(X, \widehat{\G}\bigr),\,\tfrac{\R}{2\pi\Z}\right)\,,
\end{gathered}
\end{equation}
respectively. 
Again we immediately see that $|\H(X,\G)|=|\H\bigl(X, \widehat{\G}\bigr)|$.

The $S$ transformation takes a $\G$-symmetric theory $\calZ$ to a new $\widehat{\G}$-symmetric theory $\widehat{\calZ}$.  Its partition function is given by
\begin{equation}\label{eq:fore-gauging}
    \widehat{\calZ}(X,y) = \frac{1}{N(X,\G)} 
    \sum_{\rho \in \H(X,\G)} \!\calZ(X,\rho)\,
    \exp\left(-\i\int y\cup\rho \right),\qquad y\in\H\bigl(X, \widehat{\G}\bigr)\,,
\end{equation}
where the normalization factor is given by the rather cumbersome expression%
\footnote{ Another normalization commonly used in the literature is
\begin{equation}\label{eq:N'(X,G)}
    N'(X,\G) \equiv \prod_{\bullet=0}^{d-2} 
    \frac{\bigl|H^{\bullet}(X,G_{\bullet})
    \bigr|\bigl|H^{\bullet-2}(X,G_{\bullet})\bigr|\cdots}
    {\bigl|H^{\bullet-1}(X,G_{\bullet})\bigr|\bigl|H^{\bullet-3}(X,G_{\bullet})\bigr|\cdots}\,,
\end{equation}
which differs from our choice by an Euler counterterm. Concretely, 
\begin{equation}
\frac{N'(X,\G)}{N(X,\G)}=\left(\frac{|G_0||G_2|\cdots}{|G_1||G_3|\cdots}\right)^{\chi(X)/2}\,,
\end{equation}
where $\chi(X)$ denotes $X$'s Euler characteristic.
Our choice of normalization has the advantage that it manifests Eq.~\eqref{eq:back-gauging}. 
We are grateful to Yichul Choi for comments on this point. 
The two choices agree with each other on odd-dimensional spacetimes or on spacetimes with the topology $M\!\times\!S^1$, because the Euler characteristic vanishes in these cases. 
}
\begin{equation}\label{eq:N(X,G)}
\begin{split}
    N(X,\G) \!\equiv\! \prod_{\bullet=0}^{d-2}\!
     \sqrt{\frac{\bigl|H^{\bullet}(X,G_{\bullet})
     \bigr|\bigl|H^{\bullet-2}(X,G_{\bullet})\bigr|\cdots}
     {\bigl|H^{\bullet-1}(X,G_{\bullet})\bigr|\bigl|H^{\bullet-3}(X,G_{\bullet})\bigr|\cdots}
    \frac{\bigl|H^{\bullet+1}(X,G_{\bullet})
    \bigr|\bigl|H^{\bullet+3}(X,G_{\bullet})\bigr|\cdots}
    {\bigl|H^{\bullet+2}(X,G_{\bullet})\bigr|\bigl|H^{\bullet+4}(X,G_{\bullet})\bigr|\cdots}}.
\end{split}
\end{equation} 
The dual symmetry $\widehat{\G}$ in the resulting theory $\widehat{\calZ}$ is
also anomaly-free, and gauging $\widehat{\G}$ can bring us back to the original
theory $\calZ$. A direct calculation shows the identity%
\footnote{
The calculation relies on
$N(X,\G)\,N(X,\widehat{\G})=|\H(X,\G)|=|\H\bigl(X, \widehat{\G}\bigr)|$ and the isomorphism~\eqref{eq:G(X)_pontryagin}.
}
\begin{equation}\label{eq:back-gauging}
    \calZ(X,\rho) = \frac{1}{N\!\left(X,\widehat{\G}\right)}
    \sum_{y\in\H\bigl(X, \widehat{\G}\bigr)}\!
    \widehat{\calZ}(X,y)\,\exp\left(\i\int y\cup\rho\right),\qquad \rho\in \H(X,\G)\,.
\end{equation}
Due to the supercommutativity of cup products, this operation is not exactly an $S$ transformation with respect to the symmetry $\widehat{\G}$ of the theory $\widehat{\calZ}$:  It also includes sign flips of gauge fields $\rho\to-\rho^{\vee}$, where $\vee$ is defined in a factorized way through
\begin{equation}\label{eq:vee}
    x^{\vee} \equiv (-1)^{(p+1)(d-p-1)}x\,,\qquad 
    x\in H^{p+1}(X,G)\text{ or }H^{d-p-1}(X,G)\,.
\end{equation}
We refer to the operation described by Eq.~\eqref{eq:back-gauging} as an $S^{\dag}$ transformation, and formally $S^{\dag}S=SS^{\dag}=1$.  
The operation $\vee$ is non-trivial only when both $p$ and $d$ are even. 

\subsubsection{Invertible theories and \texorpdfstring{$T$}{T} transformations}
\label{sec:T_discrete}

Given two QFTs $\calX$ and $\calY$, we can always make a new QFT
$\calX\!\otimes\!\calY$ which consists of their completely decoupled
combination, so that e.g. the partition function of the new theory is the
product of the partition functions of $\calX$ and $\calY$. The Hilbert space of
$\calX\!\otimes\!\calY$ is the tensor product of the Hilbert spaces of $\calX$
and $\calY$, and the Hamiltonian of $\calX\!\otimes\!\calY$ is the sum of the
Hamiltonians of $\calX$ and $\calY$. The common terminology is that
$\calX\!\otimes\!\calY$ is produced by ``stacking'' $\calY$ on top of $\calX$. 

A QFT is said to be \textit{invertible} if stacking with it can be undone by stacking with another QFT.
As a result, the partition function of an invertible QFT must be a pure phase factor on any manifold, and an invertible QFT must have only one state (with zero energy) on any admissible closed spatial manifold.
Any trivially gapped system is described by an invertible theory in the infrared, and
is thus said to be in an invertible phase. Although the bulk physics of
invertible QFTs is rather trivial, their behavior in the presence of boundaries
can be very rich due to anomaly inflow phenomena. Therefore, invertible theories
and invertible phases have been extensively studied in the recent literature; see
e.g.~Refs.~\cite{Kitaev_KITP,Kitaev_SCGP,Kapustin:2014tfa,Kapustin:2014dxa,
Freed:2016rqq,Witten:2015aba} for an extremely incomplete list.

Any non-anomalous symmetry allows a trivially gapped phase (essentially by
definition), and thus invertible theories with any non-anomalous global symmetry
always exist. For our purpose, we are interested in stacking the
$\widehat{\G}$-symmetric theory $\widehat{\calZ}$ with a
$\widehat{\G}$-symmetric invertible theory $\widehat{\calI}$. When we call the
invertible theory  $\widehat{\G}$-symmetric, we always assume
that pure spacetime symmetries act trivially on the theory. Namely, the
partition function of a $\widehat{\G}$-symmetric invertible theory
$\widehat{\calI}$ on a spacetime $X$ satisfies
\begin{equation}\label{eq:IQFT}
    |\widehat{\calI}(X,y)| = 1\,,\quad y\in\H\bigl(X, \widehat{\G}\bigr)\,,
    \qquad\text{ and }\qquad\widehat{\calI}(X,0)=1\,.
\end{equation}
Given that $\widehat{\G}$ is finite, $\widehat{\calI}$ is necessarily
topological. The corresponding invertible phases are called
$\widehat{\G}$-symmetry protected topological ($\widehat{\G}$-SPT)
phases~\cite{Gu:2009dr,Pollmann:2009mhk,Chen:2011bcp}. The operation 
\begin{equation}
    \otimes\,\widehat{\calI}
\end{equation}
is called a \textit{$T$ transformation}. The difference between
$\widehat{\calZ}$ and $\widehat{\calZ}\!\otimes\!\widehat{\calI}$ is actually
nothing but a consistent choice of local counterterms with respect to background
$\widehat{\G}$ gauge fields. Hence we can concisely say that a $T$
transformation is a shift of local counterterms. 

\subsubsection{\texorpdfstring{$\Theta$}{Theta} transformations and symmetry \texorpdfstring{$\theta$}{theta} angles}
\label{sec:Theta_transformation}
\label{sec:Def&Cla}

We now introduce the operation we will use to generate symmetry $\theta$ angles.
First let us consider a binary operation $\ast$ between $\G$-symmetric QFTs defined by
\begin{equation}\label{eq:convolution}
    \calX\!\ast\!\calY(X,\sigma) = \frac{1}{N(X,\G)} \sum_{\rho \in \H(X,\G)} \!\calX(X,\sigma\!-\!\rho)\,\calY(X,\rho)\,,\qquad \sigma\in \H(X,\G)\,,
\end{equation}
where the normalization factor is the one given by Eq.~\eqref{eq:N(X,G)}. 
This convolution $\ast$ is related to the stacking $\otimes$ operation described in Section~\ref{sec:T_discrete} via an
$S$ transformation:
\begin{equation}\label{eq:X*Y<->XY}
\begin{tikzcd}
    \calX\!\ast\!\calY \quad \arrow[r, rightharpoonup, yshift=0.1em, "S"] 
    & \arrow[l, rightharpoonup, yshift=-0.1em, "S^{\dag}"] \quad 
    \widehat{\calX}\!\otimes\!\widehat{\calY}
\end{tikzcd}\,.
\end{equation}
This can be viewed as a Fourier transformation identity. 
The partition functions $\calZ$ and $\widehat{\calZ}$ are functions on $\H(-,\G)$ and $\H(-,\widehat{\G})$. 
Then the $S$ and the $S^{\dag}$ transformations form a dual pair of discrete Fourier transformations. 
The stacking $\otimes$ is a point-wise multiplication while $\ast$ is a convolution.

Now, let us take $\widehat{\calY}$ to be a $\widehat{\G}$-symmetric invertible theory $\widehat{\calI}$ so that the right-hand side of Eq.~\eqref{eq:X*Y<->XY} is a $T$ transformation of $\widehat{\calX}$.
The corresponding $\calI$ on the left-hand side is then a $\G$-symmetric Dijkgraaf-Witten theory.
A TQFT is called a Dijkgraaf-Witten theory~\cite{Dijkgraaf:1989pz} if gauging some finite symmetry converts it to an invertible theory.%
\footnote{
Dijkgraaf-Witten theories are canonical examples of fully-extended TQFTs that always allow gapped boundaries and gapped defects of any dimension~\cite{Baez:1995xq,Lurie:2009keu}.
}
When saying ``$\G$-symmetric'', we mean that it is precisely gauging $\G$ that converts the Dijkgraaf-Witten theory into an invertible theory.
and explicitly we have
\begin{equation}\label{eq:Theta_transformation}
    \calZ\!\ast\!\calI(X,\sigma) = \frac{1}{|\H(X,\G)|} \sum_{\substack{\rho \in \H(X,\G) \\ y\in \H(X,\widehat{\G})}} \!\calZ(X,\sigma\!-\!\rho)\,\widehat{\calI}(X,y)\,\exp\left(\i\int y\cup\rho\right)\,,
\end{equation}
for $\sigma\in \H(X,\G)$.
We refer to this convolution with a Dijkgraaf-Witten theory,
\begin{equation}
    \ast\,\mathcal{I},
\end{equation}
as a \textit{$\Theta$ transformation}.
In short,
\begin{equation}\label{eq:Th=STS}
    \Theta = S^{\dag}TS\,,\qquad T = S\Theta S^{\dag}\,.
\end{equation}
We now propose a definition for symmetry $\theta$ angles.
\begin{definition}[\textbf{Symmetry $\theta$ angle}]
    \label{def:sym_theta}
    For a QFT with a non-anomalous finite Abelian invertible symmetry $\G$, a
    \textit{$\G$-symmetry $\theta$ angle} is defined by a $\Theta$
    transformation with respect to $\G$.
\end{definition}




How many inequivalent $\calG$-symmetry $\theta$ angles are there for a given $\calG$?
Given the definition, $\G$-symmetry $\theta$ angles are classified by the $\widehat{\G}$-symmetric invertible theories used in the $T$ transformation.
Hence the known classification of invertible theories --- a theoretical milestone achieved by physicists and mathematicians over the last decade~\cite{Kapustin:2014tfa,Kapustin:2014dxa,Freed:2016rqq,Yonekura:2018ufj,Yamashita:2021cao} --- implies that the collection of all $d$-dimensional
$\G$-symmetry $\theta$ angles naturally forms a finite Abelian group that is the
character group of a bordism group:
\begin{equation}\label{eq:classification_discrete}
    \Hom\left(\tOmega^{SO}_{d}\bigl(\calB\widehat{\G}\bigr),\tfrac{\R}{2\pi\Z}\right) 
    \quad\text{or}\quad 
    \Hom\left(\tOmega^{Spin}_{d}\bigl(\calB\widehat{\G}\bigr),\tfrac{\R}{2\pi\Z}\right)\,,
\end{equation}
in bosonic or fermionic theories, respectively. 
There are non-trivial $\G$-symmetry $\theta$ angles as long as the groups above are not trivial.

We provide a brief review of bordism groups and their role in classifying invertible theories in Appendix~\ref{app:topological_data}.
Here we just explain the meaning of the symbols above.
First of all, $\tOmega^{SO}_{\bullet}\bigl(-)$ and $\tOmega^{Spin}_{\bullet}\bigl(-)$ denote
the reduced oriented and spin bordism groups, respectively, of a topological
space. $\calB\widehat{\G}$ denotes the classifying space of $\widehat{\G}$.
Namely, $\calB\widehat{\G}$ is an auxiliary topological space that has the
defining property
\begin{equation}
    \left[X,\calB\widehat{\G}\right] \,=\, \H\left(X,\widehat{\G}\right)\,,\qquad\text{for any }X\,,
\end{equation}
where the bracket $[-,-]$ denotes the collection of homotopy classes of maps
between two topological spaces. In particular, for $\widehat{G}^{[p]}$ with
finite Abelian group $\widehat{G}$, its classifying space is the
Eilenberg-MacLane space $ B^{p+1}\widehat{G}=K\bigl(\widehat{G},p+1\bigr)$,
which satisfies
\begin{equation}
    \left[X,B^{p+1}\widehat{G}\right]\,=\, H^{p+1}\left(X,\widehat{G}\right)
\end{equation}
In general, we have $\calB\widehat{\G} = \prod_{\bullet=0}^{d-1}
B^{\bullet+1}\widehat{G}_{\bullet}$ as a direct product.

The classification in Eq.~\eqref{eq:classification_discrete} shows that the
$\G$-symmetry $\theta$ angles necessarily have a discrete moduli space.
Therefore, if $\G$ is an exact symmetry in a QFT, the $\G$-symmetry $\theta$
angle is an invariant preserved by the RG flow. If $\G$ is instead an emergent
symmetry, the $\G$-symmetry $\theta$ angle is then a discrete parameter in the
effective field theory. Different ultraviolet physics may lead to different
symmetry $\theta$ angles in a low-energy effective field theory. 

\subsection{Physical consequences}
\label{sec:topologicalWitten}

The definition of symmetry $\theta$ angles in Definition~\ref{def:sym_theta} is
quite abstract, and we now explore its physical consequences to make it more
concrete.  We will see that a very common physical consequence of a
$\G$-symmetry $\theta$ angle is a topological Witten effect on $\G$, and indeed
all possible topological Witten effects come from symmetry $\theta$ angles.

\subsubsection{Twists and charges}
\label{sec:twist&charge}

As we saw in the Maxwell example in Section~\ref{sec:Witten_reappraisal}, one particularly convenient way to highlight the topological Witten effect is to explore Hilbert spaces on closed spatial manifolds. 
In this subsection we review some well-known relations between symmetry charges and twisted sectors.  
This generalizes our discussion of twisted sectors and charges in 4-dimensional Maxwell theory from Section~\ref{sec:Witten_reappraisal}.

Let us put a $d$-dimensional QFT $\calZ$ on a ``thermal'' spacetime $M\!\times\!S^1_L$.  
We assume that $M$ is a closed $(d\!-\!1)$-dimensional oriented or spin Riemannian manifold, and $S^1_L$ is a circle of length $L$ along which bosonic and fermionic fields have periodic and anti-periodic boundary conditions, respectively. 
Then the partition function on $M\!\times\!S^1_L$ can be interpreted as 
\begin{equation}\label{eq:Hamiltonian}
    \calZ\bigl(M\!\times\!S^1_L\bigr) = \tr\exp\bigl[-\,L\,\calH(\calZ,M)\bigr]\,,
\end{equation}
where $\calH(\calZ,M)$ is the Hamiltonian of the Hilbert space on the spatial manifold $\calM$, and $L$ can be interpreted as is the inverse temperature.

On manifolds of the form $M\!\times\!S^1_L$, cohomology groups have canonical
decompositions of the form%
\footnote{ This comes from a cohomological K\"unneth formula and the
particularly simple topology of $S^1$. For a general product manifold
$M\!\times\!N$, there is only a homomorphism from
$\bigoplus_{\bullet}[H^{\bullet}(M,G)\otimes H^{k-\bullet}(N,K)]$ to
$H^k(M\!\times\!N,\,G \otimes K)$ which is not an isomorphism, and the general
K\"unneth formula is a spectral sequence.}
\begin{equation}\label{eq:decomposition-theorem}
\begin{split}
    H^{p+1}\bigl(M\!\times\! S^1_L,G\bigr) &\:=\ \Bigl[H^{p+1}(M,G)\otimes H^0\bigl(S^1_L,\Z\bigr)\Bigr] \oplus \Bigl[H^{p}(M,G)\otimes H^1\bigl(S^1_L,\Z\bigr)\Bigr] \\
    &\:\simeq\ H^{p+1}(M,G)\,\oplus\,H^{p}(M,G)\,.
\end{split}
\end{equation}
In the top line above, the first summand $H^{p+1}(M,G)$ describes $G^{[p]}$ gauge fields on $M$. 
They can trivially become $G^{[p]}$ gauge fields on $M\!\times\! S^1_L$, which is captured by tensoring with $H^0\bigl(S^1_L,\Z\bigr)=\Z$. 
The second summand on the top line $H^{p}(M,G)$ contains $G^{[p-1]}$ gauge fields on $M$. 
They can be transformed into $G^{[p]}$ gauge fields on $M\!\times\! S^1_L$ via stretching them on $S^1_L$, which is captured by tensoring them with  $H^1\bigl(S^1_L,\Z\bigr)=\Z$. 
This decomposition naturally generalizes to general finite Abelian invertible symmetries:
\begin{equation}\label{eq:decomposition_theorem'}
    \H\bigl(M\!\times\!S^1_L, \G\bigr)\ =\ \H^s\bigl(M, \G\bigr) \oplus \H^t\bigl(M, \G\bigr)\,,
\end{equation}
where
\begin{equation}\label{eq:spatial&temporal}
\begin{gathered}
    \H^s\bigl(M, \G\bigr)\ \equiv\  \bigoplus_{\bullet=0}^{d-2} \Bigl[H^{\bullet+1}(M,G_{\bullet})\otimes H^0\bigl(S^1_L,\Z\bigr)\Bigr]\ \simeq\ \bigoplus_{\bullet=0}^{d-2} H^{\bullet+1}(M,G_{\bullet})\,,\\
    \H^t\bigl(M, \G\bigr)\ \equiv\ \bigoplus_{\bullet=0}^{d-2} \Bigl[H^{\bullet}(M,G_{\bullet})\otimes H^1\bigl(S^1_L,\Z\bigr)\Bigr]\ \simeq\ \bigoplus_{\bullet=0}^{d-2} H^{\bullet}(M,G_{\bullet})\,.
\end{gathered}
\end{equation}
Hence $\H^s\bigl(M, \G\bigr)$ describes spatial $\G$ gauge fields that solely
live on $M$ while $\H^t\bigl(M, \G\bigr)$ describes temporal $\G$ gauge fields
that are stretched on $S^1_L$. 

\begin{figure}[h!tbp]
    \centering
    \includegraphics[width=0.4\textwidth]{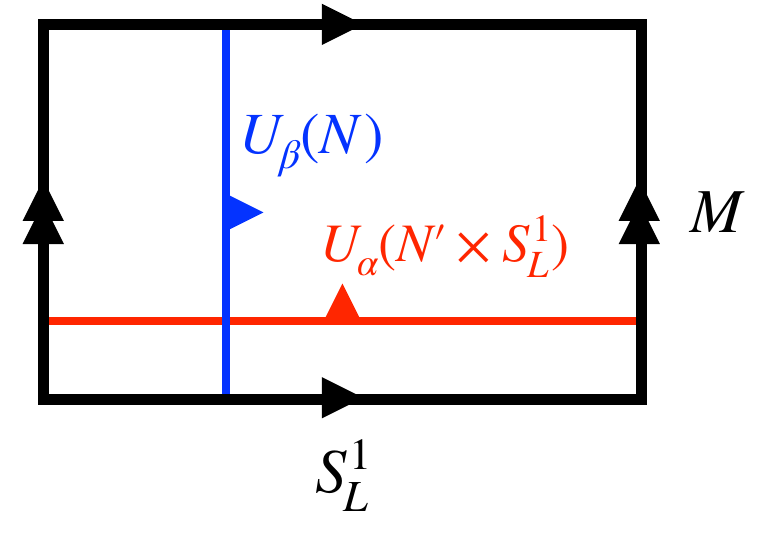}
    \caption{
    Activating a temporal gauge field is equivalent to inserting the $G^{[p]}$
    symmetry generator operator $U_{\beta}(N)$ (\textcolor{blue}{blue}), where
    $N$ is a $(d\!-\!p\!-\!1)$-dimensional submanifold of $M$.  
    Similarly, activating a spatial gauge field is equivalent to inserting a $G^{[p]}$
    symmetry generator operator $U_{\alpha}(N'\times S^1_L)$
    (\textcolor{red}{red}) on a manifold $N' \times S^1_L$ where $N'$ is a
    $(d\!-\!p\!-\!2)$-dimensional submanifold of $M$.  
    }
    \label{fig:twist_illustration}
\end{figure}

Turning on $\alpha \in \H^s\bigl(M, \G\bigr)$ and $\beta \in \H^t\bigl(M, \G\bigr)$ leads to quite different physical effects on the Hamiltonian interpretation, as illustrated in Fig.~\ref{fig:twist_illustration}.
We first recall that turning on a background gauge field $\rho\in H^{p+1}(X,G)$ is equivalent to inserting a
codimension-$(p+1)$ topological operator \emph{dual} to
$\rho$. A \emph{temporal} gauge field $\beta\in\H^t\bigl(M, \G\bigr)$ means the
insertion of \emph{spatial} topological operators inside a spatial slice. It
thus yields a unitary operator $\mathcal{U}_{\beta}$ that acts on the states on
$M$ and commutes with the Hamiltonian. Hence the group $\H^t\bigl(M, \G\bigr)$
is the concrete incarnation of the symmetry $\G$ on the Hilbert space on $M$. In
contrast, a \emph{spatial} gauge field $\alpha\in\H^s\bigl(M, \G\bigr)$ means
the insertion of \emph{temporal} topological operators parallel to $S^1_L$. It
thus introduces a static background gauge field on $M$ and modifies the Hilbert
space and the Hamiltonian. We say that the original Hamiltonian $\calH(\calZ,M)$
is in the vacuum sector and the new Hamiltonian $\calH^{\alpha}(\calZ,M)$ is in
a twisted sector. Figure~\ref{fig:twist_illustration} illustrates these
topological operators.

The above analysis implies that the partition function of theory $\calZ$ on
$M\times S^1_L$ with a background gauge field $\alpha +\beta$ has the
interpretation
\begin{equation}\label{eq:Hamiltonian_U-insertion}
    \calZ\bigl(M\times S^1_L,\, \alpha+\beta\bigr) = 
    \tr \Bigl\{\calU_{\beta}\exp\bigl[-\,L\,\calH^{\alpha}(\calZ,M)\bigr]
     \Bigr\}\,,\quad\alpha\in\H^s\bigl(M, \G\bigr),\,\beta\in\H^t\bigl(M, \G\bigr)\,.
\end{equation}
This graded partition function allows us to study how the symmetry group
$\H^t\bigl(M, \G\bigr)$ acts on the twisted sectors. In particular, we can
organize the states on $M$ into the characters of $\H^t\bigl(M, \G\bigr)$,
i.e.~irreducible unitary representations. Based on this Hamiltonian
interpretation, we introduce the following terminology:
\begin{itemize}
    \item a $\G$-\textit{twist} on $M$ is an element of the spatial
    $\H^s\bigl(M, \G\bigr)$;
    \item a $\G$-\textit{charge} on $M$ is a character of the temporal
    $\H^t\bigl(M, \G\bigr)$.
\end{itemize}
Charges and twists are dual to each other under the $S$ transformation. 

Given
the decomposition theorem~\eqref{eq:decomposition_theorem'}, the cup product
between a $\G$ gauge field and a $\widehat{\G}$ gauge field necessarily vanishes
if the gauge fields are both spatial or both temporal. This observation refines
the canonical isomorphism~\eqref{eq:G(X)_pontryagin} to finer canonical
isomorphisms, 
\begin{subequations}\label{eq:GM_pontryagin}
\begin{gather}
    \H^s\bigl(M, \widehat{\G}\bigr) 
    = \Hom\left( \H^t\bigl(M, \G\bigr),\,\tfrac{\R}{2\pi\Z}\right)\,,\\
    \H^s\bigl(M, \G\bigr) 
    = \Hom\left(\H^t\bigl(M, \widehat{\G}\bigr),\,\tfrac{\R}{2\pi\Z}\right)\,.
\end{gather}
\end{subequations}
The quantities on the right-hand side of Eq.~\eqref{eq:GM_pontryagin} are the
$\G$ and $\widehat{\G}$ charges, while the quantities on the left-hand side are
$\widehat{\G}$ and $\G$ twists, respectively.  We can summarize this as the statement
\begin{equation}\label{eq:charge_vs_twist}
    \text{a $\G$-charge }=
    \text{ a $\widehat{\G}$-twist}\,,\qquad 
    \text{a $\widehat{\G}$-charge }=\text{ a $\G$-twist}
\end{equation}

We can now decompose the Hamiltonian of each twisted sector into subsectors with
different $\G$-charges,
\begin{equation}
    \calH^{\alpha}(\calZ,M) = \!\!\!\!\!\!\! \bigoplus_{a\in\H^s\bigl(M, \widehat{\G}\bigr)} \!\!\!\!\!\!\! \calH^{\alpha}_{a}(\calZ,M)\,,\qquad \alpha\in\H^s\bigl(M, \G\bigr)\,,
\end{equation}
and reduce the insertion of $\calU_{\beta}$ in
Eq.~\eqref{eq:Hamiltonian_U-insertion} to a weighted sum of subsectors,\footnote{Here and throughout the paper, when we consider partition functions on $X=M\times S^1$ we interpret expressions like $\int \beta \cup \alpha$ as being integrated over the whole spacetime $X$. This is defined via the tensor products in Eq.~\eqref{eq:decomposition-theorem} where the gauge fields $\alpha,\beta$ supported on $M$ can be extended to gauge fields on $M\times S^1$. Note that the cohomology degree of temporal gauge fields will raise by 1 in this process, while spatial gauge fields stay the same.}
\begin{equation}\label{eq:Hamiltonian_twist+charge}
\begin{split}
    &\:\calZ\bigl(M\!\times \!S^1_L,\,\!\alpha\!+\!\beta\bigr) \\
    = &\:\!\!\!\!\!\!\!\sum_{a\in\H^s\bigl(M, \widehat{\G}\bigr)} \!\!\!\!\!\! \tr\exp\left[-\,L\,\calH^{\alpha}_a(\calZ,M) +\i\!\int\!\beta\cup a\right]
\end{split}\,,\qquad
\alpha\in\H^s\bigl(M, \G\bigr),\,\beta\in\H^t\bigl(M, \G\bigr)\,.
\end{equation}
A convenient way to visualize this decomposition is to formally organize these
subsector Hamiltonians into a spectral table, as sketched in
Fig.~\ref{fig:spectrum_table_Z}. In spectral tables we  label each row by
a $\G$-twist, label each column by a $\G$-charge, and the entries $\calH_{a}^{\alpha}$ denote Hamiltonians for states with charge $a$ and twist $\alpha$.

\begin{figure}[!ht]
\centering
\begin{tabular}{c||c|c|c|c|c}
    $\calZ$ & \multicolumn{5}{c}{$\G$ charges 
    } 
    \\\hline\hline
    \multirow{5}{*}{\begin{turn}{90}
    $\G$ twists 
    \end{turn}} & \cellcolor{Firebrick1!50} $\calH^{0}_{0}(\calZ)$ & \cellcolor{DarkGoldenrod1!50} $\calH^{0}_{a}(\calZ)$ & \cellcolor{Chartreuse1!50} $\calH^{0}_{b}(\calZ)$ & \cellcolor{RoyalBlue1!50} $\calH^{0}_{c}(\calZ)$ & \cellcolor{Purple1!50} $\calH^{0}_{d}(\calZ)$ \\\hhline{~||-----}
    & \cellcolor{Firebrick1!40} $\calH^{\alpha}_{0}(\calZ)$ & \cellcolor{DarkGoldenrod1!40} $\calH^{\alpha}_{a}(\calZ)$ & \cellcolor{Chartreuse1!40} $\calH^{\alpha}_{b}(\calZ)$ & \cellcolor{RoyalBlue1!40} $\calH^{\alpha}_{c}(\calZ)$ & \cellcolor{Purple1!40} $\calH^{\alpha}_{d}(\calZ)$ \\\hhline{~||-----}
    & \cellcolor{Firebrick1!30} $\calH^{\beta}_{0}(\calZ)$ & \cellcolor{DarkGoldenrod1!30} $\calH^{\beta}_{a}(\calZ)$ & \cellcolor{Chartreuse1!30} $\calH^{\beta}_{b}(\calZ)$ & \cellcolor{RoyalBlue1!30} $\calH^{\beta}_{c}(\calZ)$ & \cellcolor{Purple1!30} $\calH^{\beta}_{d}(\calZ)$ \\\hhline{~||-----}
    & \cellcolor{Firebrick1!20} $\calH^{\gamma}_{0}(\calZ)$ & \cellcolor{DarkGoldenrod1!20} $\calH^{\gamma}_{a}(\calZ)$ & \cellcolor{Chartreuse1!20} $\calH^{\gamma}_{b}(\calZ)$ & \cellcolor{RoyalBlue1!20} $\calH^{\gamma}_{c}(\calZ)$ & \cellcolor{Purple1!20} $\calH^{\gamma}_{d}(\calZ)$ \\\hhline{~||-----}
    & \cellcolor{Firebrick1!10} $\calH^{\delta}_{0}(\calZ)$ & \cellcolor{DarkGoldenrod1!10} $\calH^{\delta}_{a}(\calZ)$ & \cellcolor{Chartreuse1!10} $\calH^{\delta}_{b}(\calZ)$ & \cellcolor{RoyalBlue1!10} $\calH^{\delta}_{c}(\calZ)$ & \cellcolor{Purple1!10} $\calH^{\delta}_{d}(\calZ)$
\end{tabular}
\caption{Spectral table of a theory $\calZ$ on a spatial manifold $M$ with
symmetry $\G$.  The columns are labeled by $\G$-charges (=
$\widehat{\G}$-twists) while the rows are labeled by $\G$-twists (=
$\widehat{\G}$-charges).  The entries of the table are subsector Hamiltonians of the theory $\calZ$,
where $\calH^{\alpha}_{a}$ are the subsector Hamiltonians acting on states with
twist $\alpha$ and charge $a$.}
\label{fig:spectrum_table_Z}
\end{figure}

\subsubsection{Topological Witten effects}
\label{sec:topologicalWitten_derivation}

We now explore how a $\G$-symmetry $\theta$ angle acts on a spectral table, and we will see that they induce topological Witten effects.
We could find this result by just plugging Eq.~\eqref{eq:Hamiltonian_twist+charge} into the $\Theta$ transformation~\eqref{eq:Theta_transformation}. 
However, this method would obscure the underlying physics with lengthy algebraic manipulations.
Therefore, we shall instead make a detour
through $\Theta = S^{\dag}TS$ from Eq.~\eqref{eq:Th=STS}, i.e., 
\begin{equation}
\begin{tikzcd}[row sep= 2.2 em]
    \calZ \arrow[d, "\Theta"] \\
    \calZ_{\theta} = \calZ\!\ast\!\mathcal{I}
\end{tikzcd}
\qquad = \qquad
\begin{tikzcd}
    \calZ \arrow[r, "S"] & \widehat{\calZ} \arrow[d, "T"] \\
    \calZ_{\theta} = \calZ\!\ast\!\mathcal{I} & \widehat{\calZ}\!\otimes\!\widehat{\calI} \arrow[l, "S^{\dagger}"]
\end{tikzcd}\qquad\,.
\end{equation}

Let us start with $S$ transformations. On the spacetime $M\!\times\!S^1_L$,
thanks to the decomposition theorem~\eqref{eq:decomposition-theorem}, the rather
hairy normalization factor~\eqref{eq:N(X,G)} significantly simplifies to 
\begin{equation} 
    N\bigl(M\!\times\!S^1_L,\G\bigr) = \bigl|\H^t\bigl(M, \G\bigr)\bigr|\,,
    \label{eq:simplified_normalization}
\end{equation}
which simply counts the number of inequivalent temporal gauge fields. 
Then, using the Hamiltonian interpretation of $\calZ$ from Eq.~\eqref{eq:Hamiltonian_twist+charge}, we obtain
\begin{equation}\label{eq:hatZ_1}
    \widehat{\calZ}\bigl(M\!\times\!S^1_L,\,a\!+\!b\bigr) = \!\!\!\!\!\!\!\!\sum_{\alpha\in\H^s\bigl(M, \G\bigr)}\!\!\!\!\!\!\!\! \tr\exp\left[-\,L\,\calH^{\alpha}_{a^{\vee}}(\calZ,M) \,-\i\!\int b\cup\alpha\right]\,,\qquad \left\{\begin{matrix}
        a\in\H^s\bigl(M, \widehat{\G}\bigr) \\
        b\in\H^t\bigl(M, \widehat{\G}\bigr)
    \end{matrix}\right..
\end{equation}
Comparing this result with the Hamiltonian interpretation of $\widehat{\calZ}$
in the standard form like Eq.~\eqref{eq:Hamiltonian_twist+charge}, we can verify
the following isomorphisms on the subsector Hamiltonians of $\widehat{\calZ}$
and $\calZ$:
\begin{equation}
    \calH_{\alpha}^a(\widehat{\calZ},M)\,\overset{S}{=}\,\calH_{a^{\vee}}^{-\alpha}(\calZ,M)\,,\qquad \alpha\in\H^s\bigl(M, \G\bigr),\ a\in\H^s\bigl(M, \widehat{\G}\bigr) \,.
\end{equation}
This allows us to sketch the spectral table of $\widehat{\calZ}$ in terms of the
Hamiltonians of $\calZ$ in Fig.~\ref{fig:spectrum_table_Zhat}. Up to the effect of the $\vee$ transformation, which must be analyzed case by case, the $S$ transformation of a $N \times N$ spectral table first reverses the order of the bottom $N-1$ rows and then transposes the table. The $S^{\dag}$ transformation first transposes a spectral table and then reverses the order of the bottom $N-1$ rows.%

\begin{figure}[!ht]
\centering
\begin{tabular}{c||c|c|c|c|c}
    $\widehat{\calZ}$ & \multicolumn{5}{c}{$\widehat{\G}$ charges
    } \\\hline\hline
    \multirow{5}{*}{\begin{turn}{90}
    $\widehat{\G}$ twists
    \end{turn}} 
    & \cellcolor{Firebrick1!50} $\calH^{0}_{0}(\calZ)$ & \cellcolor{Firebrick1!10} $\calH^{\delta}_{0}(\calZ)$ & \cellcolor{Firebrick1!20} $\calH^{\gamma}_{0}(\calZ)$ & \cellcolor{Firebrick1!30} $\calH^{\beta}_{0}(\calZ)$ & \cellcolor{Firebrick1!40} $\calH^{\alpha}_{0}(\calZ)$ 
    \\\hhline{~||-----}
    & \cellcolor{DarkGoldenrod1!50} $\calH^{0}_{a}(\calZ)$ & \cellcolor{DarkGoldenrod1!10} $\calH^{\delta}_{a}(\calZ)$ & \cellcolor{DarkGoldenrod1!20} $\calH^{\gamma}_{a}(\calZ)$ & \cellcolor{DarkGoldenrod1!30} $\calH^{\beta}_{a}(\calZ)$ & \cellcolor{DarkGoldenrod1!40} $\calH^{\alpha}_{a}(\calZ)$ 
    \\\hhline{~||-----}
    & \cellcolor{Chartreuse1!50} $\calH^{0}_{b}(\calZ)$ & \cellcolor{Chartreuse1!10} $\calH^{\delta}_{b}(\calZ)$ & \cellcolor{Chartreuse1!20} $\calH^{\gamma}_{b}(\calZ)$ & \cellcolor{Chartreuse1!30} $\calH^{\beta}_{b}(\calZ)$ & \cellcolor{Chartreuse1!40} $\calH^{\alpha}_{b}(\calZ)$ 
    \\\hhline{~||-----}
    & \cellcolor{RoyalBlue1!50} $\calH^{0}_{c}(\calZ)$ & \cellcolor{RoyalBlue1!10} $\calH^{\delta}_{c}(\calZ)$ & \cellcolor{RoyalBlue1!20} $\calH^{\gamma}_{c}(\calZ)$ & \cellcolor{RoyalBlue1!30} $\calH^{\beta}_{c}(\calZ)$ & \cellcolor{RoyalBlue1!40} $\calH^{\alpha}_{c}(\calZ)$ 
    \\\hhline{~||-----}
    & \cellcolor{Purple1!50} $\calH^{0}_{d}(\calZ)$ & \cellcolor{Purple1!10} $\calH^{\delta}_{d}(\calZ)$ & \cellcolor{Purple1!20} $\calH^{\gamma}_{d}(\calZ)$ & \cellcolor{Purple1!30} $\calH^{\beta}_{d}(\calZ)$ & \cellcolor{Purple1!40} $\calH^{\alpha}_{d}(\calZ)$
\end{tabular}
\caption{Spectral table of $\widehat{\calZ}$ on $M$ expressed in terms of
$\calZ$'s Hamiltonians.  The columns are labeled by $\widehat{\G}$-charges (=
$\G$-twists) while the rows are labeled by $\widehat{\G}$-twists (=
$\G$-charges). }
\label{fig:spectrum_table_Zhat}
\end{figure}

To investigate $T$ transformations in the present context, we first study the Hamiltonian
interpretation of a $\widehat{\G}$-symmetric invertible theory $\widehat{\calI}$
on the spacetime $M\!\times\!S^1_L$. The only way to make the partition function
a phase factor is to let each twisted sector contain a single state with
zero energy. For the $\widehat{\G}$-twist $a\in\H^s\bigl(M, \widehat{\G}\bigr)$,
let $|a\rangle$ denote this corresponding vacuum, and let
$\theta(a)\in\H^s\bigl(M, \G\bigr)$ denote its $\widehat{\G}$-charge. Then the
partition function of $\widehat{\calI}$ is given by 
\begin{equation}\label{eq:I=cup}
    \widehat{\calI}(M\!\times\!S^1_L,a\!+\!b)\ =\ \exp\left(\i\int b\cup \theta(a)\right)\,,\qquad
    \left\{\begin{matrix}
        a\in\H^s\bigl(M, \widehat{\G}\bigr) \\
        b\in\H^t\bigl(M, \widehat{\G}\bigr)
    \end{matrix}\right..
\end{equation}
As a result, the spectral table of an invertible theory takes the form sketched
in Fig.~\ref{fig:spectrum_table_Inv}.  Note that in  Fig.~\ref{fig:spectrum_table_Inv}
we fill in the table with subsector Hilbert spaces instead of subsector
Hamiltonians (which are just all zero). The characteristic feature of a spectral
table for an invertible theory is that every \textit{row} has precisely
one non-empty entry $\C$, but a column is allowed to have multiple non-empty
entries.

\begin{figure}[!ht]
\centering
\begin{tabular}{c||c|c|c|c|c}
    $\widehat{\calI}$ & \multicolumn{5}{c}{
    $\widehat{\G}$ charges
    } \\\hline\hline
    \multirow{5}{*}{\begin{turn}{90}
    $\widehat{\G}$ twists
    \end{turn}} & \cellcolor{Firebrick1!50} $\C$ & $\varnothing$ & $\varnothing$ & $\varnothing$ & $\varnothing$ \\
    \hhline{~||-----}
    & $\varnothing$ & \cellcolor{DarkGoldenrod1!50} $\C$  & $\varnothing$ & $\varnothing$ & $\varnothing$ \\\hhline{~||-----}
    & $\varnothing$ & $\varnothing$ & \cellcolor{Chartreuse1!50} $\C$ & $\varnothing$ & $\varnothing$ \\\hhline{~||-----}
    & $\varnothing$ & $\varnothing$ & $\varnothing$ & \cellcolor{RoyalBlue1!50} $\C$ & $\varnothing$ \\\hhline{~||-----}
    & $\varnothing$ & $\varnothing$ & $\varnothing$  & $\varnothing$ & \cellcolor{Purple1!50} $\C$
\end{tabular}
\caption{Spectral table of an invertible theory $\widehat{\calI}$ on $M$.  The
columns are labeled by $\widehat{\G}$-charges while the rows are labeled by
$\widehat{\G}$-twists.  The entries are subsector Hilbert spaces rather than
subsector Hamiltonians (which are just all zero).  The characteristic feature is
that every \textit{row} has precisely one non-empty entry $\C$.}
\label{fig:spectrum_table_Inv}
\end{figure}

Under a $T$ transformation, using $\widehat{\calZ}$'s Hamiltonian interpretation
and $\widehat{\calI}$'s Hamiltonian interpretation~\eqref{eq:I=cup}, we have
\begin{equation}
\begin{split}
    &\:\ \widehat{\calZ}\!\otimes\!\widehat{\calI}\,\Bigl(M\!\times\!S^1_L,a\!+\!b\Bigr)\\
    =&\: \!\!\!\!\!\!\!\!\sum_{\alpha\in\H^s\bigl(M, \G\bigr)}\!\!\!\!\!\!\!\! \tr\exp\left\{-\,L\,\calH_{\alpha}^a(\widehat{\calZ},M) \,+\i\!\int b\cup\Bigl[\alpha +\theta(a)\Bigr]\right\}
\end{split}\,,\qquad
    \left\{\begin{matrix}
        a\in\H^s\bigl(M, \widehat{\G}\bigr) \\
        b\in\H^t\bigl(M, \widehat{\G}\bigr)
    \end{matrix}\right..
\end{equation}
This implies the following isomorphisms between subsector Hamiltonians of
$\widehat{\calZ}$ and $\widehat{\calZ}\!\otimes\!\widehat{\calI}$:
\begin{equation}
\label{eq:H->H}   \calH^a_{\alpha}\Bigl(\widehat{\calZ}\!\otimes\!\widehat{\calI},M\Bigr)\ \overset{T}{=}\ \calH^a_{\alpha-\theta(a)}\Bigl(\widehat{\calZ},M\Bigr)\,,
    \qquad \alpha\in\H^s\bigl(M, \G\bigr),\ a\in\H^s\bigl(M, \widehat{\G}\bigr)\,.
\end{equation}
We can also obtain the same result by directly inspecting how $\otimes|a\rangle$
modifies the states and the Hamiltonian of the $a$-twisted sector. The
Hamiltonian is not modified since $|a\rangle$ has zero energy. The
$\widehat{\G}$-charge of each state gets shifted by $\theta(a)$ since
$|a\rangle$ carries a $\widehat{\G}$-charge $\theta(a)$. These facts combine to
imply the transformation~\eqref{eq:H->H}. 

These remarks allow us to sketch the spectral table of
$\widehat{\calZ}\!\otimes\!\widehat{\calI}$ in terms of the subsector
Hamiltonians of $\calZ$ in Fig.~\ref{fig:spectrum_table_Zhat_otimes_Ihat}. Comparing
Fig.~\ref{fig:spectrum_table_Zhat} and Fig.~\ref{fig:spectrum_table_Zhat_otimes_Ihat}, we can
see that a $T$ transformation acts on a spectral table through a
\textit{horizontal shuffle}, i.e., it shuffles entries within each row.

\begin{figure}[!ht]
\centering
\begin{tabular}{c||c|c|c|c|c}
    $\widehat{\calZ} \otimes \widehat{\calI}$ & \multicolumn{5}{c}{$\widehat{\G}$ charges
    } \\\hline\hline
    \multirow{5}{*}{\begin{turn}{90}
    $\widehat{\G}$ twists
    \end{turn}} & \cellcolor{Firebrick1!50} $\calH^{0}_{0}(\calZ)$ & \cellcolor{Firebrick1!10} $\calH^{\delta}_{0}(\calZ)$ & \cellcolor{Firebrick1!20} $\calH^{\gamma}_{0}(\calZ)$ & \cellcolor{Firebrick1!30} $\calH^{\beta}_{0}(\calZ)$ & \cellcolor{Firebrick1!40} $\calH^{\alpha}_{0}(\calZ)$ \\\hhline{~||-----}
     & \cellcolor{DarkGoldenrod1!10} $\calH^{\delta}_{a}(\calZ)$ & \cellcolor{DarkGoldenrod1!20} $\calH^{\gamma}_{a}(\calZ)$ & \cellcolor{DarkGoldenrod1!30} $\calH^{\beta}_{a}(\calZ)$ & \cellcolor{DarkGoldenrod1!40} $\calH^{\alpha}_{a}(\calZ)$ & \cellcolor{DarkGoldenrod1!50} $\calH^{0}_{a}(\calZ)$
    \\\hhline{~||-----}
    & \cellcolor{Chartreuse1!20} $\calH^{\gamma}_{b}(\calZ)$ & \cellcolor{Chartreuse1!30} $\calH^{\beta}_{b}(\calZ)$ & \cellcolor{Chartreuse1!40} $\calH^{\alpha}_{b}(\calZ)$ & \cellcolor{Chartreuse1!50} $\calH^{0}_{b}(\calZ)$ & \cellcolor{Chartreuse1!10} $\calH^{\delta}_{b}(\calZ)$
    \\\hhline{~||-----}
    & \cellcolor{RoyalBlue1!30} $\calH^{\beta}_{c}(\calZ)$ & \cellcolor{RoyalBlue1!40} $\calH^{\alpha}_{c}(\calZ)$ & \cellcolor{RoyalBlue1!50} $\calH^{0}_{c}(\calZ)$ & \cellcolor{RoyalBlue1!10} $\calH^{\delta}_{c}(\calZ)$ & \cellcolor{RoyalBlue1!20} $\calH^{\gamma}_{c}(\calZ)$ 
    \\\hhline{~||-----}
     & \cellcolor{Purple1!40} $\calH^{\alpha}_{d}(\calZ)$ & \cellcolor{Purple1!50} $\calH^{0}_{d}(\calZ)$ & \cellcolor{Purple1!10} $\calH^{\delta}_{d}(\calZ)$ & \cellcolor{Purple1!20} $\calH^{\gamma}_{d}(\calZ)$ & \cellcolor{Purple1!30} $\calH^{\beta}_{d}(\calZ)$
\end{tabular}
\caption{Spectral table of $\widehat{\calZ}\!\otimes\!\widehat{\calI}$ on $M$
expressed in terms of $\calZ$'s Hamiltonians.  The choice of a
particular invertible theory $\widehat{\calI}$ turns out to determine a choice of symmetry $\theta$ angle from the perspective of the
original theory $\calZ$. The columns are labeled by
$\widehat{\G}$-charges (= $\G$-twists) while the rows are labeled by
$\widehat{\G}$-twists (= $\G$-charges).}
\label{fig:spectrum_table_Zhat_otimes_Ihat}
\end{figure}

Now we can assemble $\Theta=S^{\dag}TS$ and obtain
\begin{equation}\label{eq:TWE}
    \calH_{a}^{\alpha}(\calZ\!\ast\!\mathcal{I},M)\
     \overset{\Theta}{=}\ \calH_{a}^{\alpha+\theta(a^{\vee})}(\calZ,M)\,,\qquad a\in\H^s\bigl(M, \widehat{\G}\bigr),\ \alpha\in\H^s\bigl(M, \G\bigr)\,,
\end{equation}
through the following chain
\begin{equation}\label{eq:H_STS}
    \calH^{\alpha}_a(\calZ\!\ast\!\mathcal{I},M)\ \overset{S^{\dag}}{=}\ 
    \calH_{-\alpha}^{a^{\vee}}(\widehat{\calZ}\!\otimes\!\widehat{\calI},M)\ \overset{T}{=}\ 
    \calH_{-\alpha-\theta({a^{\vee}})}^{a^{\vee}}(\widehat{\calZ},M)\ \overset{S}{=}\ 
    \calH_{a}^{\alpha+\theta({a^{\vee}})}(\calZ,M)\,.
\end{equation}
This means that to find the spectral table of $\calZ_{\theta} =
\calZ\!\ast\!\mathcal{I}$, we first transpose $\calZ$'s spectral table, then perform a
horizontal shuffle, and finally transpose the table back. We can thus sketch the
spectral table of $\calZ_{\theta}$ in terms of the subsector Hamiltonians of
$\calZ$ as shown in Fig.~\ref{fig:spectrum_table_Z_theta}. Comparing
Fig.~\ref{fig:spectrum_table_Z} and Fig.~\ref{fig:spectrum_table_Z_theta}, we see
that a $\Theta$ transformation acts on a spectral table through a
\textit{vertical shuffle}, i.e., shuffling entries within each column.  This means that the $\Theta$ transformation shuffles the twists (but not the charges!) of states.

\begin{figure}[!ht]
\centering
\begin{tabular}{c||c|c|c|c|c}
    $\calZ_{\theta}$ & \multicolumn{5}{c}{$\G$ charges
    } \\\hline\hline
    \multirow{5}{*}{\begin{turn}{90}
    $\G$ twists
    \end{turn}} 
    & \cellcolor{Firebrick1!50} $\calH^{0}_{0}(\calZ)$ & \cellcolor{DarkGoldenrod1!10} $\calH^{\delta}_{a}(\calZ)$ & \cellcolor{Chartreuse1!20} $\calH^{\gamma}_{b}(\calZ)$ & \cellcolor{RoyalBlue1!30} $\calH^{\beta}_{c}(\calZ)$ & \cellcolor{Purple1!40} $\calH^{\alpha}_{d}(\calZ)$ 
    \\\hhline{~||-----}
    & \cellcolor{Firebrick1!40} $\calH^{\alpha}_{0}(\calZ)$ & \cellcolor{DarkGoldenrod1!50} $\calH^{0}_{a}(\calZ)$ & \cellcolor{Chartreuse1!10} $\calH^{\delta}_{b}(\calZ)$ & \cellcolor{RoyalBlue1!20} $\calH^{\gamma}_{c}(\calZ)$ & \cellcolor{Purple1!30} $\calH^{\beta}_{d}(\calZ)$
    \\\hhline{~||-----}
    & \cellcolor{Firebrick1!30} $\calH^{\beta}_{0}(\calZ)$ & \cellcolor{DarkGoldenrod1!40} $\calH^{\alpha}_{a}(\calZ)$ & \cellcolor{Chartreuse1!50} $\calH^{0}_{b}(\calZ)$ & \cellcolor{RoyalBlue1!10} $\calH^{\delta}_{c}(\calZ)$ & \cellcolor{Purple1!20} $\calH^{\gamma}_{d}(\calZ)$ 
    \\\hhline{~||-----}
    & \cellcolor{Firebrick1!20} $\calH^{\gamma}_{0}(\calZ)$ & \cellcolor{DarkGoldenrod1!30} $\calH^{\beta}_{a}(\calZ)$ & \cellcolor{Chartreuse1!40} $\calH^{\alpha}_{b}(\calZ)$ & \cellcolor{RoyalBlue1!50} $\calH^{0}_{c}(\calZ)$ & \cellcolor{Purple1!10} $\calH^{\delta}_{d}(\calZ)$
    \\\hhline{~||-----}
    & \cellcolor{Firebrick1!10} $\calH^{\delta}_{0}(\calZ)$ & \cellcolor{DarkGoldenrod1!20} $\calH^{\gamma}_{a}(\calZ)$ & \cellcolor{Chartreuse1!30} $\calH^{\beta}_{b}(\calZ)$ & \cellcolor{RoyalBlue1!40} $\calH^{\alpha}_{c}(\calZ)$ & \cellcolor{Purple1!50} $\calH^{0}_{d}(\calZ)$
\end{tabular}
\caption{Spectral table of $\calZ_{\theta} = \calZ\!\ast\!\mathcal{I}$ on $M$ expressed in terms of
$\calZ$'s Hamiltonians.  The columns are labeled by $\G$-charges (=
$\widehat{\G}$-twists) while the rows are labeled by $\G$-twists (=
$\widehat{\G}$-charges).  This demonstrates that the effect of the symmetry $\theta$ angle is to shuffle the twisted sectors of each given charge under the symmetry $\G$.  This is a topological Witten effect. 
}
\label{fig:spectrum_table_Z_theta}
\end{figure}

We now observe that Equation~\eqref{eq:TWE} is simply a generalization of
Eq.~\eqref{eq:spectrum_shuffle} to a general discrete non-anomalous group-like
Abelian symmetry $\G$.  This implies that $\calZ_{\theta}$ differs from $\calZ$
by a topological Witten effect on $\G$! We can obtain $\calZ_{\theta}$'s
subsector Hamiltonians from $\calZ$'s subsector Hamiltonians by shuffling
$\G$-twists while preserving $\G$-charges. Of course, while we have
described this generalization of the topological Witten effect in terms of
compact-space subsector Hamiltonians, it also affects the  operator spectra and
correlation functions on $\R^d$.  Indeed, we saw a concrete illustration of this
in our theories in Section~\ref{sec:Witten_reappraisal}. This is inevitable
because of the standard relations between states and operators in QFTs.

Before moving on, we note that the fact that stacking with an invertible theories
produces the most general possible horizontal shuffles of spectral tables
implies that our construction in this section describes the most general
possible topological Witten effect. This implies the following statement:

\begin{proposition}[\textbf{Witten effects from symmetry $\theta$ angles}]
    \label{prop:witten_effects}
   All topological Witten
effects come from symmetry $\theta$ angles.  Moreover, since all charge Witten
effects are special cases of topological Witten effects,    all
possible charge Witten effects also come from symmetry $\theta$ angles.
\end{proposition}

\subsubsection{Beyond topological Witten effects}
\label{sec:beyond-mapping-tori}

To produce a nontrivial topological Witten effect, the invertible theory used in
the $T$ transformation needs to be nontrivial on at least some spacetimes of the
form $M\!\times\!S^1$. It is interesting to observe that there are invertible
theories that have trivial partition functions on all the spacetimes of the form
$M\!\times\!S^1$, and only become non-trivial on more complicated manifolds. 
The resulting symmetry $\theta$ angles do not produce topological Witten effects,
and must therefore produce some subtler physical consequences.

Perhaps the simplest example of this phenomenon occurs in 3-dimensional theories
with a non-anomalous $\Z_n^{[1]}$ symmetry, such as $SU(n)$ Yang-Mills theories.
According to Eq.~\eqref{eq:classification_discrete}, bosonic 3-dimensional
$\Z_n^{[1]}$-symmetry $\theta$ angles are classified by
\begin{align}
    \Hom\left(\tOmega^{SO}_{3}\bigl(B\Z_n),\tfrac{\R}{2\pi\Z}\right)=\Z_n\,.
\end{align}
The associated invertible theory has the partition function
\begin{equation}
    \widehat{\calI}(X,y) = 
    \exp\left\{\frac{\i2\pi}{n}\int_{X} y\cup\beta y \right\}\,,\qquad y\in H^1(X,\Z_n)\,,
\end{equation}
where $\beta$ is the Bockstein homomorphism\footnote{Very roughly, the Bockstein homomorphism
is a discrete analog of an exterior derivative for differential forms.} with respect to
$0\to\Z_n\to\Z_{n^2}\to\Z_n\to 0$. For somewhat subtle topological reasons that
we relegate to a footnote,%
\footnote{ The image of the Bockstein $\beta: H^{\bullet}(-,\Z_n)\to
H^{\bullet+1}(-,\Z_n)$ lies in the subgroup
$\mathrm{Ext}_{\Z}\bigl(H_{\bullet}(-,\Z),\Z_n\bigr)\subseteq
H^{\bullet+1}(-,\Z_n)$ under the universal coefficient theorem. However, for any
oriented closed 2-manifold $M$, we necessarily have
$\mathrm{Ext}_{\Z}\bigl(H_{\bullet}(M\!\times\!S^1,\Z),\Z_n\bigr)=0$ because
$H_{\bullet}(M\!\times\!S^1,\Z)$ contains no torsion at all. Hence the Bockstein
$\beta$ on manifolds like $M\!\times\!S^1$ is necessarily trivial. }
this partition function is trivial on
$M\!\times\!S^1$ for any oriented closed 2-manifold $M$, i.e.
\begin{equation}
    \widehat{\calI}(M\!\times\!S^1,y)=1\,.
\end{equation}
This means that the corresponding $\Z_n^{[1]}$-symmetry $\theta$ angle does not
induce a topological Witten effect.  
Nevertheless, $\widehat{\calI}(X,y)$ becomes nontrivial on more complicated
spacetimes such as lens spaces $X=SU(2)/\Z_n$. We thus expect that this exotic
$\Z_n^{[1]}$-symmetry $\theta$ angle still influences the $\Z_n^{[1]}$-charged
objects --- such as the confining strings of 3-dimensional $SU(n)$ Yang-Mills
--- in a subtle way. We leave the analysis of this interesting example, as well
as an account of this class of symmetry $\theta$ angles, to future
work. 


\section{Distinct features of \texorpdfstring{$\Theta$}{Theta} transformations}
\label{sec:Theta_vs_others}

Symmetry $\theta$ angles for a finite symmetry $\G$ always gives a discrete parameter in the theory.
It is then natural to wonder about the status of $\G$-symmetric $\theta$ angles among all of the discrete parameters in a QFT.  
In particular, one can contemplate the large family of ``topological manipulations'' that are defined as compositions of various $S$ and $T$ transformations with respect to different non-anomalous sub-symmetries.  
What distinguishes $\Theta$ transformations from other topological transformations?

In this section we shall clarify the relationship between $\G$-symmetry $\theta$ angles and other topological manipulations.
First, in Section~\ref{sec:Theta_vs_others} we will discuss the distinguished role of $\Theta$ transformations in understanding the discrete parameters of a QFT with a \emph{fixed} symmetry.
Then in Section~\ref{sec:remark} we will outline a few general applications of $\Theta$ transformations.  This will serve to further highlight their distinct properties from other topological manipulations.

\subsection{Counting discrete parameters in a \texorpdfstring{$\G$}{G}-symmetric theory}
\label{sec:counting_params}

The topological manipulations to a $\G$-symmetric theory lead to a large discrete moduli space $\calM$ of QFTs.%
\footnote{\label{ft:degeneracy}
Whether different elements of $\calM$ are physically equivalent is an extra question.  Its answer depends on the dynamical details of the QFT.
When there are degeneracies in $\calM$, they can be interpreted as giving additional $0$-form symmetries.
This phenomenon is behind many of the known examples of non-invertible symmetries in spacetime dimension $3$ and $4$; see e.g. Refs.~\cite{Koide:2021zxj,Kaidi:2021xfk,Choi:2022zal,Sun:2023xxv,Shao:2023gho}. 
}
These theories usually have distinct global symmetries.%
\footnote{
From the viewpoint of symmetry-TFT approach (see e.g. Ref.~\cite{Freed:2022qnc}), all the theories in $\calM$ are obtained from taking different topological boundary conditions in a common symmetry-TFT.
}
Even as we focus on non-anomalous finite Abelian invertible $\G$, in $\calM$ there can emerge anomalous, non-Abelian, higher-group, and non-invertible symmetries~\cite{Tachikawa:2017gyf}. However, it is natural to decompose $\calM$ into disjoint subspaces labeled by global symmetries.  
In particular, $\calM$ always has a subspace $\calM_{\G}$ with symmetry $\G$, 
\begin{equation}
    \calM\ =\ \calM_{\G}\sqcup\cdots\cdots\,.
\end{equation}
It is then natural to interpret the moduli space $\calM_{\G}$ as the space of discrete parameters that are related to the symmetry $\G$ of a \emph{single} QFT.  This means that it is especially interesting to study $\calM_{\G}$.  

The most efficient way to explore $\calM_{\G}$ is to consider its own symmetry group,
\begin{equation}
    S(\G)\,,
\end{equation}
defined as the collection of all the physically-distinct topological manipulations that preserve the symmetry $\calG$ and leave $\calM_{\G}$ invariant.
By saying ``physically-distinct'', we mean that we quotient out trivial topological manipulation that merely shift the background gauge fields.
This means that $S(\G)$ describes the transformations that can produce all symmetry-related discrete parameters of a $\G$-symmetric QFT.

The structure of $S(\G)$ is independent of any dynamical details of the QFT.  It only depends on $\G$, the spacetime dimension, and the tangential structure necessary to define the QFT.  It always contains two special subgroups:%
\footnote{A $T$ or $\Theta$ transformation with respect to a sub-symmetry of $\G$ is always equivalent to some $T$ or $\Theta$ transformation with respect to the entire symmetry $\G$.}%
\begin{itemize}
    \item The subgroup $S_T(\G)$ is generated by $T$ transformations using $\G$-symmetric invertible theories.
    \item The subgroup $S_{\Theta}(\G)$ is generated by $\Theta$ transformations using $\widehat{\G}$-symmetric invertible theories.
\end{itemize}
For some purposes the subgroup $S_{\Theta}(\G)$ has a more significant physical role than  $S_T(\G)$, because $T$ transformations only shifts a local $\G$ counterterm, and do not affect the vacuum sector.  On the other hand, $\Theta$ transformation generically do influence vacuum-sector physics.
But for other purposes $S_T(\G)$ is also important.

Whether $S_{\Theta}(\G)$ and $S_T(\G)$ generate all of $S(\G)$ depends on the self-duality properties of $\G$.
First, let us suppose that $\G$ is not self-dual, i.e.~$\widehat{\G}\neq\G$, and contains no self-dual direct summands. Then $S_{\Theta}(\G)$ and $S_T(\G)$ indeed generate all of $S(\G)$.  
Examples of this type include most theories in three spacetime dimensions, especially 3-dimensional gauge theories (see e.g.~Sections~\ref{sec:beyond-mapping-tori} and~\ref{sec:outlook}). 
Note that the group $S(\G)$ \emph{cannot} be a free product,
\begin{equation}
    S(\G)\neq S_{\Theta}(\G)*S_{T}(\G)\,,
\end{equation}
because $\calM_{\G}$ has to be finite.\footnote{Heuristically, consider topological manipulations as permutations of a spectral table. Since a spectral table associated with a finite symmetry is necessarily finite, there are a finite number of such permutations.}
Hence there must be extra relations between $T$ and $\Theta$ transformations such that $S(\G)$ is a finite quotient of $S_{\Theta}(\G)*S_{T}(\G)$.
The simplest possibility is that $T$ and $\Theta$ transformations commute with each other, which leads to $S(\G) = S_{\Theta}(\G)\times S_{T}(\G)$.
However, although it seems difficult to find counterexamples in dimensions $\leq 4$, we see no reason why these transformations should commute in general, and we expect that there should be many examples with $T \Theta \not\simeq \Theta T$ in higher dimensions.

We must also discuss the case when $\G=\widehat{\G}$, or when $\G$ has a self-dual direct summand. 
Examples of QFTs which have this property include gauge theories in four spacetime dimensions, and almost all theories in two spacetime dimensions.  
We shall discuss examples of the former class in Sections~\ref{sec:4d_YM}.
Without loss of generality, let us consider the case of $\G=\widehat{\G}$. In this situation $S_{\Theta}(\G)$ and $S_{T}(\G)$ do not generate $S(\G)$, because the $S$ transformation with respect to $\G$ also leaves $\calM_{\G}$ invariant.
Then we can generate $S(\G)$ by either $S_T(\G)$ and $S$, or $S_{\Theta}(\G)$ and $S$.
With either choice, to keep $\calM_{\G}$ finite, there must be extra relations between $S$ and the other generators. 
Such a relation comes from the existence of $\G$-symmetric invertible Dijkgraaf-Witten theories, which leads to widely-acknowledged modular-group relations, i.e.~schematically,
\begin{equation}
    (ST)^3 \simeq 1\,,\quad\text{or equivalently}\quad (S\Theta)^3 \simeq 1\,.
    \label{eq:modular_relations}
\end{equation}
We provide a detailed analysis on this point in Appendix~\ref{sec:STSTST=1}.
As a result, either $S(\G)$ is a modular group, or at least it contains modular subgroups.

\subsection{General applications}
\label{sec:remark}

In this subsection we discuss three general applications of symmetry $\theta$ angles for a finite symmetry $\G$.
We highlight the distinct features of the $\Theta$ transformations shown in these application compared to general topological manipulations.

First, in Section~\ref{sec:act_on_infrared}, we note that changes in symmetry $\theta$ angles implement generalized Kennedy-Tasaki transformations in the infrared.  Second, in Section~\ref{sec:general_non_inv}, we show that when theories with different values of $\theta$ are degenerate, they have non-invertible $0$-form symmetries.  Finally, we discuss charge Witten effects in Section~\ref{sec:general_charge_witten}.

\subsubsection{Infrared transformations}
\label{sec:act_on_infrared}

Let us investigate the influence of a symmetry $\theta$ angle on the far
infrared physics. When a $\G$-symmetric system is in a completely trivially
gapped phase, the far-infrared limit is described by the trivial QFT. According
to Eq.~\eqref{eq:convolution}, the trivial QFT stays invariant under any
$\Theta$ transformation. As a result, symmetry $\theta$ angles do not affect the
infrared physics of infrared-trivial theories. We hasten to emphasize that this
does \emph{not} mean that they are not interesting in this class of theories,
because they generically have non-trivial effects on  excitations above the
vacuum. We saw an example of this in our discussion of AB effects for magnetic
strings in $U(1)$ Higgs phases in Section~\ref{sec:cheshire_AB_phase}.

\begin{figure}[ht]
\centering
\begin{minipage}{0.48\textwidth}
\centering
\begin{tabular}{c||c|c|c|c|c}
    $\calZ$ & \multicolumn{5}{c}{
    $\G$ charges
    } \\\hline\hline
    \multirow{5}{*}{\begin{turn}{90}
    $\G$ twists
    \end{turn}} & \cellcolor{Firebrick1!50} $\C$ & \cellcolor{DarkGoldenrod1!50} $\C$ & \cellcolor{Chartreuse1!50} $\C$ & \cellcolor{RoyalBlue1!50} $\C$ & \cellcolor{Purple1!50} $\C$ \\\hhline{~||-----}
    & $\varnothing$ & $\varnothing$ & $\varnothing$ & $\varnothing$ & $\varnothing$ \\\hhline{~||-----}
    & $\varnothing$ & $\varnothing$ & $\varnothing$ & $\varnothing$ & $\varnothing$ \\\hhline{~||-----}
    & $\varnothing$ & $\varnothing$ & $\varnothing$ & $\varnothing$ & $\varnothing$\\\hhline{~||-----}
    & $\varnothing$ & $\varnothing$ & $\varnothing$& $\varnothing$ & $\varnothing$
\end{tabular}
\caption*{(a) Spontaneously broken symmetry $\G$.}
\label{tab:broken_symm}
\end{minipage}
\hfill
\begin{minipage}{0.48\textwidth}
\centering
\begin{tabular}{c||c|c|c|c|c}
    $\calZ_{\theta}$ & \multicolumn{5}{c}{
    $\G$ charges
    } \\\hline\hline
    \multirow{5}{*}{\begin{turn}{90}
    $\G$ twists
    \end{turn}} 
    & \cellcolor{Firebrick1!50} $\C$ & $\varnothing$ & $\varnothing$ & $\varnothing$ & $\varnothing$ \\\hhline{~||-----}
    & $\varnothing$ & $\varnothing$ & $\varnothing$ & \cellcolor{RoyalBlue1!50} $\C$ & $\varnothing$ \\\hhline{~||-----}
    & $\varnothing$ & \cellcolor{DarkGoldenrod1!50} $\C$ &$\varnothing$ & $\varnothing$ & $\varnothing$ \\\hhline{~||-----}
    & $\varnothing$ & $\varnothing$ & $\varnothing$ & $\varnothing$ & \cellcolor{Purple1!50} $\C$ \\\hhline{~||-----}
    & $\varnothing$ & $\varnothing$ & \cellcolor{Chartreuse1!50} $\C$ &  $\varnothing$ & $\varnothing$
\end{tabular}
\caption*{(b) $\G$-symmetric SPT.}
\label{tab:SPT}
\end{minipage}
\caption{On the left we show a spectral table for a QFT with a spontaneously
broken symmetry $\G$, and a vanishing $\G$-symmetry $\theta$ angle.  The columns
are labeled by $\G$-charges (= $\widehat{\G}$-twists), the rows by $\G$-twists
(= $\widehat{\G}$-charges), and the entries denote the Hilbert spaces, which are
either one-dimensional or empty. On the right, we show the spectral table when
the $\G$-symmetry $\theta$ angle is turned on.  The table on the right describes
a $\G$-symmetry SPT phase, so varying the symmetry $\theta$ angle relates SPT
phases and symmetry-broken phases.  The infrared effect of symmetry $\theta$
angles is thus to induce generalized Kennedy-Tasaki transformations. }
\label{fig:Kennedy_Tasaki}
\end{figure}

On the other hand, when the infrared limit is not trivial, a symmetry $\theta$
angle can influence the far-infrared physics. Let us consider non-trivial gapped
phases of a $\G$-symmetric system. In general, they are described by non-trivial
$\G$-symmetric TQFTs. 
According to Eq.~\eqref{eq:convolution}, $\Theta$ transformations can transform one such gapped phase to another.
However in general, not all gapped phases can be connected by $\Theta$ transformations.
In particular, $\Theta$ transformations connect all $\G$-symmetric Dijkgraaf-Witten theories, but do not mix them with other TQFTs.
For example, all $3$-dimensional topological orders that allow gapped boundaries and have a maximal non-anomalous subsymmetry $\Z_N^{[1]}$ are connected through $\Z_N^{[1]}$-symmetry $\theta$ angles.

As we mentioned in Section~\ref{sec:counting_params}, when $\calG$ is self-dual the $S$ and $T$ transformations obey modular group relations.  As discussed in Appendix~\ref{sec:STSTST=1}, this comes from the appearance of $\G$-symmetric \emph{invertible} Dijkgraaf-Witten theories.  As a result, when $\calG$ is self-dual, $\Theta$ transformations can transform a phase where $\G$ is spontaneously broken to a $\G$-SPT phase, and vice versa.
This transformation is illustrated in Fig.~\ref{fig:Kennedy_Tasaki} at the level of spectral tables. 
This feature is the hallmark feature of a Kennedy-Tasaki transformation~\cite{Kennedy:1992abc,Kennedy:1992ifl,oshikawa_hidden:1992}. 
See the recent paper~\cite{Li:2023ani} for a detailed discussion of this point (see also Ref.~\cite{Bhardwaj:2023bbf,Li:2023knf,ParayilMana:2024txy}).

\subsubsection{Induced non-invertible symmetries}
\label{sec:general_non_inv}

Some QFTs can stay invariant under shifts of a symmetry $\theta$ angle.  When this happens, extra non-invertible 0-form symmetries appear as a consequence.
This is a special case of Footnote~\ref{ft:degeneracy}, and is also a generalization of Section~\ref{sec:non_inv_chiral}.  
More precisely, let us consider a QFT $\calZ$ with symmetry $\G$ such that
shifting the $\G$ symmetry $\theta$ angles (which naturally form an Abelian
group)  by elements of some subgroup $H$ results in an equivalent QFT. 
Then there are extra symmetry transformations corresponding to elements of $H$, which form an extra non-invertible $0$-form symmetry in the theory.

\begin{figure}[ht]
\centering
\includegraphics[width=0.8\textwidth]{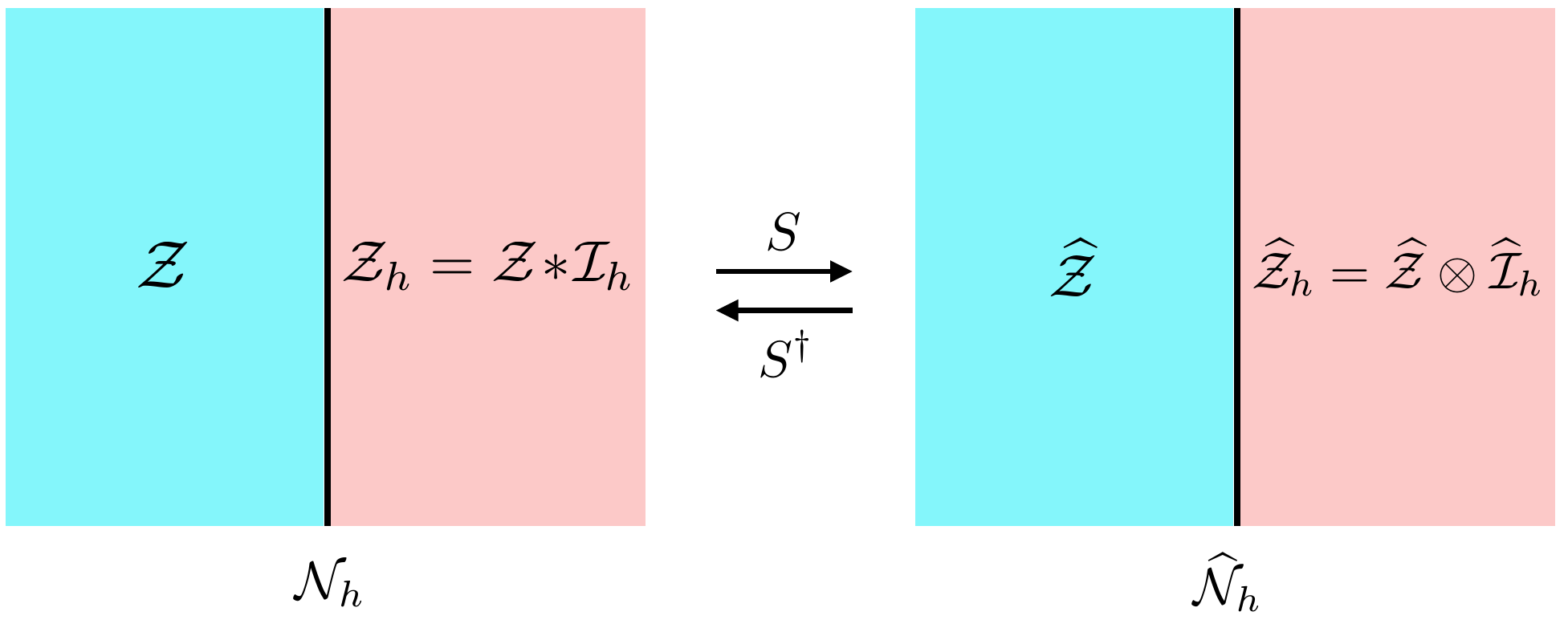}
\caption{We illustrate the $0$-form symmetry defects associated with a
$\G$-symmetry $\theta$ angle when a QFT is invariant under a shift of
$\theta$ angle from $0$ to some non-zero value $\theta = h$. On the
left, we show the non-invertible codimension-$1$ topological operator
$\calN_{h}$ that appears on the boundary.  The operator $\calN_h$ is non-invertible because it does not preserve the locality properties of charged operators, since passing through it can take an operator from the vacuum sector to a twisted sector.  On the right, we show the
related invertible codimension-$1$ defect $\widehat{\calN}_{h}$ in the
dual theory $\hat{\calZ}$. }
\label{fig:non_inv_defect}
\end{figure}

A topological operator $\calN_h$ for the symmetry transformation labeled by $h\in H$ can be constructed by the now-standard half-space construction~\cite{Koide:2021zxj,Choi:2021kmx,Choi:2022zal,Shao:2023gho}.
Namely, we shift the symmetry $\theta$ angle on half of the spacetime and obtain a codimension-$1$ topological operator $\calN_h$ on the interface.
We illustrate this construction on the left-hand side of Fig.~\ref{fig:non_inv_defect}.
Although the symmetry transformations are labeled by $h \in H$ with $H$ being an Abelian group, they are not invertible and do not generate an $H^{[0]}$ symmetry. 
Hence the $\calN_h$ operators have non-invertible fusion rules that involve $\G$, and the collection $\{\calN_h|h\in H\}$ together with $\G$ spans a fusion $(d\!-\!1)$-category.

It is interesting to examine the fate of the non-invertible symmetry from the Fourier point of view (see Section~\ref{sec:Theta_transformation}).  
When $\calZ$ is invariant under $\Theta$ transformations $\in H$, the dual theory $\widehat{\calZ}$ should be invariant under $T$ transformations $\in H$.
Hence $\widehat{\calZ}$ also has an extra 0-form symmetry whose symmetry transformations are also labeled by elements of $H$, as also a special case of Footnote~\ref{ft:degeneracy}.
However, this time the extra symmetry is indeed an invertible symmetry $H^{[0]}$, since $T$ transformations are invertible.

A topological operator $\widehat{\calN_h}$ for this symmetry can be constructed by a $T$ transformation on half of the spacetime,
as illustrated on the right-hand side of Fig.~\ref{fig:non_inv_defect}. 
Hence $\widehat{\calN_h}$ receives an anomaly inflow from the stacked $\widehat{\G}$-symmetric invertible theory on one side.
This means that in $\widehat{\calZ}$, the symmetries $\widehat{\G}$ and $H^{[0]}$ have a mixed 't Hooft anomaly.
If the $\widehat{\G}$-symmetric invertible theory $\widehat{\calI}$ has the partition function on a $d$-dimensional spacetime $X$ given by
\begin{equation}
    \widehat{\calI}(\rho) = \exp\left(\i\int_{X} \omega(\rho)\right)\,,\qquad \rho\in\H(X,\G)\,,
\end{equation}
the mixed anomaly is described by the inflow from the $(d\!+\!1)$-dimensional invertible theory
\begin{equation}
    \calA(\rho,\sigma) = \exp\left(\i\int_{W} \sigma\cup\omega(\rho)\right)\,,\qquad \rho\in\H(W,\G)\,,\ \sigma\in H^1(W,H)\,,\ X=\partial W\,.
\end{equation}
Hence we see that the non-invertible symmetry and the mixed anomaly are exchanged by the $S$ transformation,
\begin{equation}
\begin{tikzcd}
    \text{non-invertible }\calN_{h} \quad \arrow[r, rightharpoonup, yshift=0.1em, "S"] 
    & \arrow[l, rightharpoonup, yshift=-0.1em, "S^{\dag}"] \quad 
    \text{invertible }\widehat{\calN_{h}}\text{ with mixed anomaly}
\end{tikzcd}\,.
\end{equation}
This sort of relation between a mixed 't Hooft anomaly and a non-invertible symmetry was first explained in Ref.~\cite{Kaidi:2021xfk}.

\subsubsection{Charge Witten effects}
\label{sec:general_charge_witten}

When there is an extra non-anomalous symmetry $\calK$ that has a mixed 't Hooft anomaly with $\G$, a $\G$-symmetry $\theta$ angle induces a charge Witten effect on $\calK$, just as in Maxwell theory.  This means that $\calK$ charges get fractionalized in a way that depends on $\G$ charges.
However, unlike in Maxwell theory, the charge Witten effect for finite symmetries can alter the global structure of the symmetry.

For any value of $\theta$, the finite symmetry $\G$ is a non-anomalous sub-symmetry of the full symmetry of the system, and $\calK$ is a non-anomalous quotient of the full symmetry. However, the details of the extension between them and their mixed anomaly can change as a function of $\theta$.  Therefore dialing $\theta$ can change the full global symmetry of the system in this situation. 

This symmetry rewriting caused by charge Witten effects echoes our comments in Section~\ref{sec:act_on_infrared}, where we said that $\Theta$ transformation connect all the $\G$-symmetric Dijkgraaf-Witten theories (almost by construction).
These Dijkgraaf-Witten theories  differ from each other precisely by their total global symmetry and anomaly, and share the common non-anomalous sub-symmetry $\G$.

These ideas were used in Ref.~\cite{Hsin:2020nts} to construct finite 2-group symmetries in quantum field theories.
We direct the readers to Ref.~\cite{Hsin:2020nts} for many examples of charge Witten effects in the finite-symmetry realm.
Here we just give one simple example to illustrate the phenomenon.
Let us consider 4-dimensional Yang-Mills theory with gauge group $SU(4)/\Z_2$.
It then has a $(\Z_2^{[1]})_e$ symmetry that comes from the center of $SU(4)/\Z_2$ and a $(\Z_2)^{[1]}_m$ that comes from the fundamental group of $SU(4)/\Z_2$.
The two symmetries have a standard mixed 't Hooft anomaly.
If we turn on a $(\Z_2)^{[1]}_m$-symmetry $\theta$ angle, then the charge Witten effect rewrites the global symmetry into $\Z^{[1]}_4$, which is the extension of $(\Z_2)^{[1]}_e$ by $(\Z_2)^{[1]}_m$.

\section{Symmetry \texorpdfstring{$\theta$}{theta} angles: the \texorpdfstring{$U(1)$}{U(1)} case}
\label{sec:general_discussion_continuous}

In this section we generalize symmetry $\theta$ angles to non-anomalous $U(1)$ symmetries.  This allows us to interpret our original motivating example, the Maxwell $\theta$ term, as a symmetry $\theta$ angle.   We consider $U(1)$ symmetries that are direct sums of $p$-form $U(1)$ symmetries for various choices of $0 \le p \le d-2$:
\begin{equation}\label{eq:U(1)-type}
    \U = U(1)^{[0]}\oplus \cdots\oplus U(1)^{[1]}\oplus \cdots\oplus U(1)^{[d-2]}\,.
\end{equation}
We will refer to this sort of symmetry as a $U(1)$-type symmetry. We have intentionally labeled it by the symbol $\U$ rather than the symbol $\G$, because we reserve $\G$ for finite invertible Abelian symmetries.
Again, we do not take the mixing between internal and spacetime symmetry into account.

$\G$ gauge fields must have a vanishing curvature, so they can only accommodate global topological data.   
On the other hand, the non-vanishing curvature of $\U$ gauge fields encodes local geometric data.  This has no analog in the discrete case. 
The presence of the geometric data complicates the treatment of $U(1)$-type symmetries.
First in Section~\ref{sec:S&T_U(1)}, we review how the local geometry modifies the standard $S$ and $T$ transformations.
Then in Section~\ref{sec:construction_U(1)}, we explain the construction of $U(1)$-type symmetry $\theta$ angles. 
Eventually in Section~\ref{sec:U(1)_remark}, we briefly remark on an easy-to-think-of but profound generalization of $U(1)$-type symmetry $\theta$ angles, which can show up thanks to the presence of extra geometric data.

\subsection{\texorpdfstring{$S$}{S} and \texorpdfstring{$T$}{T}
transformations in the \texorpdfstring{$U(1)$}{U(1)} realm}
\label{sec:S&T_U(1)}

In the $U(1)$ realm, many of the ideas we relied on in Section~\ref{sec:general_discussion_finite} become more subtle due to the presence of local geometric data in $U(1)$-type gauge fields.
We first review the notion of deformation classes and dual symmetries in Section~\ref{sec:defor&dual}, and then discuss the properties of $S$ and $S^{\dag}$ transformations in Section~\ref{sec:S_continuous}.
Finally we clarify the notion of invertible theories in Section~\ref{sec:T_continuous}.

\subsubsection{Deformation classes and dual symmetries}
\label{sec:defor&dual}

Locally, a $U(1)^{[p]}$ gauge field is a differential $(p\!+\!1)$-form.
Globally, its deformation classes on a manifold $X$ are classified by $H^{p+2}(X,\Z)$, which physically counts the magnetic monopole charges on different $(p\!+\!2)$-cycles in the spacetime.
The deformation class of a $U(1)^{[p]}$ gauge field $a$ will be denoted by
\begin{equation}
    \left[\frac{\d a}{2\pi}\right]\in H^{p+2}(X,\Z)\,.
\end{equation}
This can be interpreted as a $\Z^{[p+1]}$ gauge field.
This generalizes to general $\U$ gauge fields in a factorized way, so that the deformation class of a $\U$ gauge field can be interpreted as a gauge field of the symmetry
\begin{equation}
    [\calU]\equiv\Z^{[1]}\oplus\cdots\oplus\Z^{[2]}\oplus\cdots\oplus\Z^{[d-1]}\,.
\end{equation}

With the definition of the $S$ transformation that we discuss below, the dual symmetry of a $U(1)^{[p]}$ symmetry will be a $U(1)^{[d-p-3]}$ symmetry.
The deformation class of a $U(1)^{[d-p-3]}$ gauge field is a $\Z^{[d-p-2]}$ gauge field.
This generalizes to general $\U$ gauge fields in a factorized way as well.
Namely, the dual symmetry of a $\calU$ symmetry is
\begin{equation}
    \widehat{\U} \equiv U(1)^{[d-3]}\oplus \cdots\oplus U(1)^{[d-4]}\oplus \cdots\oplus U(1)^{[-1]}\,,
\end{equation}
and the deformation classes of $\widehat{\U}$ can be interpreted as gauge fields of the symmetry
\begin{equation}
    \Bigl[\widehat{\calU}\Bigr]\equiv\Z^{[d-2]}\oplus\cdots\oplus\Z^{[d-3]}\oplus\cdots\oplus\Z^{[0]}\,.
\end{equation}

We now make a remark on the notion of $U(1)^{[-1]}$ symmetry (see e.g.~\cite{Cordova:2019jnf,Cordova:2019uob,Tanizaki:2019rbk,Heidenreich:2020pkc,Aloni:2024jpb}) which appears in the dual symmetry $\widehat{\U}$ above.
A $U(1)^{[-1]}$ gauge field is simply a $2\pi$-periodic compact scalar field.
Having a $U(1)^{[-1]}$ symmetry is defined to mean having a $2\pi$-periodic parameter $\varphi$ in the theory, and turning on a background $U(1)^{[-1]}$ gauge field means promoting the parameter $\varphi$ to a spacetime-dependent field $\varphi(x)$.
Allowing $U(1)^{[-1]}$ symmetry as a dual symmetry can produce interesting symmetry $\theta$ angles that involve $U(1)^{[d-2]}$ symmetries; see the discussion around Eq.~\eqref{eq:Omega_3_U(1)} as well as Section~\ref{sec:cheshire_examples}.

We should emphasize that while we allow $U(1)^{[-1]}$ symmetries to appear in $\widehat{\calU}$, we \emph{do not} allow $U(1)^{[-1]}$ symmetry to be a part of $\calU$ itself.  
This is because a $U(1)^{[-1]}$ symmetry just means having a parameter $\varphi$ in the theory.
If we apply the construction in the section below  to such a $U(1)^{[-1]}$ symmetry, we will simply rediscover the parameter $\varphi$ itself.  This means that a ``$U(1)^{[-1]}$-symmetry $\theta$ angle'' is tautologically equivalent to having a  ``$U(1)^{[-1]}$ symmetry'' in the first place.

\subsubsection{\texorpdfstring{$S$}{S} transformations}
\label{sec:S_continuous}

Let us consider a $d$-dimensional theory $\calZ$ with an anomaly-free global $U(1)^{[p]}$ symmetry on a closed oriented spacetime $X$.
For a background $U(1)^{[p]}$ gauge field $A$ on $X$, let $\calZ(A)$ denote its partition function.
The $S$ transformation from $\calZ$ to the dual theory $\widehat{\calZ}$ is given by~\cite{Witten:2003ya} 
\begin{equation}\label{eq:S_continuous}
    \widehat{\calZ}(B) \equiv \int\!\calD a\ \calZ(a)\,\exp\left( -\frac{\i}{2\pi}\int  {\d B \wedge a}   \right)\,,
\end{equation}
where $B$ is a background gauge field for the $U(1)^{[d-p-3]}$ symmetry of $\calZ$.
To see the physics of this $S$ transformation, it is illuminating to first turn on a kinetic term for the dynamical gauge field:
\begin{equation}
    \widehat{\calZ}_{g}(B) = \int\!\calD a\ \calZ(a)\,\exp\left(-\frac{1}{2g^2}\int\d a\wedge\star\d a\   {+(-1)^{d-p}} \frac{\i}{2\pi}\int B\wedge\d a \right)\,,
\end{equation}
and then take the formal strong-coupling limit $g\to\infty$.
Here the dual symmetry $U(1)^{[d-p-3]}$ can be recognized as the magnetic symmetry associated with the topological current $\star j =\frac{\d a}{2\pi}$.
Thanks to the absence of a kinetic term in  Eq.~\eqref{eq:S_continuous}, the $S$ transformation  preserves a QFT's metric-dependence pattern (e.g., it transforms conformal theories to conformal theories).
In contrast to a finite-symmetry $S$ transformation, which modifies only the global structure of a QFT, a $U(1)$-type $S$ transformation has a profound influence on the local physics.
For example, the $S$ transformation implements the celebrated particle-vortex duality on 3-dimensional $U(1)^{[0]}$-symmetric conformal theories~\cite{Karch:2016sxi,Seiberg:2016gmd,Senthil:2018cru}.

We should make a technical remark here.
Since we allow the spacetime $X$ to be a very general closed oriented $d$-manifold, there are in general non-trivial deformation classes for $U(1)$ gauge fields.
As a result, the level-1 BF coupling we used in the $S$ transformation~\eqref{eq:S_continuous}, 
\begin{equation}\label{eq:adb}
    \exp\left(-\i\int \frac{ {\d B\wedge A }}{2\pi}\right)\,,
\end{equation}
is not quite well-defined in general.
It is well-defined only when $[\frac{\d A}{2\pi}]=0$, meaning that $A$ sits in the trivial deformation class.
The reason is that this is the only situation in which we can express $A$ as a global differential $(p\!+\!1)$-form.  The normalization of the exponent of Eq.~\eqref{eq:adb} then ensures the independence of the many gauge-equivalent choices of the global $(p\!+\!1)$-form $A$.
Similarly, when instead $[\frac{\d B}{2\pi}]=0$, we can define
\begin{equation}\label{eq:bda}
    \exp\left(\i\, {(-1)^{d-p}}\!\!\int \frac{B \wedge \d A}{2\pi} \right)\,,
\end{equation}
using partial integration.
But when simultaneously $[\frac{\d A}{2\pi}]\neq0$ and $[\frac{\d B}{2\pi}]\neq0$, neither Eq.~\eqref{eq:adb} nor Eq.~\eqref{eq:bda} are well-defined.

The solution is to give up on defining the integrand in the exponential of the BF theory as a global differential $d$-form. 
Instead, the integrand turns out to be well-defined as a $U(1)^{[d-1]}$ gauge field. 
It turns out that there is a notion of multiplication $\boxtimes$ of gauge fields such that a $U(1)^{[p]}$ gauge field $A$ and a $U(1)^{[d-p-3]}$ gauge field $B$ can multiply to a $U(1)^{[d-1]}$ gauge field $A \boxtimes B$. The operation $\boxtimes$ is known as the Beilinson-Deligne cup product on ordinary differential cohomology~\cite{Deligne:1971pmi,Beilinson:1985mct,Gajer:1997nla,Hopkins:2002rd}.
Then the exponential integral 
\begin{equation}\label{eq:AxB} 
    \exp\left( -\i \int A \boxtimes B\right)
\end{equation}
is understood as the holonomy of the $U(1)^{[d-1]}$ gauge field $A \boxtimes B$ on $X$. 
In the rest of this paper we will keep using expressions like  Eqs.~\eqref{eq:adb} and~\eqref{eq:bda} to keep our presentation more accessible, but they should always be viewed as shorthands for globally-well-defined expressions like Eq.~\eqref{eq:AxB}.

For our discussion on $S$ transformations, we must also verify that gauging the dual symmetry will bring us back to the original theory.
The key is to apply the identity
\begin{equation}\label{eq:delta_functional}
    \int\!\calD a\,\calD b\,\exp\left(\frac{\i}{2\pi}\int a \wedge \d b \right)\,F(a) \ =\ F(0)\,,
\end{equation}
for any gauge-invariant functional $F(a)$.
Following Ref.~\cite{Witten:2003ya}, we can understand this identity via the following steps.
First, the $b$ path integral over the trivial deformation class of $b$ --- in the form of Eq.~\eqref{eq:bda} --- forces $a$ to be flat, i.e.~$\d a=0$.
Second, the summation over deformation classes $[\frac{\d b}{2\pi}]\in H^{d-p-1}(X,\Z)$ --- in the form of Eq.~\eqref{eq:adb} --- forces $a$ to have trivial holonomies.
Hence the path integral $\propto F(0)$.
Finally, to see that  ``$\propto$'' can be replaced by ``$=$'' we can use what we know about the case $F(a)=1$.
Specifically, we know that
\begin{equation}
    \int\!\calD a\,\calD b\,\exp\left(\frac{\i}{2\pi}\int a \wedge \d b \right) \ = 1\,,
\end{equation}
because of the triviality of the level-1 BF theory.%
\footnote{
To be more accurate, the partition function of the level-1 BF theory is forced to be trivial only in odd dimensions.
An Euler counterterm can appear in even dimensions, but one can always choose it to be trivial.
}
This demonstrates the identity Eq.~\eqref{eq:delta_functional}, using which we can show that
\begin{equation}\label{eq:S^dag_continuous}
    \calZ(A) = \int\!\calD b\ \widehat{\calZ}(b)\,\exp\left( \frac{\i}{2\pi}\int  {\d b \wedge A}\right)\,.
\end{equation}
Due to the supercommutativity of the wedge product, this operation is an $S$ transformation on $\widehat{\calZ}$ followed by flipping the sign of the gauge field $A\to  {(-)^{d(p+1)}}A$.
We then refer to the operation in Eq.~\eqref{eq:S^dag_continuous} as the $S^{\dag}$ transformation.  One can check that $S^{\dag}S=SS^{\dag}=1$.

Finally, all the  discussion above naturally generalizes to generic $U(1)$-type symmetries $\U$ in the obvious factorized way.

\subsubsection{Invertible theories}
\label{sec:T_continuous}

Due to the presence of local geometric data, $\widehat{\U}$-symmetric invertible theories are more complicated than $\widehat{\G}$-symmetric invertible theories.
While $\widehat{\U}$-symmetric invertible theories are independent%
\footnote{
For a continuous symmetry, non-topological terms that depend on the spacetime metric (e.g.~a kinetic term) may survive in the infrared limit of the partition function with background gauge fields.
A trivially-gapped system is described by an invertible theory in the infrared only up to such non-topological terms.
}
of the spacetime metric, they can have a dependence on the local geometric data of background gauge fields. 
Probably the most well-known example is the classical Chern-Simons theory for a background $U(1)^{[0]}$ gauge field $C$:
\begin{align}\label{eq:CS}
    \widehat{\calI}_k(C) = \exp\left(\frac{i k}{4\pi} \int C \wedge\d C \right) \,, \qquad  k \in \Z\,.
\end{align}
This gives a 3-dimensional fermionic
or bosonic $U(1)^{[0]}$-symmetric invertible theory for each odd or even integer $k$, respectively.
$\widehat{\calI}_k(C)$ does not depend on the spacetime metric, but it changes smoothly under smooth deformations of $C$.
These invertible theories appear in the description of integer quantum Hall states.

The presence of local physics makes general $\widehat{\U}$-symmetric invertible theories quite interesting.
However, for our application, we only need a simpler class of invertible theories that stay invariant under continuous deformations of background gauge fields.
We shall refer to them as \textit{strictly-topological} $\widehat{\U}$-symmetric invertible theories.
In essence, they are $[\widehat{\calU}]$-symmetric invertible field theories, with background $[\widehat{\calU}]$ gauge fields given by the deformation classes of $\widehat{\U}$ gauge fields.

For example, let us consider 4-dimensional strictly-topological $U(1)^{[0]}$-symmetric invertible theories, with a background $U(1)^{[0]}$ gauge field $C$.
In essence, they are $\Z^{[1]}$-symmetric invertible theories with the background $\Z^{[1]}$ gauge field given by $[\frac{\d C}{2\pi}]$.
Their partition functions are given by
\begin{align}\label{eq:theta_4D}
    \widehat{\calJ}_{\theta}\left(\left[\frac{\d C}{2\pi}\right]\right) = \exp\left(\frac{i \theta}{2} \int \left[\frac{\d C}{2\pi}\right] \cup \left[\frac{\d C}{2\pi}\right] \right) = \exp\left(\frac{i \theta}{8\pi^2} \int \d C \wedge\d C\right)\,,
\end{align}
with a continuous parameter $\theta$.
The periodicity of $\theta$ is $2\pi$ on spin manifolds and $4\pi$ on other oriented manifolds.


\subsection{Strictly-topological \texorpdfstring{$\Theta$}{Theta} transformations}
\label{sec:construction_U(1)}

In principle, we could use \emph{any} $\widehat{\U}$-symmetric invertible theory to define a $T$ transformation and in turn a $\Theta$ transformation via $\Theta=S^{\dag}TS$.
For example, Witten's $T$ transformation in Ref.~\cite{Witten:2003ya} corresponds to stacking with the classical Chern-Simons theory~\eqref{eq:CS}.
However, in this paper we focus on $\Theta$ transformations that use strictly-topological $\widehat{\U}$-symmetric invertible theories because, as we will see shortly, they are topological manipulations and do not change the local physics.

$\Theta$ transformations that use non-strictly-topological invertible theories are also very interesting in their own right, and are briefly discussed in Section~\ref{sec:U(1)_remark}.  We save a more detailed treatment of non-strictly-topological transformations for future work.

\subsubsection{Local physics is not affected}
\label{sec:topologicalness_of_Theta}

Since $S$ transformations in the $U(1)$ realm modify the local physics, it is natural to wonder whether $\Theta$ transformations modify the local physics as well.       
We now demonstrate that the wildness of the $S$ transformations is tamed by the strictly-topological property of our $T$ transformations, so that the transformations we call ``strictly-topological $\Theta = S^{\dag} T S$ transformations'' do not affect the local physics, just as their name is meant to suggest.
To minimize the notational mess, we will focus on a $U(1)^{[p]}$ symmetry in this section.  
It is straightforward to generalize the argument to generic $U(1)$-type symmetries.

Let us consider a $U(1)^{[p]}$-symmetric theory $\calZ$ on a $d$-dimensional spacetime $X$.
The dual symmetry of $U(1)^{[p]}$ is $U(1)^{[d-p-3]}$, and the associated strictly-topological $U(1)^{[d-p-3]}$-symmetric invertible theory is
\begin{equation}
    \widehat{\calJ}_{\theta}\left(\left[\frac{\d C}{2\pi}\right]\right)\,.
\end{equation}
The corresponding $\Theta$ transformation of $\calZ$ leads to 
\begin{equation}\label{eq:Theta_U(1)}
    \calZ_{\theta}(X,A) \equiv \int\!\calD b\,\calD c\ \calZ(X,A-b)\ \widehat{\calJ}_{\theta}\left(\left[\frac{\d c}{2\pi}\right]\right)\,\exp\left(\frac{\i}{2\pi}\int  {\d c\wedge b}\right)\,,
\end{equation}
where $A$ and $b$ are $U(1)^{[p]}$ gauge fields and $c$ is a $U(1)^{[d-p-3]}$ gauge field.  
Now, for each $\rho\in H^{d-p-1}(X,\Z)$, let us pick a reference $U(1)^{[d-p-3]}$ gauge field $c_{\rho}$ such that $[\frac{\d c_{\rho}}{2\pi}]=\rho$.  
Then any $U(1)^{[d-p-3]}$ gauge field in the deformation class $\rho$ can be written as $c_{\rho}+\xi$, where $\xi$ is a global differential $(d\!-\!p\!-\!2)$-form.
As a result, we can express Eq.~\eqref{eq:Theta_U(1)} as
\begin{equation}
\begin{split}
    \calZ_{\theta}(X,A)
    \,\,\,&= \sum_{\rho\in H^{d-p-1}(X,\Z)}\int\!\calD b\ \calZ(X,A-b)\ \widehat{\calJ}_{\theta}(\rho)\ \exp\left( \frac{\i}{2\pi}\int  {\d c_{\rho} \wedge b} \right)\\
    &\qquad\qquad\qquad\qquad\quad \times\int\!\calD \xi\exp\left( {-(-1)^{d-p}\frac{\i}{2\pi}\int \xi\wedge\d b }\right)\,.
\end{split}
\end{equation}
Since $\xi$ is a global differential form, the path integral over $\xi$ just sets $\d b=0$, forcing  $b$ to be a flat gauge field.
Hence we obtain
\begin{align}
    \calZ_{\theta}(X,A) \,\,\,&= \sum_{\rho\in H^{d-p-1}(X,\Z)}\int_{\text{flat}}\!\!\!\!\calD b\ \calZ(X,A-b)\ \widehat{\calJ}_{\theta}(\rho)\ \exp\left( \i\int  {\frac{\d c_{\rho}}{2\pi} \wedge b}\right)\,.
    \label{eq:flat_b}
\end{align}
We explain the precise meaning of $\int_{\rm flat} \calD b$ below. The exponentiated integral of the wedge product depends on the deformation class $\rho$, but it is independent of the particular choice of reference $c_{\rho}$.  Geometrically, it simply evaluates the holonomy of $b$ along the $(p\!+\!1)$-cycle that is the Pontryagin dual of $[\frac{\d c_{\rho}}{2\pi}]=\rho$.  

We now explain the precise meaning of $\int_{\textrm{flat}} \calD b$, and also eliminate the explicit appearance of the reference $c_{\rho}$.
Exactly like the finite-symmetry cases, gauge equivalence classes of flat $U(1)^{[p]}$ gauge fields are classified by $H^{p+1}\bigl(X,\tfrac{\R}{2\pi\Z}\bigr)$.
This is because
\begin{equation}\label{eq:UCT_U(1)}
    H^{\bullet}\left(X,\tfrac{\R}{2\pi\Z}\right) = \Hom\left( H_{\bullet}\left(X,\Z\right),\,\tfrac{\R}{2\pi\Z}\right)
\end{equation}
specifies the holonomies on each $\bullet$-cycle.
Then for $b\in H^{p+1}\bigl(X,\tfrac{\R}{2\pi\Z}\bigr)$ and $\rho\in H^{d-p-1}(X,\Z)$, there is a cup product $ {\rho\cup b}\in H^{d}\bigl(X,\tfrac{\R}{2\pi\Z}\bigr)$ which evaluates the holonomy of $b$ along the cycle $\tilde{\rho}\in H_{p+1}(X,\Z)$ that is the Pontryagin dual of $\rho$.
Therefore, to explicitly eliminate the appearance of $c_{\rho}$ in Eq.~\eqref{eq:flat_b}, we can replace the wedge product with this cup product, and replace $\int_{\rm flat}$ with a ``path integral'' over $H^{p+1}\left(X,\tfrac{\R}{2\pi\Z}\right)$.  This leads to
\begin{align}\label{eq:flat_cup}
    \calZ_{\theta}(X,A) \,\,\,&= \sum_{\rho\in H^{d-p-1}(X,\Z)}\underset{H^{p+1}\left(X,\tfrac{\R}{2\pi\Z}\right)}{\int}\!\!\!\!\!\!\!\!\!\!\! {\calD b}\quad \calZ(X,A-b)\ \widehat{\calJ}_{\theta}(\rho)\ \exp\left( \i\int  {\rho\cup b} \right)\,.
\end{align}
The ``path integral'' over $H^{p+1}\bigl(X,\tfrac{\R}{2\pi\Z}\bigr)$ has the following definition.
Equation~\eqref{eq:UCT_U(1)} implies that, in general, cohomology groups with coefficients in $\tfrac{\R}{2\pi\Z}$  can be decomposed as 
\begin{equation}\label{eq:U(1)cohomology}
    H^{\bullet}\left(X,\tfrac{\R}{2\pi\Z}\right)\ =\ \left(\tfrac{\R}{2\pi\Z}\right)^{\beta_{\bullet}}\ \oplus\ G^{\bullet}\,.
\end{equation}
Here $\beta_{\bullet}$ denotes the $\bullet$-th Betti number, which is just the rank of $H_{\bullet}(X,\Z)$.
The finite group $G^{\bullet}$ is the character group of the torsional subgroup of $H_{\bullet}(X,\Z)$.
As a result, when we write the ``path-integral'' over $H^{\bullet}\bigl(X,\tfrac{\R}{2\pi\Z}\bigr)$, what we actually mean is a finite sum and a finite-dimensional integral
\begin{equation}
    \underset{H^{\bullet}\left(X,\tfrac{\R}{2\pi\Z}\right)}{\int}\!\!\!\!\!\!\!\! {\calD b}\ f(b)\ \equiv\ \frac{1}{|G^{\bullet}|}\sum_{b_0\in G^{\bullet}}\ \int_{\left(\tfrac{\R}{2\pi\Z}\right)^{\beta_{\bullet}}}\frac{\d^{\beta_{\bullet}}\boldsymbol{b}}{(2\pi)^{\beta_{\bullet}}}\ f(\boldsymbol{b}\oplus b_{0})\,.
\end{equation}
Above we have decomposed $b=\boldsymbol{b}\oplus b_{0}$ according to Eq.~\eqref{eq:U(1)cohomology}.
The key point of this discussion is that the equation defining $\calZ_{\theta}$, Eq.~\eqref{eq:flat_cup}, does not involve a true path integral.  It involves only a finite sum and a finite-dimensional integral, which both involve only topological data.  The local physics is mainfestly left invariant.  This should be contrasted with the ``raw'' $S$ transformation in Eq.~\eqref{eq:S_continuous}, which certainly changes the local physics unless it is part of the strictly-topological $\Theta$ transformation. 

\subsubsection{Symmetry \texorpdfstring{$\theta$}{theta} angles}
\label{sec:Theta_continuous}

We can now propose a definition for symmetry $\theta$ angles.

\begin{definition}[\textbf{Symmetry $\theta$ angle}]
    \label{def:sym_theta_continuous}
    For a QFT $\calZ$ with a non-anomalous $U(1)$-type global symmetry $\U$, a \textit{$\U$-symmetry $\theta$ angle} is defined as the result of applying a strictly-topological $\Theta$ transformation with respect to $\U$.
\end{definition}

Given this definition, $\U$-symmetry $\theta$ angles are classified by the strictly-topological $\widehat{\U}$-symmetric invertible theories used in the constituent $T$ transformations.
Therefore, just as in the finite-symmetry case, the known classification of invertible theories implies that the collection of all $d$-dimensional $\U$-symmetry $\theta$ angles naturally forms a finite Abelian group:
\begin{equation}\label{eq:classification_continuous}
    \Hom\left(\tOmega^{SO}_{d}\bigl(\calB\widehat{\U}\bigr),\tfrac{\R}{2\pi\Z}\right) \quad\text{or}\quad \Hom\left(\tOmega^{Spin}_{d}\bigl(\calB\widehat{\U}\bigr),\tfrac{\R}{2\pi\Z}\right)\,,
\end{equation}
in bosonic or fermionic theories, respectively.
Here $\calB\widehat{\U}$ still denotes the classifying space of $\widehat{\U}$. 
Namely, $\calB\widehat{\U}$ is an auxiliary topological space such that for any $X$, $[X,\calB\widehat{\U}]$ classifies all the deformation classes of $\widehat{\U}$ gauge fields.
In particular, the classifying space of a $U(1)^{[p]}$ symmetry is the Eilenberg-MacLane space
\begin{equation}
    B^{p+1}U(1)\ =\ B^{p+2}\Z\ =\ K(\Z,p+2)\,,
\end{equation}
which satisfies
\begin{equation}
    \left[X,B^{p+1}U(1)\right]\ =\ H^{p+2}\left(X,\Z\right)\,.
\end{equation}
This classifying space again reminds us that the deformation class of a $U(1)^{[p]}$ gauge field can be represented by a $\Z^{[p+1]}$ gauge field.
The space $\calB\widehat{\U}$ is just a direct product of spaces like $B^{p+1}U(1)$.

The moduli space of $\U$-symmetry $\theta$ angles can be either continuous or discrete.
More precisely, a $\Z$ summand in one of the relevant bordism groups in Eq.~\eqref{eq:classification_continuous} contributes a continuous $U(1)$ modulus, while a $\Z_n$ summand contributes a discrete $\Z_n$ modulus. 
For example, to classify 4-dimensional fermionic $U(1)^{[1]}$-symmetry $\theta$ angles, given the dual symmetry $U(1)^{[0]}$, we should look at the bordism group
\begin{equation}
    \tOmega^{Spin}_{4}\bigl(BU(1)\bigr) = \Z\,.
\end{equation}
This bordism group corresponds to Eq.~\eqref{eq:theta_4D}, and produces a continuous $U(1)$ moduli space in 4-dimensional fermionic QFTs with $U(1)^{[1]}$ symmetries.
However, if we instead want to classify 3-dimensional fermionic $U(1)^{[1]}$-symmetry $\theta$ angles, given the dual symmetry $U(1)^{[-1]}$, we should look at the bordism group
\begin{equation}\label{eq:Omega_3_U(1)}
    \tOmega^{Spin}_{3}\bigl(U(1)\bigr) = \Z_2\,.
\end{equation}
This bordism group contributes to a discrete $\Z_2$ moduli space in 3-dimensional fermionic QFTs with $U(1)^{[1]}$ symmetries.

If $\U$ is an exact symmetry, then continuous $\U$-symmetry $\theta$ angles can flow along an RG trajectory.  In contrast, discrete $\U$-symmetry $\theta$ angles  are invariant on any RG flow.
If $\U$ is an emergent symmetry in an EFT, then $\U$-symmetry $\theta$ angle are parameters of the EFT.  Different short-distance physics may lead to different symmetry $\theta$ angles in the EFT.
Discrete $\theta$ angles label sharply distinct phases of matter described by these EFTs.

\subsubsection{Topological Witten effects}
\label{sec:Witten_continuous}

Following the same analysis as in Sec.~\ref{sec:topologicalWitten_consequences}, we can explore the physical consequence of $\calU$-symmetry $\theta$ angles and demonstrate topological Witten effects.
To minimize the notational mess, we will focus on a $U(1)^{[p]}$ symmetry just like Sec.~\ref{sec:topologicalness_of_Theta}.
It is straightforward to generalize the argument to generic $U(1)$-type symmetries.

We first repeat the treatment in Sec.~\ref{sec:twist&charge} for a $U(1)^{[p]}$ symmetry.
Let us consider a thermal spacetime $X=M\!\times\!S^1_L$ that supports  flat $U(1)^{[p]}$ gauge fields and $\Z^{[d-p-2]}$ gauge fields.
We then distinguish the spatial and the temporal parts of the gauge fields based on the decomposition theorem~\eqref{eq:decomposition-theorem}:
\begin{subequations}
\begin{align}
    H^{p+1}\left(M\!\times\!S^1,\tfrac{\R}{2\pi\Z}\right)\ &\simeq\  H^{p+1}\left(M,\tfrac{\R}{2\pi\Z}\right)\ \oplus\ H^{p}\left(M,\tfrac{\R}{2\pi\Z}\right)\,,\\
    H^{d-p-1}\left(M\!\times\!S^1,\Z\right) \ &\simeq\ H^{d-p-1}\left(M,\Z\right)\ \oplus\ H^{d-p-2}\left(M,\Z\right)\,.
\end{align}
\end{subequations}
A temporal flat $U(1)^{[p]}$ gauge field $\beta\in H^{p}\left(M,\tfrac{\R}{2\pi\Z}\right)$ inserts a topological operator $\calU_{\beta}$ in the thermal partition function. 
A spatial flat $U(1)^{[p]}$ gauge field $\alpha\in H^{p+1}\left(M,\tfrac{\R}{2\pi\Z}\right)$ moves the Hamiltonian from the vacuum-sector $\calH$ to a twisted-sector $\calH^{\alpha}$.
With a flat background $U(1)^{[p]}$ gauge field $\alpha + \beta$, the partition function of a $U(1)^{[p]}$-symmetric theory $\calZ$ has the interpretation,
\begin{equation}
    \calZ(M\times S^1_L,\alpha+\beta) = 
    \tr \Bigl\{\calU_{\beta}\exp\bigl[-\,L\,\calH^{\alpha}(\calZ,M)\bigr]
     \Bigr\}\,,\qquad
\begin{cases}
    \alpha\in H^{p+1}\left(M,\tfrac{\R}{2\pi\Z}\right)\\
    \beta\in H^{p}\left(M,\tfrac{\R}{2\pi\Z}\right)
\end{cases}\,.
\end{equation}
Similarly, a $\Z^{[d-p-2]}$ gauge field $\rho$ also decomposes into spatial and temporal parts, i.e.~$\rho=\mu+\nu$ with spatial $\mu\in H^{d-p-1}(M,\Z)$ and temporal $\nu\in H^{d-p-2}(M,\Z)$.
Then the partition function of a $\Z^{[d-p-2]}$-symmetric invertible theory $\widehat{\calJ}$ has the interpretation
\begin{equation}
    \widehat{\calJ}(M\times S^1_L,\mu+\nu) = \tr \Bigl\{\calU_{\nu}\exp\bigl[-\,L\,\calH^{\mu}(\widehat{\calJ},M)\bigr] \Bigr\}\,,\qquad 
    \begin{cases}
    \mu\in H^{d-p-1}\left(M,\Z\right)\\
    \nu\in H^{d-p-2}\left(M,\Z\right)
\end{cases}\,.
\end{equation}
Based on these Hamiltonian interpretations, we again introduce the following terminology: 
\begin{itemize}
        \item a $U(1)^{[p]}$ twist on $M$ is an element of the spatial $H^{p+1}\left(M,\tfrac{\R}{2\pi\Z}\right)$;
        \item a $\Z^{[d-p-2]}$ twist on $M$ is an element of the spatial $H^{d-p-1}(M,\Z)$;
        \item a $U(1)^{[p]}$ charge on $M$ is a character of the temporal $H^{p}\left(M,\tfrac{\R}{2\pi\Z}\right)$;
        \item a $\Z^{[d-p-2]}$ charge on $M$ is a character of the temporal $H^{d-p-2}(M,\Z)$;
\end{itemize}
Just as before, thanks to Poincar\'e duality, the cup product on $M\!\times\!S^1_L$ implies the correspondence
\begin{equation}
    \text{a $U(1)^{[p]}$-charge }=
    \text{ a $\Z^{[d-p-2]}$-twist}\,,\qquad 
    \text{a $\Z^{[d-p-2]}$-charge }=\text{ a $U(1)^{[p]}$-twist}\,.
\end{equation}

We can now decompose the Hamiltonian of each twisted sector into subsectors with different charges.
For a $U(1)^{[p]}$-symmetric theory $\calZ$, we have the decomposition
\begin{equation}
    \calH^{\alpha}(\calZ,M) = \!\!\!\!\!\!\! \bigoplus_{\sigma\in H^{d-p-1}(M,\Z)} \!\!\!\!\!\!\! \calH^{\alpha}_{\sigma}(\calZ,M)\,,\qquad \alpha\in H^{p+1}\left(M,\tfrac{\R}{2\pi\Z}\right)\,.
\end{equation}
We can thus reduce the insertion of $\calU_{\beta}$ to a weighted sum of subsectors:\footnote{Recall that $\int \beta \cup \sigma$ is integrated on $M\times S^1_L$, so we interpret $\beta$ in that integral as being stretched on $S^1_L$ via the tensor product in Eq.~\eqref{eq:decomposition-theorem}.}
\begin{equation}\label{eq:Hamiltonian_U(1)}
\begin{split}
    \calZ\bigl(M\times S^1_L,\alpha\!+\!\beta\bigr) 
    = \!\!\!\!\!\!\!\!\!\!\! \sum_{\sigma\in H^{d-p-1}(M,\Z)} \!\!\!\!\!\!\!\!\!\! \tr\exp\left[-\,L\,\calH^{\alpha}_{\sigma}(\calZ,M) +\i\!\int\!\beta\cup \sigma\right]
\end{split}\,,\,
\begin{cases}
    \alpha\in H^{p+1}\left(M,\tfrac{\R}{2\pi\Z}\right)\\
    \beta\in H^{p}\left(M,\tfrac{\R}{2\pi\Z}\right)
\end{cases}\!\!\!\!,
\end{equation} 
As for the $\Z^{[d-p-2]}$-symmetric invertible theory $\widehat{\calJ}$, invertibility implies that for each $\Z^{[d-p-2]}$-twist $\mu$, there is only a single state $|\mu\rangle$, which has zero energy.
Let $\theta(\mu)$ denote the $\Z^{[d-p-2]}$-charge of this state $|\mu\rangle$.
Then we have
\begin{equation}\label{eq:Hamiltonian_Z}
    \widehat{\calJ}(M\times S^1_L,\mu+\nu) = \exp\left\{ \i\int \nu \cup \theta(\mu) \right\}\,,\qquad 
    \begin{cases}
    \mu\in H^{d-p-1}\left(M,\Z\right)\\
    \nu\in H^{d-p-2}\left(M,\Z\right)
\end{cases}\,.
\end{equation}
Plugging Eqs.~\eqref{eq:Hamiltonian_U(1)} and~\eqref{eq:Hamiltonian_Z} in Eq.~\eqref{eq:flat_cup}, we obtain
\begin{equation}
\begin{split}
    &\: \calZ\!*\!\calJ(M\times S^1_L, A_s+A_t) \\
    = &\: \!\!\!\!\!\!\!\!\!\!\!\sum_{\sigma\in H^{d-p-1}(M,\Z)} \!\!\!\!\!\!\!\!\!\! \tr\exp\left[-\,L\,\calH^{A_s+ {\theta(\sigma^\vee)}}_{\sigma}(\calZ,M) +\i\!\int\!A_t\cup \sigma\right],\qquad
\end{split}
\begin{cases}
    A_s\in H^{p+1}\left(M,\tfrac{\R}{2\pi\Z}\right)\\
    A_t\in H^{p}\left(M,\tfrac{\R}{2\pi\Z}\right)
\end{cases}\!\!\!\!\!\!\!.
\end{equation}
This teaches us that
\begin{equation}\label{eq:TWE_continuous}
    \calH^{A_s}_{\sigma}(\calZ\!*\!\calJ,M)\ =\ \calH^{A_s+ {\theta(\sigma^\vee)}}_{\sigma}(\calZ,M)\,,\qquad
    \begin{cases}
        A_s\in H^{p+1}\left(M,\tfrac{\R}{2\pi\Z}\right)\\
        \sigma\in H^{d-p-1}(M,\Z)
    \end{cases}.
\end{equation}
This subsector mapping has precisely  the same form as Eq.~\eqref{eq:TWE} in the finite-symmetry case.
Hence we obtain the topological Witten effect.

\subsection{Non-strictly-topological \texorpdfstring{$\Theta$}{Theta} transformations}
\label{sec:U(1)_remark}

In Definition~\ref{def:sym_theta_continuous} of $\U$-symmetry $\theta$ angles, we only use strictly-topological $\Theta$ transformations.  However, it is also interesting to ask what happens if we allow ourselves to use more general $\widehat{\U}$-symmetric invertible theories in $\Theta$ transformations.
In contrast to strictly-topological $\Theta$ transformations, the resulting generalized $\Theta$ transformations produce QFTs with different local physics compared to the original QFT.  These transformations are thus very interesting in their own right, and deserve a detailed exploration in a separate paper. 
Here we briefly discuss two examples that serve to illustrate what makes generalized $\Theta$ transformations so interesting.

First, let us consider a $3$-dimensional theory with a $U(1)^{[0]}$ symmetry.  One example of such a theory is $3$-dimensional Maxwell theory, with its $U(1)^{[0]}$ magnetic symmetry.  If we turn on a background $U(1)^{[0]}$ gauge field $B$, the path integral looks like
\begin{align}
    \calZ(B) &= \int\!\calD a\, \exp\left( -\frac{1}{2 g^2} \int \d a\wedge\star\d a + \frac{\i}{2\pi} \int B\wedge\d a \right)\,.
    \label{eq:3d_maxwell}
\end{align}
Abelian duality implies that this theory has an equivalent description in terms of a compact scalar $\phi$, with the magnetic $U(1)^{[0]}$ symmetry mapping to the shift symmetry of $\phi$:
\begin{align}
    \calZ(B) &= \int\!\calD\phi\,\exp\left[ -\frac{g^2}{4\pi} \int (\d\phi - B)\wedge\star(\d\phi - B) \right] \,.
    \label{eq:3d_compact_scalar}
\end{align}
If we apply the $\Theta$ transformation using the invertible but not strictly-topological theory~\eqref{eq:CS} to $\calZ$, then working with Eq.~\eqref{eq:3d_maxwell} we get Maxwell-Chern-Simons theory:
\begin{align}
    \calZ_{k}(B) =&\: \int\!\calD a\,\exp\left( -\frac{1}{2 g^2} \int \d a\wedge\star\d a + \frac{\i k}{4\pi}\int a\wedge\d a + \frac{\i}{2\pi} \int B\wedge\d a \right) \,.
    \label{eq:MCS}
\end{align}
On the other hand, working with Eq.~\eqref{eq:3d_compact_scalar}, the same $\Theta$ transformation yields
\begin{align}
    \label{eq:scalar_dual_MCS}
      \calZ_{k}(B) &= \!\! \int\!\calD\phi\calD b\calD c\,\exp\left[ -\frac{g^2}{4\pi} \int (\d\phi - b)\wedge\star(\d\phi - b)  
     \right. \\
    &\qquad\qquad\qquad\quad\ \left.
    + \frac{\i k}{4\pi} \int c\wedge\d c - \frac{\i}{2\pi}\int (B-b)\wedge\d c \right]\,. \nonumber
\end{align}
Since Eq.~\eqref{eq:scalar_dual_MCS} is equivalent to Eq.~\eqref{eq:MCS}, we conclude that Eq.~\eqref{eq:scalar_dual_MCS} is the compact-scalar dual of Maxwell-Chern-Simons theory.  The same result was recently found by Armoni~\cite{Armoni:2022xhy} using a different method.

It is also interesting to consider the even broader class of $\Theta$ transformations obtained by using an \textit{arbitrary} $\widehat{\U}$-symmetric QFT.
For example, consider the 4-dimensional Maxwell theory and its two $U(1)^{[1]}$ symmetries, given by Eq.~\eqref{eq:Maxwell_repeat}.
In four spacetime dimensions, the dual symmetry of $U(1)^{[1]}$ is given by $U(1)^{[0]}$.
So let us consider a generic $4$-dimensional $U(1)^{[0]}$-symmetric QFT with background $U(1)^{[0]}$ gauge field $A$,
\begin{equation}
    \calY(A) = \int\!\calD\Psi\,\e^{-\calS(A,\Psi)}\,,
\end{equation}
where $\Psi$ schematically denotes the collection of dynamical fields.
For the sake of concreteness, one may regard this QFT as e.g.~a Dirac fermion or a complex scalar.
Then a $U(1)^{[1]}_m$ $\Theta$ transformation with $\calY$ used in the stacking step just gives us QED:
\begin{equation}
    \calZ(B_m) = \!\int\!\calD a\calD\Psi\,\exp\left[ -\frac{1}{2 g^2} \int \d a\wedge\star\d a -\calS(a,\Psi) + \frac{\i}{2\pi} \int B_m\wedge\d a \right].
\end{equation}
Therefore, such a generalized $\Theta$ transformation with respect to $U(1)^{[1]}_m$ couples the Maxwell to electric matter fields, and explicitly breaks the $U(1)^{[1]}_e$ symmetry.
We thus see that generalized $\Theta$ transformations can do very wild things.

Things become even more intriguing if we preform a generalized $\Theta$ transformation with respect to $U(1)^{[1]}_e$.
Then reciprocally, this construction instead couples Maxwell theory to magnetic matter, explicitly breaking the $U(1)^{[1]}_m$ symmetry!
The resulting partition function is given by
\begin{align}\label{eq:magnetic_matter}
    \calZ(B_e) = \!\!\int\!\!\calD a\calD\Psi\calD b\calD c\,\exp\Bigg[ &-\frac{1}{2 g^2}\!\! \int (\d a - b)\wedge\star(\d a - b) - \calS(c,\Psi)\\
    &+ \frac{\i}{2\pi} \!\!\int (B_e-b) \wedge\d c \Bigg]\,,\nonumber
\end{align}
where $b$ and $c$ are auxiliary $U(1)^{[1]}$ and $U(1)^{[0]}$ gauge fields, respectively. 
Equation~\eqref{eq:magnetic_matter} is particularly intriguing because it describes the coupling to magnetic matter in the \emph{electric} duality frame for the gauge field.
\section{Symmetry \texorpdfstring{$\theta$}{theta} angles versus Lagrangian \texorpdfstring{$\theta$}{theta} angles}
\label{sec:lagrangian_theta}

\begin{figure}[h!tbp]
    \centering
    \includegraphics[width=0.9\textwidth]{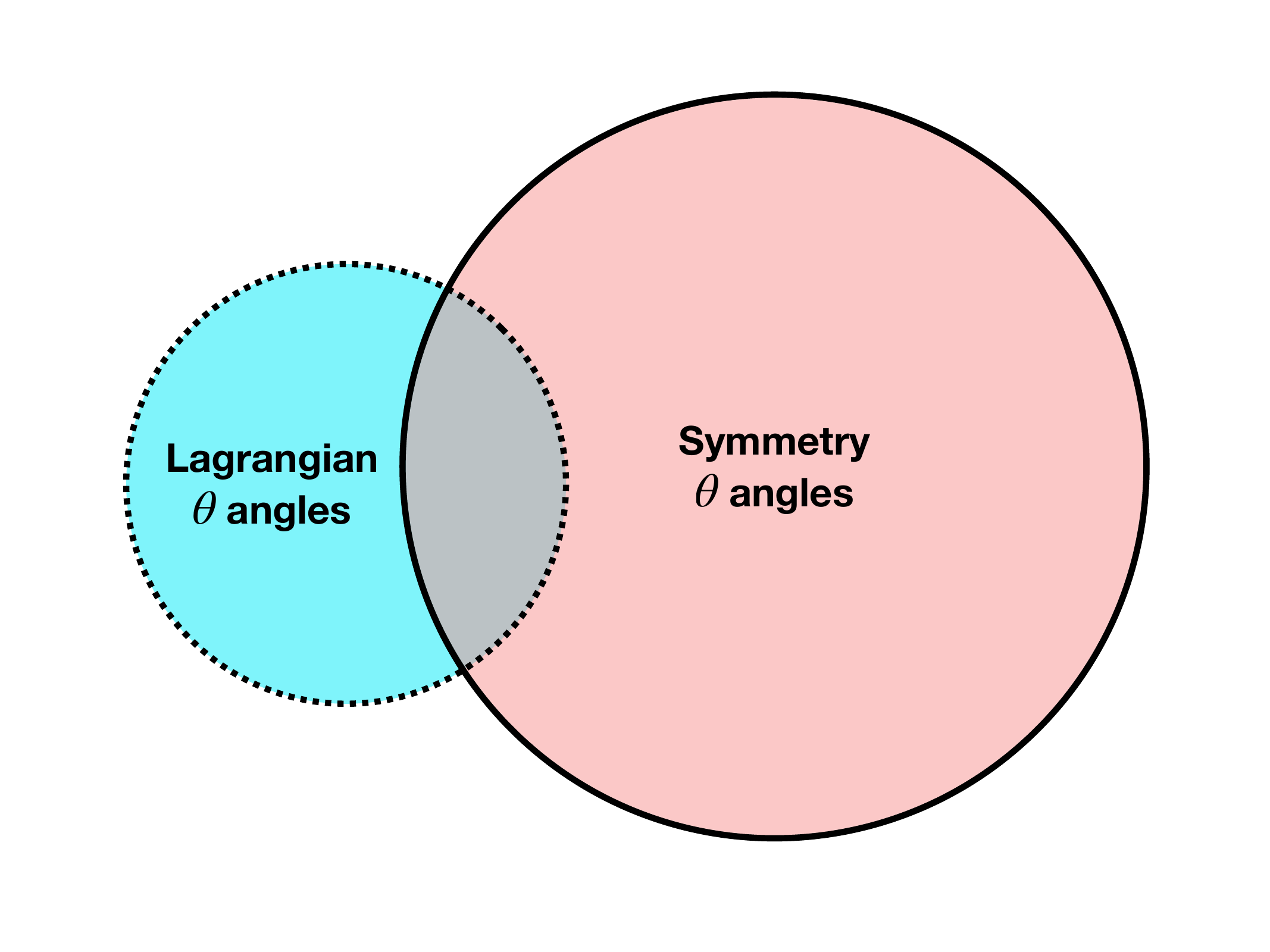}
    \caption{A Venn diagram that illustrates the relationship between Lagrangian
    $\theta$ angles and symmetry $\theta$ angles.  The definition of  Lagrangian $\theta$ angles depends on the
    choice of Lagrangian for a QFT. In contrast, symmetry $\theta$ angles, which we introduce in this paper, are defined using only the
    universal symmetry data of a QFT. In special cases, certain symmetry
    $\theta$ angles can also be interpreted as Lagrangian symmetry angles,
    leading to overlap between the two types of $\theta$ angles shown in the
    diagram. }
    \label{fig:venn_theta}
\end{figure}

So far we have focused on defining symmetry $\theta$ angles, but of course there is also a much more traditional notion of $\theta$ angles in QFTs described by path integrals.  
For narrative convenience, we will refer to this traditional class of $\theta$ angles as \emph{Lagrangian $\theta$ angles} in this paper. 
In this section, we discuss the relationship between Lagrangian and symmetry $\theta$ angles, which is illustrated in Fig.~\ref{fig:venn_theta}.  

\subsection{Lagrangian \texorpdfstring{$\theta$}{theta} angles}

Lagrangian $\theta$ angles appear when the field configuration space in a path integral has multiple topological sectors.  
In this situation, the path integral yields a sum over these topological sectors, and a Lagrangian $\theta$ angle controls the relative weight of the topological sectors.  Schematically, 

\begin{equation}
    \calZ=\sum_{n}\calZ_n\qquad \longrightarrow \qquad \calZ_{\theta}=\sum_{n}\e^{\i\theta_n}\calZ_n\,.
\end{equation}
All of the standard $\theta$ angles in gauge theories and nonlinear sigma models are Lagrangian $\theta$ angles.  
The phase factors associated with a Lagrangian $\theta$ angle cannot be chosen randomly, because they must satisfy appropriate locality constraints.  The resulting classification problem has been carefully studied in the literature, and it turns out that Lagrangian $\theta$ angles are also classified by suitable bordism groups, see e.g.~Refs.~\cite{Freed:2006mx,Seiberg:2010qd,Kapustin:2014gua,Freed:2016rqq,Freed:2017rlk,Cordova:2017vab,Yonekura:2018ufj,Hsin:2018vcg,Hsin:2019fhf,Hsin:2020nts,Chen:2022cyw,Chen:2023czk}.

The bordism classification of Lagrangian $\theta$ angles involves the notion of the homotopy target space.
For a path integral, the homotopy target space $Y$ is an auxiliary topological space such that on any closed spacetime $X$,
\begin{equation}
    [X,Y] 
\end{equation}
classifies the topological sectors, i.e.~the deformation classes of all the field configurations.
For example, in a nonlinear sigma model with a $M$-valued scalar, we simply have $Y\simeq M$, while in a pure Yang-Mills theory, $Y\simeq BG$ is the classifying space of the gauge group $G$.%
\footnote{When there are more fields, the space $Y$ becomes more complicated. For example, if we couple a $G$ gauge field to a $M$-valued scalar, then $Y$ fits into a fibration $M\to Y\to BG$ whose structure is determined by the $G$-action on $M$.
For example, $Y\simeq M/G$ when $G$ acts on $M$ freely, and $Y\simeq BG$ when $M$ is contractible (for whatever $G$-action).
}
Lagrangian $\theta$ angles of a $d$-dimensional path integral are then classified by
\begin{equation}\label{eq:classification_lagrangian}
    \Hom\left(\tOmega^{SO}_{d}\bigl(Y\bigr),\tfrac{\R}{2\pi\Z}\right) \quad\text{and}\quad \Hom\left(\tOmega^{Spin}_{d}\bigl(Y\bigr),\tfrac{\R}{2\pi\Z}\right)\,,
\end{equation}
in the bosonic and fermionic cases, respectively.

\begin{definition}[\textbf{Lagrangian $\theta$ angle}]
    \label{def:lagrangian_theta}
   A Lagrangian $\theta$ angle in a particular path integral description of a QFT is the coefficient of a topological term associated with the fields appearing in the path integral.  Mathematically, it is an element of
\[
  \Hom\!\Bigl(\widetilde\Omega^{\mathrm{SO}}_{d}(Y),\,
              \tfrac{\R}{2\pi\Z}\Bigr)
  \quad\text{or}\quad
  \Hom\!\Bigl(\widetilde\Omega^{\mathrm{Spin}}_{d}(Y),\,
              \tfrac{\R}{2\pi\Z}\Bigr).
\]

\end{definition}

\subsection{A conceptual distinction}
\label{sec:lag_vs_sym_distinction}

On the surface, it may seem that symmetry $\theta$ angles and Lagrangian $\theta$ angles are quite similar.  
The bordism classification~\eqref{eq:classification_lagrangian} has a similar look to Eqs.~\eqref{eq:classification_discrete} and~\eqref{eq:classification_continuous}.
Also, there are terms that are simultaneously Lagrangian $\theta$ angles and symmetry $\theta$ angles, such as the standard Maxwell $\theta$ term (see Section~\ref{sec:Maxwell_revisit} for the details).
However, at the conceptual level, Lagrangian and symmetry $\theta$ angles are \emph{radically different} from each other:
\begin{itemize}
    \item Lagrangian $\theta$ angles are defined for \emph{path integrals}.
    \item Symmetry $\theta$ angles are defined for \emph{quantum field theories}.
\end{itemize}

To discuss Lagrangian $\theta$ angles of a QFT, one must first specify a path-integral representation of the QFT.  
It is well-known that this is not always possible. 
Even when a QFT has a path integral description, it is often not universal.
Many QFTs have multiple equivalent path integral descriptions, for example thanks to dualities.  As a result, their Lagrangian $\theta$ angles vary with the choices of different path-integral representations.
These standard observations imply that Lagrangian $\theta$ angles do not necessarily have any universal meaning in QFTs. 

The above observation implies that, when a QFT has multiple path-integral descriptions, a Lagrangian $\theta$ angle in one of the descriptions will become something else in another description.
For example, in Section~\ref{sec:cheshire_theta_3d} we will discuss a QFT with four different path-integral representations.
As we will see there, a Lagrangian $\theta$ angle in one of its path-integral descriptions becomes something drastically different in all the other three descriptions.
Actually, the Maxwell $\theta$ term itself is also an example of this phenomenon; see Section~\ref{sec:Maxwell_revisit} for the details.

We also want to mention another less-toy-model example, the long-distance behavior of QCD.  
This system can of course be described by a QCD path integral over gluons and $N_f$ flavors of quarks, and this path integral has a standard $\theta$ angle that controls the phases of the weights of instanton sectors.  
However, this system can \emph{also} be described by the chiral EFT path integral,  
which is a four-dimensional $SU(N_f)$ non-linear sigma model, where $N_f$ is the number of light quark flavors. 
Rather than appearing as the coefficient of some topological term in the action of the chiral EFT, the QCD $\theta$ angle gets  encoded in a phase of the quark mass spurion field.

In drastic contrast, symmetry $\theta$ angles are \emph{intrinsic} parameters of a QFT.  
Their definition only involves the symmetry data of a QFT, and does not refer to path integrals at all. 
The choice of symmetry $\theta$ angles for a QFT is therefore part of the universal data specifying the QFT, just like the choice of symmetries and their anomalies.
This is precisely what motivates us to seek the notion of symmetry $\theta$ angles.

Let us now discuss the relation between Lagrangian and symmetry $\theta$ angles in a \emph{fixed} path-integral description for a given QFT.
Even in this case, they are still two distinct notions, as is highlighted in Fig.~\ref{fig:venn_theta}.  There are Lagrangian $\theta$ angles that are not symmetry $\theta$ angles.  This includes the familiar $\theta$ angle of 4-dimensional Yang-Mills theories.%
\footnote{
If we allow the somewhat peculiar notion of $(-\!1)$-form symmetry (see e.g.~\cite{Cordova:2019jnf,Cordova:2019uob,Tanizaki:2019rbk,Heidenreich:2020pkc,Aloni:2024jpb}), we can amusingly recast such Lagrangian $\theta$ angles into symmetry $\theta$ angles for $(-\!1)$-form symmetries as well.
Though alluring, this viewpoint is circular; see our comment at the end of Section~\ref{sec:defor&dual}.
}
Second, there are symmetry $\theta$ angles that are not Lagrangian $\theta$ angles.  
A simple example of such a $\theta$ angle that we are aware of is the ``Cheshire $\theta$ angle'' that appeared in our recent paper with Gongjun Choi~\cite{Chen:2024tsx}.  
The present paper actually originated as an attempt to find a more fundamental interpretation of the Cheshire $\theta$ angle. 
We explain how to think about the Cheshire $\theta$ angle as a symmetry $\theta$ angle in Section~\ref{sec:cheshire_theta_4d}.

\subsection{Connection via solitonic symmetry}
\label{sec:solitonic_symmetry}

When is a term simultaneously a symmetry $\theta$ angle and a Lagrangian $\theta$ angle?
A symmetry theta angle is also a Lagrangian $\theta$ angle if and only if the symmetry under consideration is a \emph{solitonic symmetry} that acts solely on defect operators (see~\cite{Chen:2022cyw,Chen:2023czk} for some recent discussions). 
For example, the standard Maxwell $\theta$ term is simultaneously a symmetry $\theta$ angle and a Lagrangian $\theta$ angle because $U(1)_m^{[1]}$ is a solitonic symmetry that acts on 't Hooft defects (see Section~\ref{sec:Maxwell_revisit} for the details).

Whether a given symmetry is solitonic depends on the path integral description, and it is therefore not a universal property of the symmetry.
For example, in Section~\ref{sec:cheshire_theta_3d} we will discuss a QFT with four different path-integral representations.
As we will see, its symmetry $\theta$ angle is also a Lagrangian $\theta$ angle in one of the descriptions, but is something else far removed from Lagrangian $\theta$ angles in all the other three descriptions.

Reciprocally, as long as the symmetry under consideration is not solitonic, a symmetry $\theta$ angle is never a Lagrangian $\theta$ angle.
If one moves to another path-integral description, then there might be a chance to make it a Lagrangian $\theta$ angle, or might not.
For example, for the symmetry $\theta$ angle we mentioned in Section~\ref{sec:beyond-mapping-tori}, as well as some examples given in Ref.~\cite{Chen:2024tsx}, a path-integral representation where they appear as Lagrangian $\theta$ angles is very unlikely to exist. 

In general and more formally, if either finite $\G$ or continuous $\U$ is realized as a solitonic symmetry, then by definition~\cite{Chen:2022cyw,Chen:2023czk} there is  a natural fibration $Y\to \calB\widehat{\G}$ or $\calB\widehat{\U}$, whose pullback homomorphisms
\begin{subequations} \label{eq:sym_to_L_theta_classification}
\begin{gather}
    \Hom\left(\tOmega^{SO}_{d}\bigl(\calB\widehat{\G}\text{ or }\calB\widehat{\U}\bigr),\tfrac{\R}{2\pi\Z}\right)  \to \Hom\left(\tOmega^{SO}_{d}\bigl(Y\bigr),\tfrac{\R}{2\pi\Z}\right)\\
    \Hom\left(\tOmega^{Spin}_{d}\bigl(\calB\widehat{\G}\text{ or }\calB\widehat{\U}\bigr),\tfrac{\R}{2\pi\Z}\right)  \to \Hom\left(\tOmega^{Spin}_{d}\bigl(Y\bigr),\tfrac{\R}{2\pi\Z}\right)
\end{gather}
\end{subequations}
tell us how $\G$-symmetry or $\U$-symmetry $\theta$ angles are mapped to $Y$-Lagrangian $\theta$ angles.  
In general, these homomorphisms do not have to be either injective or surjective.
On the one hand, the absence of surjectivity means that there are Lagrangian $\theta$ angles that cannot be understood as symmetry $\theta$ angles, as we explained a moment ago in Section~\ref{sec:lag_vs_sym_distinction}.
On the other hand, the absence of injectivity implies that it is possible for two symmetry $\theta$ angles to degenerate to one Lagrangian $\theta$ angle.  When this happens one gets an extra 0-form non-invertible symmetry, as we discussed in Section~\ref{sec:general_non_inv}.


\section{Examples}
\label{sec:examples}

Having explained the abstract construction of symmetry $\theta$ angles established in the previous sections, we now discuss three classes of concrete examples.
First, in Section~\ref{sec:4d_gauge} we discuss continuous and discrete symmetry $\theta$ angles in 4-dimensional gauge theories.  
Second, in Section~\ref{sec:cheshire_examples} we discuss a class of symmetry $\theta$ angles that were dubbed `Cheshire $\theta$ angles' when they first appeared in our paper with Gongjun Choi \cite{Chen:2024tsx}.
Finally, in Section~\ref{sec:cubic_examples} we discuss `cubic $\theta$ angles' constructed using cubic invertible theories, which lead to amusingly unusual versions of 
Witten effects.

We will see that in our examples, the symmetry $\theta$ angles sometimes reduce to known parameters, while in other cases they amount to new parameters.
Our choices of examples is motivated by our two goals, which are to make contact with the existing literature, as well as to show that symmetry $\theta$ angles are not pathological and have interesting implications. 
There are many other very interesting examples of symmetry $\theta$ angles that we will not discuss here.  We mentioned a few of them in Section~\ref{sec:outlook}.  They deserve much more detailed investigation that goes beyond the scope of this paper.  


\subsection{4D gauge theories}
\label{sec:4d_gauge}
In this section we will focus on examples of symmetry $\theta$ angles in $4$-dimensional gauge theories with $1$-form symmetries.  
As we shall see shortly, the symmetry $\theta$ angles in this class of QFTs are closely related to well-known examples of conventional $\theta$ angles.  
This shows that symmetry $\theta$ angles are not particularly strange, and cannot simply be ignored, even in a very well-studied class of QFTs.


\subsubsection{Maxwell theory revisited}
\label{sec:Maxwell_revisit}


The simplest $4$-dimensional theory that has a continuous $1$-form symmetry is simply Maxwell theory.
It has an electric $U(1)^{[1]}_e$ symmetry and a magnetic $U(1)^{[1]}_m$ symmetry with background gauge fields $B_e, B_m$ respectively. This discussion generalizes Section~\ref{sec:Witten_reappraisal}.  The partition function of Maxwell theory can be written as
\begin{subequations}\label{eq:Maxwell_repeat}
\begin{gather}
    \calZ(B_e,0) = \int\!\calD a\, \exp\left[ -\frac{1}{2 g^2} \int (\d a-B_e)\wedge\star(\d a-B_e) \right]\,,\label{eq:Maxwell_repeat_e}\\
    \calZ(0,B_m) = \int\!\calD a\, \exp\left( -\frac{1}{2 g^2} \int \d a\wedge\star\d a + \frac{\i}{2\pi} \int B_m\wedge\d a \right)\,.\label{eq:Maxwell_repeat_m}
\end{gather}
\end{subequations}
If we simultaneously turn on both $B_e$ and $B_m$, background gauge invariance is lost due to the mixed 't Hooft anomaly between $U(1)^{[1]}_e$ and $U(1)^{[1]}_m$.
Nevertheless, both $U(1)^{[1]}_e$ and $U(1)^{[1]}_m$ are anomaly-free on their own.
Hence we can construct symmetry $\theta$ angles for either of these symmetries.

The dual symmetry of a $U(1)^{[1]}$ symmetry in a 4-dimensional QFT is a $U(1)^{[0]}$ symmetry. We already mentioned in Eq.~\eqref{eq:theta_4D} that $U(1)^{[0]}$-symmetric invertible theories are parameterized by a continuous parameter $\theta$.
According to Eq.~\eqref{eq:classification_continuous}, 4-dimensional $U(1)^{[0]}$-symmetric invertible theories are classified by
\begin{equation}\label{eq:BU(1)_classification_4}
    \Hom\left(\tOmega^{SO}_{4}\bigl(BU(1)\bigr),\tfrac{\R}{2\pi\Z}\right)= \tfrac{\R}{4\pi\Z}\,, \qquad \Hom\left(\tOmega^{Spin}_{4}\bigl(BU(1)\bigr),\tfrac{\R}{2\pi\Z}\right)=\tfrac{\R}{2\pi\Z}
\end{equation}
in the bosonic and the fermionic cases, respectively.
The natural map from the fermionic group to the bosonic group is a double cover.
Hence $\theta$ is $2\pi$-periodic in the fermionic case, and $4\pi$-periodic in the bosonic case.

Let us first construct the $U(1)^{[1]}_m$-symmetry $\theta$ angle.
Plugging Eqs.~\eqref{eq:Maxwell_repeat_m} and~\eqref{eq:theta_4D} into the $\Theta$ transformation~\eqref{eq:Theta_U(1)} and integrating out the auxiliary gauge fields, we obtain
\begin{equation}\label{eq:Maxwell_m}
\begin{split}
    \calZ_{\theta_m}(0,B_m) =&\: \int\!\calD a\, \exp\left( -\frac{1}{2g^2} \int \d a\wedge\star\d a + \frac{i \theta_m}{8\pi^2} \int \d a \wedge\d a + \frac{\i}{2\pi} \int B_m\wedge\d a \right)\,.
\end{split}
\end{equation}
We thus see that the $U(1)^{[1]}_m$-symmetry $\theta$ angle is precisely the standard Maxwell $\theta$ term.
The periodicity of $\theta_m$ prescribed by Eq.~\eqref{eq:BU(1)_classification_4} is the same as what one expects from the standard Lagrangian approach.
This exemplifies the claim we made in Section~\ref{sec:solitonic_symmetry} that a solitonic symmetry $\theta$ angle reduces to a Lagrangian $\theta$ angle.
This also shows that, just as we suggested in Section~\ref{sec:Witten_reappraisal}, the standard Maxwell $\theta$ term is intrinsically associated with the $U(1)^{[1]}_m$ symmetry of the theory.

Next, let us construct the $U(1)^{[1]}_e$-symmetry $\theta$ angle.
Plugging Eqs.~\eqref{eq:Maxwell_repeat_e} and~\eqref{eq:theta_4D} into the $\Theta$ transformation~\eqref{eq:Theta_U(1)}, we obtain
\begin{equation}\label{eq:Maxwell_e}
\begin{split}
    \calZ_{\theta_e}(B_e,0) =&\: \int\!\calD a\calD b\calD c\, \exp\left[ -\frac{1}{2g^2} \int (\d a-b)\wedge\star(\d a-b) \right.\\
    &\: \left. + \frac{i \theta_e}{8\pi^2} \int \d c \wedge\d c + \frac{\i}{2\pi} \int {\d c \wedge (B_e - b)} \right]\,,
\end{split}
\end{equation}
where $b$ is a $U(1)^{[1]}$ gauge field and $c$ is a $U(1)^{[0]}$ gauge field.
Eq.~\eqref{eq:Maxwell_e} exhibits an unconventional parameter $\theta_e$ in Maxwell theory.%
\footnote{  
Some aspects of the electric $\theta$ terms in Maxwell theory have been recently discussed in Ref.~\cite{Heidenreich:2023pbi}.
}
Thanks to  Eq.~\eqref{eq:BU(1)_classification_4}, this additional $\theta_e$ parameter is also $2\pi$-periodic if Maxwell theory is viewed as a fermionic theory.  It is $4\pi$-periodic if Maxwell theory is viewed as a bosonic theory.

\begin{figure}[t]
\centering
\begin{subfigure}[t]{0.47\textwidth}
\centering
\includegraphics[width = 0.9 \textwidth]{theta_nonzero_lattice.pdf}
\caption{$\theta_m \neq 0$}
\end{subfigure}
\hfill
\begin{subfigure}[t]{0.47\textwidth}
\centering
\includegraphics[width = 0.78 \textwidth]{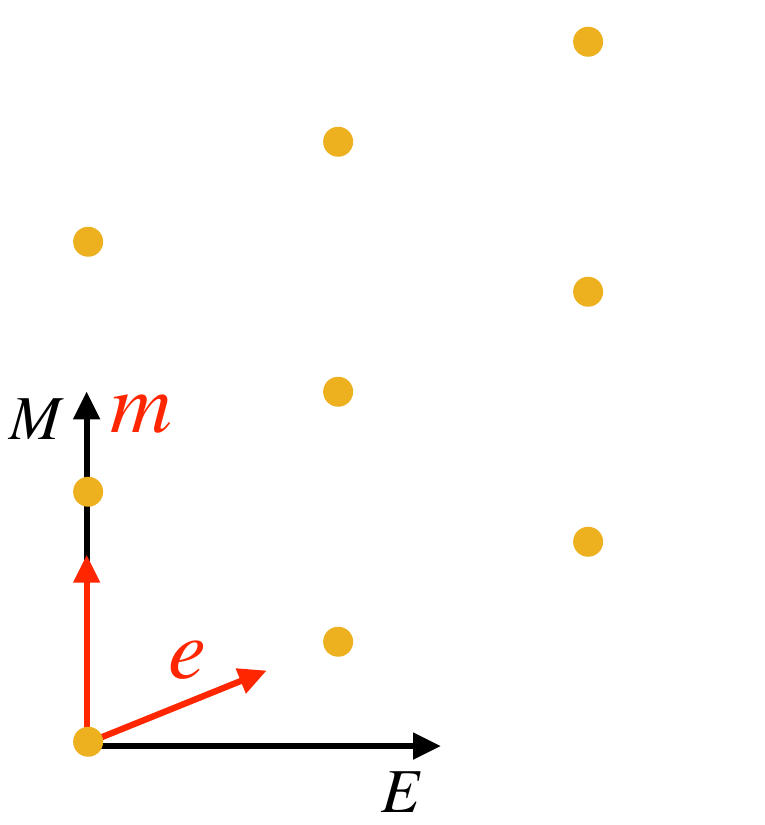}
\caption{$\theta_e \neq 0$}
\end{subfigure}
\caption{These physical charge $E,M$ lattices of 4d $U(1)$ Maxwell gauge theory illustrate its two charge Witten effects summarized in Eq.~\eqref{eq:two_charge_Witten_effects}.
Left: The magnetic symmetry $\theta$ angle, $\theta_m$, gives a fractional electric charge to 't Hooft line operators.  
This is the familiar Witten effect of Ref.~\cite{Witten:1979ey}. 
Right: The electric symmetry $\theta$ angle, $\theta_e$, gives a fractional magnetic charge to Wilson line operators.  The figures also include  the $U(1)^{[1]}_{e,m}$ symmetry charges $e, m \in \Z$.}
\label{fig:charge-lattice_two_theta}
\end{figure}

The two symmetry $\theta$ angles described above lead to different physical consequences.
In particular, they yield different charge Witten effects on line operators: 
\begin{equation}
\begin{gathered}
    \theta_m:\qquad (e,m)\quad\mapsto\quad (E,M)= \left(e+\frac{\theta_m}{2\pi}m\,,\, m \right)\,,\\
    \theta_e:\qquad (e,m)\quad\mapsto\quad (E,M)= \left( e\,,\, m+\frac{\theta_e}{2\pi}e \right) \,,
\end{gathered}    
\label{eq:two_charge_Witten_effects}
\end{equation}
where $(e,m)$ labels the $U(1)^{[1]}_e\times U(1)^{[1]}_m$ charge of the line operator, while $(E,M)$ denotes the physical electromagnetic charges that determines the physical observables.  For example, $(E,M)$ determine quantities like Coulomb forces between line operators, as well as the tensions of string excitations in finite volume.
We see that the conventional $U(1)^{[1]}_m$-symmetry $\theta$ angle assigns fractional electric charges to 't Hooft lines, whereas the exotic $U(1)^{[1]}_e$-symmetry $\theta$ angle instead assigns fractional magnetic charges to Wilson lines. These charge Witten effects are illustrated in Fig.~\ref{fig:charge-lattice_two_theta}.
The two symmetry $\theta$ angles are interchanged under electromagnetic duality.%
\footnote{
The electromagnetic duality referenced in the main text should not be confused with the other version of electromagnetic duality --- which is (unfortunately) also usually called ``the S transformation'' --- that sends 
\begin{equation}
    \tau \ \ \to\ \ -\frac{1}{\tau}\,,\qquad \qquad\text{ where }\tau\equiv \frac{\theta_m}{2\pi} + \i\frac{2   \pi}{g^2}\,.
\end{equation}
The two versions of electromagnetic duality differ from each other because the ``electric direction'' in one of these dualities is an ``oblique'' direction in the other duality.
}
As a result, the $U(1)^{[1]}_e$-symmetry $\theta$ angle does not extend the physical moduli space of inequivalent Maxwell theories.
Nevertheless, the $U(1)^{[1]}_e$-symmetry $\theta$ angle provides a way of writing down the conventional $\theta$ angle in the electromagnetically-dual frame.

The two symmetry $\theta$ angles can also exist in the presence of matter fields.
Coupling Maxwell theory to electric matter breaks $U(1)^{[1]}_e$, but leaves $U(1)^{[1]}_m$ untouched. 
Therefore the $U(1)^{[1]}_m$-symmetry $\theta$ angle (i.e.~the standard Maxwell $\theta$ term) survives in this case.
Similarly, coupling the Maxwell theory to magnetic matter%
\footnote{
In Section~\ref{sec:U(1)_remark}, we present how to do this \emph{without} moving to the electromagnetic-dual frame.
}
breaks $U(1)^{[1]}_m$, but leaves $U(1)^{[1]}_e$ untouched.  
In this case, it is the unconventional $U(1)^{[1]}_e$-symmetry $\theta$ angle that survives.

It is worth mentioning that, although we cannot introduce $\theta_e$ and $\theta_m$ simultaneously due to the presence of mixed anomaly, we can still turn them on \emph{sequentially}.
For example, we can first introduce $\theta_m$ with respect to the $U(1)_m^{[1]}$ symmetry.
The resulting theory~\eqref{eq:Maxwell_m} has both a $U(1)_m^{[1]}$ symmetry and a $U(1)_e^{[1]}$ symmetry.
We can then perform a $\Theta$ transformation to this $U(1)_e^{[1]}$ symmetry to obtain a $\theta_e$ parameter.
Let us denote the overall construction by
\begin{equation}
    \theta_m \vartriangleright \theta_e\,.
\end{equation}
We can also run this program in the opposite order to obtain
\begin{equation}
    \theta_e \vartriangleright \theta_m\,.
\end{equation}
The two resulting theories are different from each other, with a different spectrum of line operators, which reflects the mixed anomaly.  
If both $\theta_m$ and $\theta_e$ are irrational, all non-trivial line operators generically become dyonic. 
However, we should emphasize that neither $\theta_m \vartriangleright \theta_e$ nor $\theta_e \vartriangleright \theta_m$ extends the physical moduli space of inequivalent Maxwell theories, because we can always rotate the charge spectrum via continuous $SO(2)$ electromagnetic dualities.
Nevertheless, the electric and magnetic symmetry $\theta$ angles provide ways of constructing unconventional charge spectra without using electromagnetic dualities.

\subsubsection{Relation between electric and magnetic \texorpdfstring{$\theta$}{theta} angles}
\label{sec:electric_theta}

Let us take a closer look into the $U(1)^{[1]}_e$-symmetry $\theta$ angle we introduce above.
When $\theta_e$ is irrational, the $U(1)^{[1]}_e$-symmetry $\theta$ angle realizes an unconventional charge spectrum of line operators that can never be reached by any $U(1)^{[1]}_m$-symmetry $\theta$ angle in a fixed duality frame. 

When $\theta_e$ is rational, the $U(1)^{[1]}_e$-symmetry $\theta$ angle is almost equivalent to a $U(1)^{[1]}_m$-symmetry $\theta$ angle --- or equivalently, a Lagrangian $\theta$ angle --- without using electromagnetic dualities.
To see this, let us consider
\begin{equation}
    \theta_e = 2\pi\frac{q}{p}\,,\qquad\qquad p\in\Z_+\,,\quad q\in\Z_p\,,\quad \mathrm{gcd}(p,q)=1\,.
\end{equation}
Then the physical charges $(E,M)$ of line operators form a lattice spanned by the basis 
\begin{equation}
    (0,1)\quad\text{ and }\quad (1,q/p)\,.
\end{equation}
The basis consists of an 't Hooft line and a dyonic line.
But a lattice has infinitely many different bases.
Because there is always an integer $s\in\Z$ such that $sq=1\mod p$ (due to B\'ezout's lemma), the charge lattice can also be spanned by another basis,
\begin{equation}
    (p,0)\quad\text{ and }\quad (s,1/p)\,,
\end{equation}
This new basis consists of a Wilson line and a dyonic line, which implies that the same spectrum of line operators can also be produced by a $U(1)^{[1]}_m$-symmetry $\theta$ angle, along with a rescaling of the coupling constant.
More precisely, starting with pure Maxwell theory without any $\theta$ angle, if we rescale the coupling constant $g\to gp$ and introduce a magnetic $\theta$ angle with
\begin{equation}
    \theta_m = 2\pi\frac{s}{p}\,,
\end{equation}
then the resulting theory produces the same spectrum of genuine line operators.
Therefore, in pure Maxwell theory, a rational electric $\theta$ angle is closely related to a magnetic $\theta$ angle, without using electromagnetic duality.

To explain why we wrote ``closely related'' rather than ``equivalent'' above, we now look at the above phenomenon from a different angle.
A rational $U(1)^{[1]}_e$-symmetry $\theta$ angle is in essence a $\Z_p^{[1]}$-symmetry $\theta$ angle for the subsymmetry $\Z_p^{[1]}\subset U(1)^{[1]}_e$.
In dimension 4, the dual symmetry of $\Z_p^{[1]}$ is again $\Z_p^{[1]}$.
Then as we mentioned in Section~\ref{sec:Theta_vs_others}, the topological manipulations about $\Z_p^{[1]}$ satisfy a modular-group structure (see Appendix~\ref{sec:STSTST=1} for more details).
As a result, the $\Theta$ transformation corresponding to $\theta_e=2\pi/p$ satisfies the relation 
\begin{equation}
    \Theta\ \simeq\ T^{-1}\Theta^{-1} S\,.
\end{equation}
This means that introducing $\theta_e=2\pi/p$ is equivalent to the composition of the following three operations.
First, we gauge the electric $\Z_p^{[1]}$.
Second, we introduce a conventional magnetic $\theta_m=-2\pi/p$.
Third, for the magnetic $\Z_p^{[1]}\subset U(1)^{[1]}_m$, we stack a $\Z_p^{[1]}$-symmetric invertible theory $\calI^{-1}$ defined in Appendix~\ref{sec:quadratic_refinement}.
This last step rearranges the $\Z_p^{[1]}$-charge assignments of the non-genuine line operators in the twisted sectors.  This effectively transfers $\Z_p^{[1]}$ from the magnetic side to the electric side. Without this last step, the equivalence between the two theories is not complete.



The discussion above generalizes to 4-dimensional $U(1)$ gauge theories with charge-$p$ electric matter.
Rational electric $\theta$ angles $\theta_e=2\pi q/p$ survive in charge-$p$ QED theories because these theories have a $(\Z_p)^{[1]}_e$ symmetry.
If the electric matter condenses and the system is in a gapped phase, $(\Z_p)^{[1]}_e$ is spontaneously broken.
Then turning on the electric $\theta$ angle $\theta_e=2\pi/p$ restores the $(\Z_p)^{[1]}_e$ symmetry but pushes the system into a $(\Z_p)^{[1]}_e$-SPT phase.  (Its infrared partition function appears in Appendix~\ref{sec:quadratic_refinement}.) 
This yields an example of our discussion from Section~\ref{sec:act_on_infrared}, where we argued that symmetry $\theta$ angles for discrete symmetries can implement Kennedy-Tasaki transformations.

\subsubsection{\texorpdfstring{$\mathfrak{su}(2)$}{su(2)} Yang-Mills: re-reading between lines}
\label{sec:4d_YM}

Let us consider fermionic pure Yang-Mills theories with a gauge Lie algebra $\mathfrak{su}(2)$, which have a $\Z_2^{[1]}$ symmetry.
By `fermionic', we mean that we always put these theories on spinnable manifolds.
Aharony, Seiberg, and Tachikawa studied these theories based on their spectra of line operators~\cite{Aharony:2013hda}.%
\footnote{
In Ref.~\cite{Aharony:2013hda}, there appears another set of $S$ and $T$ operations related to electromagnetic duality.
They should not be confused with the topological manipulations $S$ and $T$, although in some cases they lead to coincident results.
}
Line operators are labeled by their electric and magnetic symmetry charges $(e,m)\in\Z_2\oplus\Z_2$.
To satisfy locality constraints, the charge spectrum of line operators can only be a $\Z_2$ subgroup of $\Z_2\!\oplus\!\Z_2$ generated by one of
\begin{equation}
    (1,0)\,,\qquad (0,1)\,,\qquad (1,1)\,.
\end{equation}
In Ref.~\cite{Aharony:2013hda} these three spectra were interpreted as  three different Yang-Mills theories:
\begin{itemize}
    \item The QFT with the line operator $(1,0)$ was identified as Yang-Mills theory with gauge group $SU(2)$.  Its $1$-form symmetry was interpreted as being electric.
    \item The QFTs with the line operators $(0,1)$ and $(1,1)$ were identified as Yang-Mills theories with gauge group $SO(3)$, which differ from each other by a shift of a Lagrangian $\theta$ angle. These  theories were named $SO(3)_+$ and $SO(3)_-$ YM theories, respectively, in Ref.~\cite{Aharony:2013hda}.  The $1$-form symmetry was interpreted as being magnetic. 
\end{itemize}

\begin{figure}[h!]
\centering
\begin{adjustbox}{center,clip,raise=0cm,scale=0.8}

\includegraphics[width=0.6\textwidth]{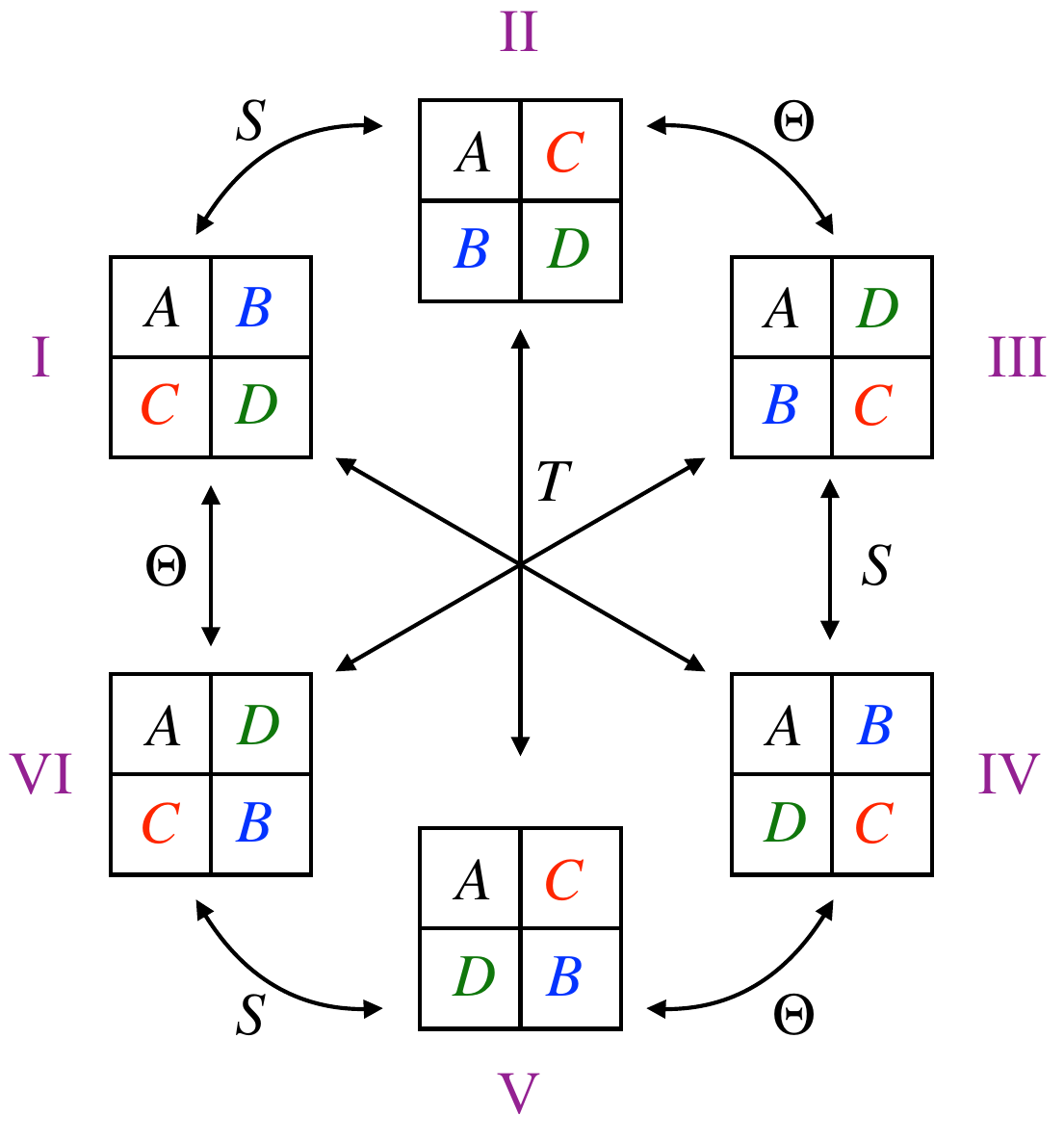}

\end{adjustbox}
\caption{Six inequivalent $\mathfrak{su}(2)$ Yang-Mills theories visualized via their spectral tables.
The columns are labeled by $\Z_2^{[1]}$-charges and the rows are labeled by $\Z_2^{[1]}$-twists.
These theories are related by gauging $\Z_2^{[1]}$ symmetries ($S$ arrows), $\Z_2^{[1]}$-symmetry $\theta$ angles ($\Theta$ arrows), and stacking with the $\Z_2^{[1]}$-symmetric invertible theory ($T$ arrows).  
The discussion in the classic work of Ref.~\cite{Aharony:2013hda} focused on the top three theories I, II, and III, and called them $SU(2), SO(3)_{+},$ and $SO(3)_{-}$ YM theories, respectively.
In all six theories, the sector B always contains Wilson lines, the sector C always contains 't Hooft lines, and the sector D  always contains dyonic lines. 
}
\label{fig:SU2_relations}
\end{figure}

We now discuss $\mathfrak{su}(2)$ Yang-Mills theories from a different approach centered on topological manipulations.%
\footnote{
This approach has been discussed in many other papers, see Ref.~\cite{Duan:2024xbb} for a very nice recent example.
}
With a $\Z_2^{[1]}$ symmetry, physically inequivalent topological manipulations form the order-6 modular group (see the discussions in Section~\ref{sec:counting_params} and Appendix~\ref{sec:STSTST=1})
\begin{equation}
    PSL(2,\Z_2)\,.
\end{equation}
Hence we have 6 inequivalent theories.
To distinguish them, it is illuminating to put the theories on a spacetime of the form $S^2\!\times\!S^1_{L_s}\!\times\!S^1_{L_t}$, as we did for Maxwell theory in Section~\ref{sec:Witten_reappraisal}.
$S^1_{L_t}$ is viewed as the temporal direction and $S^1_{L_s}$ is a spatial direction along which string excitations can stretch.
According to our discussion in Section~\ref{sec:twist&charge}, on such a spacetime each theory has two different sectors labeled by the $\Z_2^{[1]}$-twist on the spatial $S^2$, and each sector decomposes into two subsectors with different $\Z_2^{[1]}$-charges.
For each theory, we can arrange the four subsectors in a $2\times 2$ spectral table, with rows labeled by $\Z_2^{[1]}$-twists and columns labeled by $\Z_2^{[1]}$-charges.
The topological manipulations then shuffle the four subsectors.  We show the spectral tables of the $6$ inequivalent $\mathfrak{su}(2)$ gauge theories in Fig.~\ref{fig:SU2_relations}.
Based on Section~\ref{sec:topologicalWitten_derivation} and Appendix~\ref{sec:quadratic_refinement}, the $S$ transformation transposes the spectral table, the $T$ transformation switches the two entries in the second row, and the $\Theta$ transformation switches the two entries in the second column.

For the sake of reference, let us view Theory I in Fig.~\ref{fig:SU2_relations} as the conventional $SU(2)$ Yang-Mills theory.
It has an electric $\Z_2^{[1]}$ symmetry from the center of the group $SU(2)$.  
This symmetry is believed to be unbroken, which means that the charged line operators have an area law.
There are four classes of line operators classified by their $\Z_2^{[1]}$-twists and $\Z_2^{[1]}$-charges.
When they are stretched along $S^1_{L_s}$, they (together with point operators) create and annihilate states in the four subsectors in the spectral table.
We summarize the physics of $SU(2)$ Yang-Mills in Table~\ref{tab:SU2}.

\begin{table}[!ht]
    \centering
    \begin{tabular}{c||c|c||c|c|c|}
        \textbf{(twist,\,charge)} & \multicolumn{2}{c||}{\textbf{lines}} & \multicolumn{3}{c|}{\textbf{subsector}} \\\hline\hline
        $(0,0)$ & trivial & perimeter law & A & particles & $E\sim 0$ \\\hline
        $(0,1)$ & Wilson & area law & B & strings & $E\sim\sigma L_s$ \\\hline
        $(1,0)$ & 't Hooft & perimeter law & C & particles & $E\sim 0$ \\\hline
        $(1,1)$ & dyonic & area law & D & strings & $E\sim\sigma L_s$ \\\hline
    \end{tabular}
    \caption{Twists and charges in standard $SU(2)$ gauge theory, along with line operators, confinement behavior, and vacuum energy scaling. The quantity $E$ denotes the ground-state energy of each subsector in the large-$S^2$-size limit,
    and the right-most column gives the large-$L_s$ behavior of the ground state energy.
    The coefficient $\sigma$ is the tension of the ground-state confining strings.}
    \label{tab:SU2}
\end{table}

Although we obtained Table \ref{tab:SU2} from our knowledge about $SU(2)$ Yang-Mills theory, topological manipulations only modify the left-most column in Table~\ref{tab:SU2}, and leave everything else unaltered.
We can read off the $\Z_2^{[1]}$-twists and $\Z_2^{[1]}$-charges of these lines in the other five theories from Fig.~\ref{fig:SU2_relations}.
Checking which types of lines are charged under $\Z_2^{[1]}$, we can identify the nature of each theory's $\Z_2^{[1]}$ --- whether it is electric or magnetic --- and identify the global structure of each theory's gauge group. This gives us the following classification.
\begin{itemize}
    \item I \& VI: Wilson and dyonic lines charged $\Rightarrow$ electric $\Z_2^{[1]}$ and gauge group $SU(2)$.
    \item II \& III: 
    't Hooft and dyonic lines charged $\Rightarrow$ magnetic $\Z_2^{[1]}$ and gauge group $SO(3)$.
    \item IV \& V: Wilson and 't Hooft lines charged $\Rightarrow$ dyonic $\Z_2^{[1]}$ and no conventional assignment of the gauge group.
\end{itemize}
In summary, we find three pairs of theories with different gauge groups and different types of $\Z_2^{[1]}$ symmetries.  

The two theories within each pair are exchanged by a $\Z_2^{[1]}$ $\Theta$ transformations, meaning that they differ by a shift of a symmetry $\theta$ angle.  This shuffles the twists of the two charged-line-operator subsectors.
This is just a topological Witten effect.
More explicitly, the $SU(2)$ theories I and V differ from each other by an electric symmetry $\theta$ angle, which is analogous to the electric $\theta$ angle discussed in Section~\ref{sec:electric_theta}.
The $SO(3)$ theories II and III are the $SO(3)_+$ and $SO(3)_-$ theories introduced in Ref.~\cite{Aharony:2013hda}, and differ from each other by a magnetic symmetry $\theta$ angle, which is also a Lagrangian $\theta$ angle as we discussed in Section~\ref{sec:solitonic_symmetry}. 
While $\Theta$ preserves the global structure, both of $S$ and $T$ change the global structure.

\begin{figure}[!ht]
\centering
\begin{adjustbox}{center,clip,raise=0cm,scale=0.8}

\includegraphics[width=0.7\textwidth]{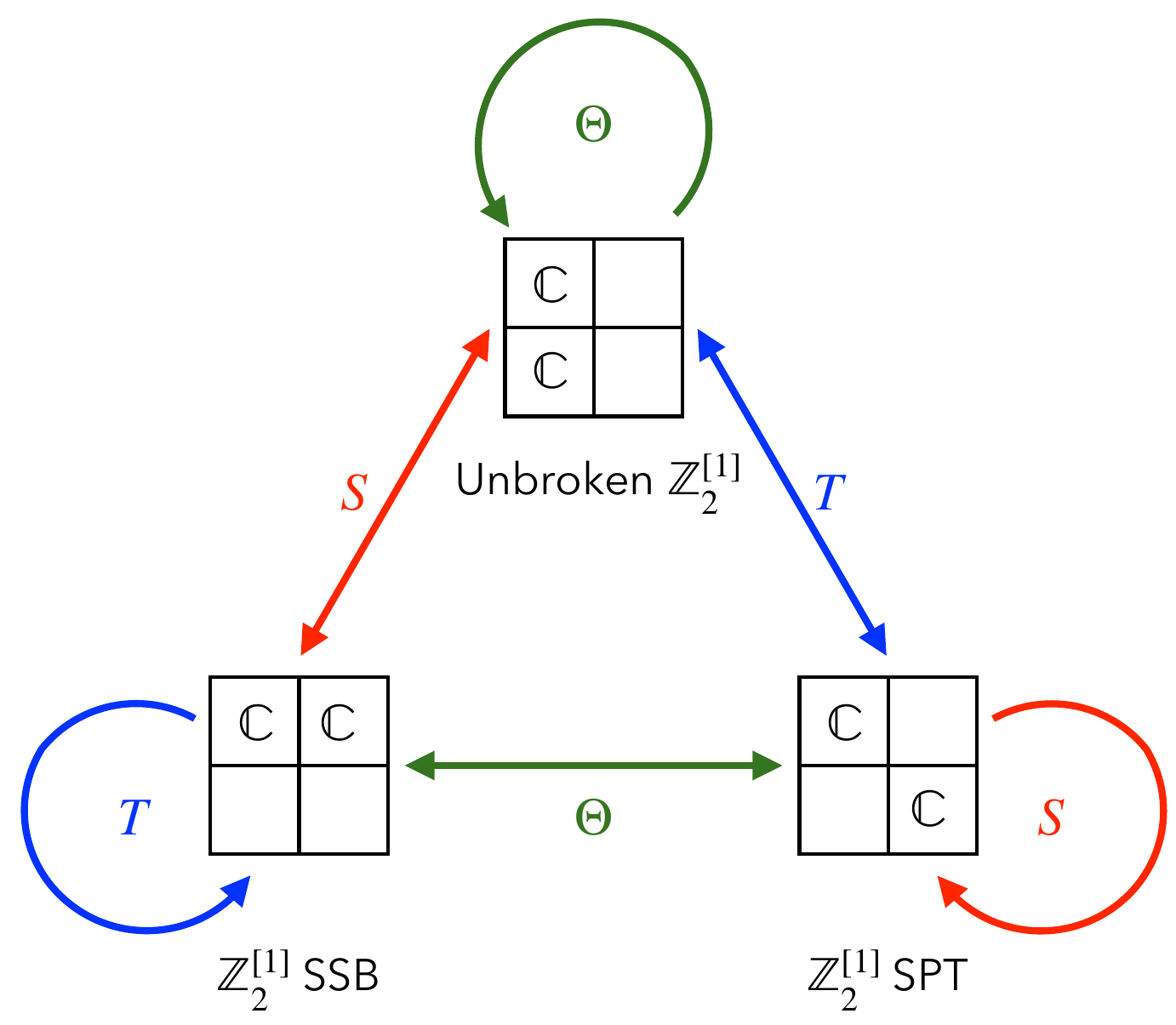}

\end{adjustbox}
\caption{
The three possible phases of $\mathfrak{su}(2)$ Yang-Mills theories and their relationships under the topological manipulations $S, T$, and $\Theta$.  The $\Theta$ transformation takes a phase with a spontaneously broken $\Z_2^{[1]}$ symmetry (SSB) to a $\Z_2^{[1]}$ SPT phase, so that it can be interpreted as a Kennedy-Tasaki transformation.
}
\label{fig:triangle}
\end{figure}

In the deep infrared, the stringy subsectors $B$ and $D$, which contain Wilson and dyonic lines, respectively, are completely gapped out.
The particle subsectors $A$ and $C$, which contain the trivial lines and 't Hooft lines, respectively, reduce to their vacua in the deep infrared.  This tells us the infrared structure of the six $\mathfrak{su}(2)$ theories: 
\begin{itemize}
    \item I$_{SU(2)}$ \& VI$_{SU(2)}$: 
    $\Z_2^{[1]}$-unbroken trivial phase, with the infrared partition function
    \begin{equation}
        \calZ(\rho)=1\,,\qquad \rho\in H^2(X,\Z_2)\,.
    \end{equation} 
    \item II$_{SO(3)}$ \& V$_{\text{dyonic}}$: 
    Spontaneous $\Z_2^{[1]}$-broken phase, with the infrared partition function
    \begin{equation}
        \calZ(\rho)=\delta(\rho)\,\sqrt{|H^2(X,\Z_2)|}\,,\qquad \rho\in H^2(X,\Z_2)\,.
    \end{equation} 
    \item III$_{SO(3)}$ \& IV$_{\text{dyonic}}$: $\Z_2^{[1]}$-unbroken SPT phase, with the infrared partition function (see Appendix~\ref{sec:quadratic_refinement} for the precise meaning of this expression)
    \begin{equation}
        \calZ(\rho)=\exp\left(\frac{\i\pi}{2}\int \calP(\rho)\right)\,,\qquad \rho\in H^2(X,\Z_2)\,.
    \end{equation}
\end{itemize}
These phases and the topological manipulations that transform between them are summarized in Fig.~\ref{fig:triangle}.
We see that $\Theta$ can indeed implement a Kennedy-Tasaki-type transformation, as expected in Section~\ref{sec:act_on_infrared}.
Finally, while the various phases above have different infrared realizations of the $\Z_2^{[1]}$ symmetry, all of them are confining in the sense that Wilson and dyonic lines have an area law, and subsectors $B$ and $D$ consist of stringy excitations.

\subsection{Cheshire \texorpdfstring{$\theta$}{theta} angles}
\label{sec:cheshire_examples}

In our recent paper with Gongjun Choi~\cite{Chen:2024tsx}, we introduced Cheshire $\theta$ angles that exist in $d$-dimensional QFTs with a non-anomalous $U(1)^{[0]}$ symmetry and a compact scalar $\varphi$.
In Section~\ref{sec:cheshire_theta_4d}, we show that they are actually $U(1)^{[0]}\!\times\!U(1)^{[d-2]}$ symmetry $\theta$ angles, where $U(1)^{[d-2]}$ is the winding symmetry of $\varphi$.
Then in Section~\ref{sec:cheshire_theta_3d}, we investigate some interesting properties of Cheshire $\theta$ angles using a simple toy model.
Finally in Section~\ref{sec:largeN_QCD}, we comment on the intriguing possibility of an infrared-emergent Cheshire $\theta$ angle in $4$-dimensional large-$N_c$ QCD.



\subsubsection{Cheshire \texorpdfstring{$\theta$}{theta} terms are symmetry \texorpdfstring{$\theta$}{theta} angles}
\label{sec:cheshire_theta_4d}

Consider a non-anomalous $U(1)^{[0]}\!\times\!U(1)^{[d-2]}$ symmetry. 
The dual symmetry is $U(1)^{[d-3]} \times U(1)^{[-1]}$; see Section~\ref{sec:defor&dual} for the meaning of $U(1)^{[-1]}$.
Hence according to Eq.~\eqref{eq:classification_continuous}, $d$-dimensional $U(1)^{[0]}\!\times\!U(1)^{[d-2]}$-symmetry $\theta$ angles that are not individual $U(1)^{[0]}$-symmetry or $U(1)^{[d-2]}$-symmetry $\theta$ angles are classified by%
\footnote{\label{ft:smash}
Here $\wedge$ is the smash product between (pointed) topological spaces.
Roughly, $X\times Y$ contains information about individual $X$ and $Y$, while $X\wedge Y$ extracts out the information purely about their mixture. 
}
\begin{align}
\begin{gathered}
    \Hom\left(\tOmega^{Spin}_{d}\Bigl(B^{d-2}U(1)\wedge S^1\Bigr),\tfrac{\R}{2\pi\Z}\right) = \Hom\left(\tOmega^{SO}_{d}\Bigl(B^{d-2}U(1)\wedge S^1\Bigr),\tfrac{\R}{2\pi\Z}\right) = \tfrac{\R}{2\pi\Z}\,.
\end{gathered}
\end{align}
The associated strictly-topological $U(1)^{[d-3]}\!\times\!U(1)^{[-1]}$-symmetric invertible theory, essentially a $\Z^{[d-2]}\!\times\!\Z^{[0]}$-symmetric invertible theory, is
\begin{align}
    \widehat{\calJ}_{\theta}\left(\left[\frac{\d C}{2\pi}\right],\left[\frac{\d\phi}{2\pi}\right]\right) = \exp \left(\i \theta \int \left[\frac{\d C}{2\pi}\right]\cup \left[\frac{\d\phi}{2\pi}\right]\right) = \exp \left( \frac{\i \theta}{4\pi^2} \int \d C \wedge \d \phi\right)\,,
\end{align}
where $C$ is a $U(1)^{[d-3]}$ gauge field and $\phi$ is a $2\pi$-periodic compact scalar. 
Performing a $\Theta$ transformation to a $U(1)^{[0]}\!\times\!U(1)^{[d-2]}$-symmetric theory $\calZ$, namely applying Eq.~\eqref{eq:Theta_U(1)}, we find the theory with a symmetry $\theta$ angle given by 
\begin{equation}
\label{eq:general_cheshire}
\begin{split}
     \calZ_{\theta}(A,B) 
     &= \int \calD\phi\,\calD b \ \calZ\left(A+\frac{\theta}{2\pi}\d\phi,\,B-b\right)\,\exp\left( \frac{\i}{2\pi} \int \d \phi \wedge b \right)\,,
\end{split}
\end{equation}
where $B$ and $b$ are $U(1)^{[d-2]}$ gauge fields, $A$ is a $U(1)^{[0]}$ gauge field, and $\phi$ is a $2\pi$-periodic compact scalar.

To make contact with the discussion of Ref.~\cite{Chen:2024tsx}, we take $\calZ$ to contain a dynamical $2\pi$-periodic compact scalar $\varphi$ and let $U(1)^{[d-2]}$ be the winding symmetry of $\varphi$, i.e.
\begin{equation}
    \calZ(A,B) = \int\calD\varphi\,\exp\left\{-\calS(A,\varphi) + \frac{\i}{2\pi} \int  {B \wedge \d \varphi} \right\}\,.
\end{equation}
Plugging this partition function into Eq.~\eqref{eq:general_cheshire}, we obtain
\begin{equation}\label{eq:cheshire_from_STS}
    \calZ_{\theta}(A,B) = \int\calD\varphi\,\exp\left\{-\calS\left(A+ {(-1)^{d-1}}\frac{\theta}{2\pi}\d\varphi,\,\varphi \right) + \frac{\i}{2\pi} \int  {B \wedge \d \varphi }\right\}\,.
\end{equation}
Hence we confirmed that the Cheshire $\theta$ angle of Ref.~\cite{Chen:2024tsx} is precisely a $U(1)^{[0]}\!\times\!U(1)^{[d-2]}$-symmetry $\theta$ angle, with $U(1)^{[d-2]}$ being the winding symmetry of a dynamical compact scalar in the theory. 
In fact, the physical consequences of the Cheshire $\theta$ angle discussed in Ref.~\cite{Chen:2024tsx} is precisely the topological Witten effect associated to the $U(1)^{[0]}\!\times\!U(1)^{[d-2]}$-symmetry $\theta$ angle.
An advantage of the construction in this paper is that we can write down the Cheshire $\theta$ angle even if the compact scalar $\varphi$ appears in its dual form, a $U(1)^{[d-3]}$ gauge field; we just need to plug things into Eq.~\eqref{eq:general_cheshire}. 

Finally, we should emphasize that despite the fact that the Cheshire $\theta$ angle in Eq.~\eqref{eq:cheshire_from_STS} appears as a term in the Lagrangian, it is actually \emph{not} a Lagrangian $\theta$ angle under Definition~\ref{def:lagrangian_theta}.
According to Section~\ref{sec:solitonic_symmetry}, the Cheshire $\theta$ angle reduces to a Lagrangian $\theta$ angle only when both $U(1)^{[0]}$ and $U(1)^{[d-2]}$ are solitonic symmetries.
In fact, this paper started as an attempt to understand what framework for $\theta$ angles might include general Cheshire $\theta$ angles.


\subsubsection{Quadratic Abelian theories}
\label{sec:cheshire_theta_3d}

We now consider Cheshire $\theta$ angles in a simple class of QFTs with two different fields. While we will focus on free field theories below for concreteness, everything below (with the exception of exact partition function evaluations) also applies to generic interacting QFTs with the same symmetries. 
Let us consider a dynamical $U(1)^{[d-3]}$ gauge field $a_1$ and a dynamical $2\pi$-periodic compact scalar $\varphi_2$, with kinetic terms only,
\begin{equation}\label{eq:2A1}
    \int\!\calD a_1\calD \varphi_2\ \exp\left( - \frac{1}{4\pi R_1^2} \int \d a_1 \wedge \star \d a_1 - \frac{R_2^2}{4\pi} \int \d \varphi_2 \wedge \star \d \varphi_2 \right)\,.
\end{equation}
This field content is self-dual, in the sense that Abelian duality implies that $a_1$ is equivalent to a $2\pi$-periodic compact scalar $\varphi_1$, while $\varphi_2$ is equivalent to a $U(1)^{[d-3]}$ gauge field $a_2$.
As a result, this theory has three additional equivalent path-integral representations,
\begin{subequations}
\begin{align}
    \int\!\calD \varphi_1\calD\varphi_2\ &\exp\left( - \frac{R_1^2}{4\pi} \int \d\varphi_1 \wedge \star \d\varphi_1 - \frac{R_2^2}{4\pi} \int \d \varphi_2 \wedge \star \d \varphi_2 \right)\,,\label{eq:2A2}\\
    \int\!\calD a_1\calD a_2\ &\exp\left( - \frac{1}{4\pi R_1^2} \int \d a_1 \wedge \star \d a_1 - \frac{1}{4\pi R_2^2} \int \d a_2 \wedge \star \d a_2 \right)\,,\label{eq:2A3}\\
    \int\!\calD \varphi_1\calD a_2\ &\exp\left( - \frac{R_1^2}{4\pi} \int \d \varphi_1 \wedge \star \d \varphi_1 - \frac{1}{4\pi R_2^2} \int  \d a_2 \wedge \star \d a_2 \right)\,.\label{eq:2A4}
\end{align}
\end{subequations}
This theory has two non-anomalous global symmetries 
\begin{align}
    U(1)^{[0]}_{1} \times U(1)^{[d-2]}_{2}\,,\qquad U(1)^{[d-2]}_{1}\times U(1)^{[0]}_{2}\,,
    \label{eq:quadratic_symm}
\end{align} 
as well as a mixed 't Hooft anomaly between $ U(1)^{[0]}_{1}$ and $ U(1)^{[d-2]}_{1}$, as well as another one between $ U(1)^{[0]}_{2}$ and $ U(1)^{[d-2]}_{2}$.
Since the two symmetries in Eq.~\eqref{eq:quadratic_symm} are not anomalous, we can construct Cheshire $\theta$ angles for either of them.
Since these two symmetries are exchanged under $1\leftrightarrow 2$, we just focus on one of them, $U(1)^{[0]}_{1}\!\times\! U(1)^{[d-2]}_{2}$. 

The nature of the symmetry $U(1)^{[0]}_{1}\!\times\! U(1)^{[d-2]}_{2}$ --- whether it is solitonic or not --- varies among different path-integral representations.
We thus want to examine the form of the Cheshire $\theta$ angle in the four different path-integral representations.
In the first representation~\eqref{eq:2A1}, Eq.~\eqref{eq:general_cheshire} leads us to 
\begin{equation}
    \int\!\calD a_1\calD \varphi_2\ \exp\left( - \frac{1}{4\pi R_1^2} \int \d a_1 \wedge \star \d a_1 - \frac{R_2^2}{4\pi} \int \d \varphi_2 \wedge \star \d \varphi_2 + \frac{\i\theta}{4\pi^2}\int \d a_1 \wedge \d\varphi_2  \right)\,.
\end{equation}
Here the Cheshire $\theta$ term appears as a Lagrangian $\theta$ angle, and the $2\pi$-periodicity of the parameter $\theta$ is due to the quantization of topological charges.
As we discussed in Section~\ref{sec:solitonic_symmetry}, this is because in the first representation, $U(1)^{[0]}_{1} \!\times\! U(1)^{[d-2]}_{2}$ is a solitonic symmetry that acts on defects only.
In the second representation~\eqref{eq:2A2}, Eq.~\eqref{eq:general_cheshire} leads us to
\begin{equation}\label{eq:2A2_theta}
    \int\!\!\calD \varphi_1\calD\varphi_2\,\exp\left[ \!- \frac{R_1^2}{4\pi} \!\int\!\!\left(\d\varphi_1\!-  {(-1)^{d-1}}\!\frac{\theta}{2\pi}\d\varphi_2 \right)\!\wedge\!\star\! \left(\d\varphi_1 \!-\!\frac{\theta}{2\pi}\d\varphi_2 \right) - \frac{R_2^2}{4\pi} \!\!\int\!\!\d \varphi_2 \wedge \star \d \varphi_2 \right]\!,
\end{equation}
while in the third representations~\eqref{eq:2A3}, Eq.~\eqref{eq:general_cheshire} leads us to 
\begin{equation}
    \int\!\!\calD a_1\calD a_2\,\exp\left[ - \frac{1}{4\pi R_1^2} \!\int\!\!\d a_1 \wedge \star \d a_1 - \frac{1}{4\pi R_2^2} \!\int\!\!\left(\d a_2\!-\!\frac{\theta}{2\pi}\d a_1 \right)\!\wedge\!\star\! \left(\d a_2 \!-\!\frac{\theta}{2\pi}\d a_1 \right) \right]\!.
\end{equation}
Hence the Cheshire $\theta$ term now appears via kinetic mixing contributions, and the $2\pi$-periodicity of the parameter $\theta$ comes from the field-periodicity-preserving variable changes $\varphi_1\to\varphi_1+  {(-1)^{d-1}}\varphi_2$ or $a_2\to a_2+a_1$.
In each of the second and the third representations, one of $U(1)^{[0]}$ and $U(1)^{[d-2]}$ is solitonic, while the other is not.
In the last representation where neither symmetry is solitonic, Eq.~\eqref{eq:general_cheshire} leads us to
\begin{equation}
\begin{split}
    \int\!\calD \varphi_1\calD a_2\calD\phi\calD b\ \exp\left[ - \frac{R_1^2}{4\pi} \int \left( \d\varphi_1-\frac{\theta}{2\pi}\d\phi \right)\wedge\star \left(\d\varphi_1 -\frac{\theta}{2\pi}\d\phi \right) \right. \\
    \left. - \frac{1}{4\pi R_2^2} \int\left(\d a_2   {+} b \right)\wedge\star \left(\d a_2    {+} b \right)   {+} \frac{\i}{2\pi} \int \d \phi \wedge b \right]\,,
\end{split}
\end{equation}
where $\phi$ is an auxiliary compact scalar and $b$ is an auxiliary $U(1)^{[d-2]}$ gauge field.
While the definition of the Cheshire $\theta$ angle as a symmetry $\theta$ angle is completely independent of the choice of path integral representation, the way it actually appears in path integrals can look drastically different depending on the choice of duality frame.

Working in whichever path-integral representation one likes, one always finds the same physical consequence of the Cheshire $\theta$ angle, because symmetry $\theta$ angles are intrinsic structures associated to QFTs and their symmetries.
For example, the evaluations of the four path integrals on the spacetime $S^{d-1}\!\times\!S^1_L$ will always lead to the same partition function, 
\begin{align}\label{eq:Cheshire_partition}
    \calZ_\theta = \calF^2 \!\!\sum_{\ell_1,\ell_2\in\mathbb{Z}}\!\!\!\e^{-L E_{\theta}(\ell_1,\ell_2)}\,,\quad\  E_{\theta}(\ell_1,\ell_2) = \frac{\pi}{|S^{d-1}|} \left[  \frac{\ell_1^2}{R_1^2}  + \frac{\left(\ell_2+\frac{\theta}{2\pi}\ell_1\right)^2}{ R_2^2} \right],
\end{align}
where $|S^{d-1}|$ denotes the volume of $S^{d-1}$.
Here $\calF$ does not depend on $\theta$, $R_1$, or $R_2$, and is the temperature-$1/L$ thermal partition function of a free massless spin-0 boson moving on $S^{d-1}$.
The parameters $\ell_1$ and $\ell_2$ are the $U(1)^{[0]}_1$-charge and the $U(1)^{[0]}_2$-charge, respectively.
Since we are working in a compact geometry, none of the symmetries are spontaneously broken. If we interpret our $d$-dimensional QFT as the quantum mechanics of a $(d\!-\!1)$-dimensional membrane, the $\ell_1,\ell_2$ sector describes the motion of the center of mass of this membrane on a non-rectangular torus defined by
\begin{equation}
    \frac{\R^2}{\Z\left(R_1,0\right)+\Z\left(\frac{\theta}{2\pi}R_1,R_2\right)}\,,
\end{equation}
as we expect from the second path-integral representation~\eqref{eq:2A2_theta}.

\subsubsection{Application to large-\texorpdfstring{$N_c$}{Nc} QCD}
\label{sec:largeN_QCD}

A Cheshire $\theta$ angle may appear in the low-energy effective theory of 4-dimensional large-$N_c$ QCD with light fundamental quarks.
For the sake of argument, let us consider
the $N_f=2$ chiral Lagrangian for large-$N_c$ QCD.  
This Lagrangian includes a light pseudoscalar $\eta'$ meson~\cite{Witten:1980sp}.  
The $U(1)_B$ symmetry of QCD is realized as a solitonic symmetry in the chiral EFT. The $U(1)_B$ conserved current is the Skyrme current~\cite{Witten:1983tw,Witten:1983tx}
\begin{align}
    j_B = -\frac{1}{24\pi^2} \star \tr (U^{\dag}dU)^3 \,,
\end{align}
where the matrix $U \in SU(2)$ contains the pion fields. 
In principle, the chiral Lagrangian can include a Cheshire $\theta$ term
\begin{align}
    \i \, \theta \int \frac{d\eta'}{2\pi f_{\pi}} \wedge \frac{1}{24\pi^2} \tr (U^{\dag}dU)^3\,,
    \label{eq:cheshire_skyrme}
\end{align}
where we take a normalization where the $\eta'$ periodicity is $2\pi f_{\pi}$, and $f_{\pi}$ is the pion decay constant.  
Since the Skyrme current is tautologically conserved due to a Bianchi identity, this Cheshire $\theta$ angle~\eqref{eq:cheshire_skyrme} in the large-$N_c$ QCD chiral Lagrangian is also a Lagrangian $\theta$ angle.  
Such a $\theta$ term may induce interesting physical effects.
The simplest effect might be that a Skyrmion will carry a $U(1)_A$ charge $\theta/2\pi$.  Also, there should be an AB effect for Skyrmions in the presence of $\eta'$ strings. However, the presence of a CS theory on $\eta'$ domain walls~\cite{Gaiotto:2017tne} and the proposed relationship of $\eta'$ `pancakes' to high-spin baryons~\cite{Komargodski:2018odf,Ma:2019xtx,Karasik:2020pwu,Karasik:2020zyo,Bigazzi:2022luo} means that this should be analyzed carefully in future works.

The only two values of the large-$N_c$ chiral EFT Cheshire $\theta$ angle that are consistent with parity symmetry  are $\theta = 0$ or $\theta = \pi$.  Therefore these are the only two values of the Cheshire $\theta$ angle that are possible when the YM $\theta$ angle takes the parity-preserving values $\theta_{YM} = 0 \textrm{ or } \pi$. 
There are no symmetry restrictions on the possible values of the Chershire $\theta$ angle when $\theta_{YM} \neq 0 \textrm{ or } \pi$.  

The $U(1)^{[2]}$ symmetry that counts $\eta'$ strings appears at the QCD scale $\Lambda_{\rm QCD}$.  This means that one should not attempt to match Eq.~\eqref{eq:cheshire_skyrme} to any interactions in the quark and gluon QCD Lagrangian.  One should instead think of the Cheshire $\theta$ angle as a topological interaction term that is generated by the non-perturbative confining dynamics during the RG flow toward the IR.  Indeed, given that generic values of $\theta$ are not excluded by any symmetry principle when $\theta_{\rm YM} \neq 0 \textrm{ or } \pi$, standard EFT naturalness arguments imply that we should find $\theta = \mathcal{O}(1)$ when $\theta_{\rm YM} = \mathcal{O}(1)$. 
It would be very interesting to find a method to pin down the value of this $\theta$ parameter.
\subsection{Cubic \texorpdfstring{$\theta$}{theta} angles}
\label{sec:cubic_examples}

So far, the $\Theta$ transformations in all of the examples we have discussed involve invertible theories that are quadratic in background gauge fields.
We now introduce examples of a more exotic type of symmetry $\theta$ angles that involves cubic invertible theories. 
We will start with a discussion of an example of such cubic $\theta$ angles in a QFT in $3$ spacetime dimensions in Section~\ref{sec:U1_cubed_3d}, and then give an example in a QFT in $4$ spacetime dimensions in Section~\ref{sec:U1_cubed_4d}.  While the cubic $\theta$ angles in our examples will be continuous, they also have discrete counterparts with very similar consequences.
Cubic $\theta$ angles turn out to lead to quite unusual Witten effects. It would be very interesting to realize cubic $\theta$ angles in e.g. condensed-matter systems or BSM models in particle physics.


\subsubsection{3D cubic \texorpdfstring{$\theta$}{theta} angles}
\label{sec:U1_cubed_3d}


Let us start with three copies of $3$-dimensional Maxwell theory.
Namely, we consider three $U(1)^{[0]}$ gauge fields $(a_1,a_2,a_3)$ with the path integral
\begin{align}
    \int\!\calD a_1\calD a_2\calD a_3\ \exp\left( -\sum_{k=1}^{3} \frac{1}{4\pi R_k^2} \int \d a_k \wedge \star \d a_k \right)\,.
\end{align}
Each gauge field $a_k$ is equivalent to a $2\pi$-periodic compact scalar $\varphi_k$ under Abelian duality.
Therefore, there are in total $2^3 = 8$ different path-integral representations of the same theory, just as in Section~\ref{sec:cheshire_theta_3d}. 
The all-scalar representation is especially useful for our purposes:
\begin{align}
    \int\!\calD \varphi_1\calD \varphi_2\calD \varphi_3\ \exp\left( -\sum_{k=1}^{3} \frac{R_k^2}{4\pi} \int \d \varphi_k \wedge \star \d \varphi_k \right)\,.
\end{align}
The model has two sets of non-anomalous global symmetries
\begin{equation}
    U^{[0]}_1 \times U^{[0]}_2 \times U^{[0]}_3\,,\qquad U^{[1]}_1 \times U^{[1]}_2 \times U^{[1]}_3\,,
\end{equation}
as well as a mixed 't Hooft anomaly between each pair of $U(1)^{[0]}_k$ and $U(1)^{[1]}_k$.
Each $U(1)^{[0]}_k$ acts on local monopole operators associated to $a_k$, 
while each $U(1)^{[1]}_k$ acts on Wilson lines associated to $a_k$.

Let us consider the electric $\U = U(1)^{[1]}_1 \times U(1)^{[1]}_2 \times U(1)^{[1]}_3$ symmetry.   We will call the symmetry $\theta$ angle associated to $\U$ a \textit{cubic} $\theta$ angle.
Since the dual symmetry is $\widehat{\U} = U(1)^{[-1]}_1 \times   U(1)^{[-1]}_2 \times  U(1)^{[-1]}_3$,   
the cubic $\theta$ angles are classified by
\begin{align}
    \Hom\left(\tOmega^{SO}_{3}\bigl(S^3\bigr),\tfrac{\R}{2\pi\Z}\right)= \Hom\left(\tOmega^{Spin}_{3}\bigl(S^3\bigr),\tfrac{\R}{2\pi\Z}\right)= \tfrac{\R}{2\pi\Z} \,,
\end{align}
where $S^3=S^1\wedge S^1\wedge S^1$.
The associated strictly-topological invertible theory is
\begin{align}
    \widehat{\calJ}_{\theta}\left(\left[\frac{\d\phi_1}{2\pi}\right],\left[\frac{\d\phi_2}{2\pi}\right],\left[\frac{\d\phi_3}{2\pi}\right]\right) = \exp\left(\i\theta \int \left[\frac{\d\phi_1}{2\pi}\right] \cup  \left[\frac{\d\phi_2}{2\pi}\right] \cup 
     \left[\frac{\d\phi_3}{2\pi}\right]\right)\,,\quad \theta \in \tfrac{\R}{2\pi\Z}\,.
\end{align}
Applying a $\Theta$ transformation using $\widehat{\calJ}_{\theta}$ generates a continuous symmetry $\theta$ angle.  It can look more or less exotic depending on the choice of duality frame, and looks simplest in the all-scalar duality frame:
\begin{align}
    \int\!\calD \varphi_1\calD \varphi_2\calD \varphi_3\ \exp\left( -\sum_{k=1}^{3} \frac{R_k^2}{4\pi} \int \d \varphi_k \wedge \star \d \varphi_k   {+}\frac{\i\theta}{8\pi^3} \int \d\varphi_1 \wedge \d\varphi_2 \wedge \d\varphi_3 \right)\,.
\end{align}
In terms of the compact scalars, $U(1)^{[1]}_1 \times U(1)^{[1]}_2 \times U(1)^{[1]}_3$ is solitonic, and thus the cubic $\theta$ angle is also a Lagrangian $\theta$ angle. The cubic $\theta$ angle is not a Lagrangian $\theta$ angle in any of the other $7$ path-integral descriptions; c.f.~Section~\ref{sec:cheshire_theta_3d}.
The scalars $\varphi_k$ are odd under spacetime reflections, so the $\theta$ term above does not break reflection symmetry.  This means that it is an allowed term in the effective field theory description of e.g. the $U(1)^3$ Coulomb phase of 3d reflection-symmetric $SU(4)$ gauge theory. 

One of the most direct ways to understand the effect of this cubic $\theta$ angle is to compute the partition function of the theory on $S^1_{X} \times S^1_{Y}\times S^1_{L}$.  None of the symmetries are spontaneously broken in this compact-spacetime setting.
Taking $S^1_{L}$ as the temporal direction and $S^1_{X}\!\times\! S^1_{Y}$ as the spatial manifold, we can take advantage of the fact that the path integral is one-loop exact to write the partition function as 
\begin{equation}\label{eq:cubic_magnetic_Witten}
    \calZ_{\theta} = \calF^3\!\!\! \sum_{\Vec{m},\Vec{\mu},\Vec{\nu}\,\in\,\Z^3}\!\!\! \exp\left\{-\,L\, \Bigl[\,E_{\theta}(\Vec{m},\Vec{\mu},\Vec{\nu})\, +\, X\,\calV(\Vec{\mu}) \,+\, Y\,\calW(\Vec{\nu})\, \Bigr]\right\}\,,
\end{equation}
with
\begin{subequations}
\begin{gather}
    E_{\theta}(\Vec{m},\Vec{\mu},\Vec{\nu}) = \frac{\pi}{XY} \sum_{k=1}^3 \frac{1}{ R_k^2}\left(\!m_k+\frac{\theta}{2\pi}\sum_{i,j=1}^3\epsilon_{ijk}\,\nu_j\,\mu_i \! \right)^2\,,\label{eq:cubic_energies}\\
    \calV(\Vec{\mu}) = \frac{\pi}{Y} \sum_{k=1}^3 {R_k^2}\mu_k^2\,,\qquad\quad \calW(\Vec{\nu}) = \frac{\pi}{X} \sum_{k=1}^3 {R_k^2} \nu_k^2\,.\label{eq:cubic_strings}
\end{gather}
\end{subequations}
The physical interpretation of the quantities appearing in Eq.~\eqref{eq:cubic_magnetic_Witten} is as follows:
\begin{itemize}
    \item The function $\calF$ is the temperature-$1/L$ thermal partition function of a free massless spin-$0$ boson moving on $S^1_{X}\!\times\! S^1_{Y}$, and does not depend on $\theta$ or $R_k$. 
    \item The sum over  $\mu_k\in\Z$ in Eq.~\eqref{eq:cubic_magnetic_Witten} is a sum over the $U(1)^{[1]}_k$-charges on $S^1_X$, and counts electric-flux strings stretched along $S^1_X$. The function $\calV(\Vec{\mu})$ gives the tension of these strings. 
    \item Similarly, $\nu_k\in\Z$ labels the $U(1)^{[1]}_k$ charges of electric-flux of strings stretched along $S^1_Y$, and $\calW(\Vec{\nu})$ is the tension of these strings.
    \item The final summation index $m_k$ labels the magnetic-monopole $U(1)^{[0]}_k$-charge on $S^1_X\!\times\!S^1_Y$.  It can be interpreted as describing the center-of-mass mode of the 2-dimensional membrane $S^1_X\!\times\!S^1_Y$, similarly to our discussion below Eq.~\eqref{eq:Cheshire_partition}.
\end{itemize} 


Equation~\eqref{eq:cubic_energies} exhibits an unusual charge Witten effect.
It implies that the intersection line between an $a_1$-electric-flux string and an $a_2$-electric-flux string carries a fractional $a_3$-magnetic-monopole charge.
In order to see the same effects on $\R^3$, we need to get $a_1$ and $a_2$ to confine via a Polyakov mechanism, which breaks $U(1)^{[0]}_1\!\times\!U(1)^{[0]}_2$ but preserves $U(1)^{[1]}_1\!\times\!U(1)^{[1]}_2$.
Then intersections of the two types of confining strings carry a fractional magnetic-monopole charge under the third photon.
The same conclusions hold if we permute the three fields.

\begin{figure}[htbp]
    \centering
    \begin{subfigure}[t]{0.45\textwidth}
        \centering
        \includegraphics[width=0.9\textwidth]{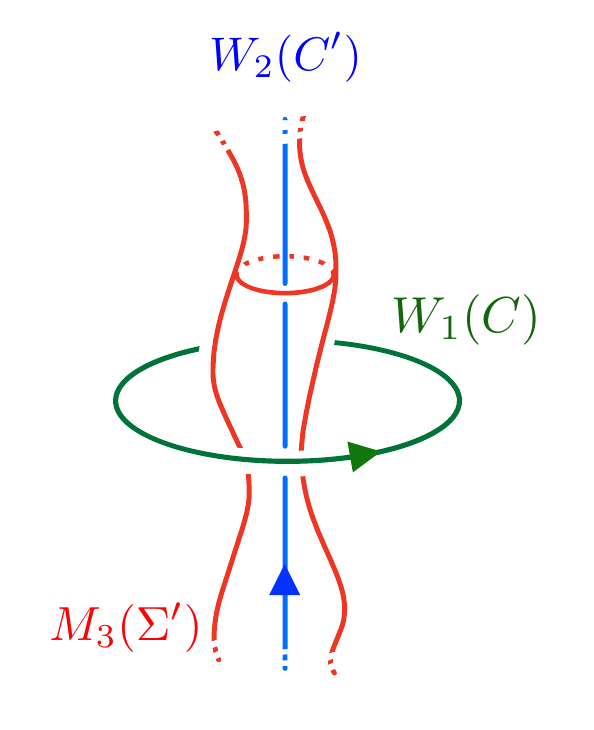}
        \caption{Charge Witten effect}
        \label{fig:cubic_CWE}
    \end{subfigure}
    \hfill
    \begin{subfigure}[t]{0.45\textwidth}
        \centering
        \includegraphics[width=0.9\textwidth]{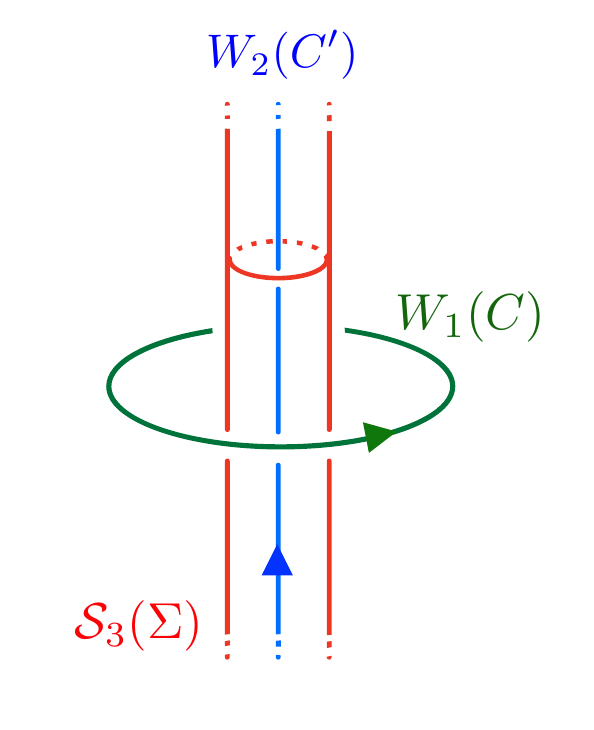}
        \caption{Topological Witten effect}
        \label{fig:cubic_TWE}
    \end{subfigure}
    \caption{The cubic symmetry $\theta$ angle induces topological and charge Witten effects in $U(1)_1 \times U(1)_2 \times U(1)_3$ gauge theory in three spacetime dimensions. Figure~\ref{fig:cubic_CWE} shows the charge Witten effect, which appears in the presence of two linked Wilson loops (blue and green curves).  In the situation displayed in the figure, the physical magnetic charge $M_3$ measured on a surface $\Sigma'$ can take a fractional value, $M_3(\Sigma') = \theta/(2\pi)$. Figure~\ref{fig:cubic_TWE} shows a situation in which the topological Witten effect leads to an AB effect.  Confining string excitations $\calS_3(\Sigma)$ with unit charge under the $U(1)_3$ gauge group can pick up an AB phase $e^{i\theta}$ when their worldsheet $\Sigma$ has the topology of a torus which appropriately encircles two linked Wilson loops with unit charge under the other two gauge groups.   }
    \label{fig:cubic_Witten_3d}
\end{figure}

We can also see this charge Witten effect at the operator level by inspecting the topological operator of $U(1)^{[0]}_3$ on a closed surface $\Sigma$:
\begin{align}
    U_3(\alpha;\Sigma) = \exp \left[\i \alpha\int_{\Sigma} \left(  {-} \frac{\i R_3^2}{2\pi} \star\!\d\varphi_3 
    - \frac{\theta}{
    8\pi^3}\, d\varphi_2\wedge d\varphi_3 \right)\right]\,.
\end{align}
Because the $U(1)^{[0]}_3$ charges $m_3$ measured by $U_3(\alpha;\Sigma)$ are always integers, the third physical magnetic charge 
\begin{align}
    M_3(\Sigma) \equiv \int_{\Sigma}   {-}\frac{\i R_3^2}{2\pi} \star\!\d\varphi_3
\end{align}
becomes generically fractional when $\theta \neq 0$ and $\Sigma$ has an appropriate topology.  
For example, such a fractional magnetic charge appears when an $a_1$ Wilson loop is linked with an $a_2$ Wilson loop,
as illustrated in Fig.~\ref{fig:cubic_CWE}.
Again, this discussion stays true if we permute the three fields.

Let us now instead confine $a_3$ through a Polyakov mechanism.
Then an $a_3$ confining string sees an exotic AB effect when it passes a pair of linked $a_1$ and $a_2$ Wilson loops.
For the sake of argument, let us work on the spacetime $\R^3$.  First we put a unit-charge $a_2$ Wilson line $W_2(C')$ along the temporal $t$ axis at the origin $(x,y) = (0,0)$ of the spatial coordinates.
We then put a unit-charge $a_1$ Wilson loop $W_1(C)$ circling this static probe $W_2(C')$ at the moment of $t=0$.
Then let us consider the worldsheet of a unit-charge $a_3$ confining string that encloses the static probe $W_2(C')$.
If the string passes through the inner side of $W_1(C)$ as illustrated in Fig.~\ref{fig:cubic_TWE}, then this process 
picks up an AB phase of $e^{i\theta}$.
Once again, this discussion stays true if we permute the three fields.

In analogy to the discussion of Section~\ref{sec:electric_theta}, the rational cubic $\theta$ angle with $\theta=2\pi q/p$ survives when we couple to electric matter with charges that are multiples of $p$.
This is because in essence, they are symmetry $\theta$ angles for the $(\Z_{p}^{[1]})_1\!\times\! (\Z_{p}^{[1]})_2\!\times\! (\Z_{p}^{[1]})_3$ subsymmetry of $U(1)^{[1]}_1\!\times\!U(1)^{[1]}_2\!\times\!U(1)^{[1]}_3$, which survives after coupling to charge $p$ matter.

\subsubsection{4D cubic \texorpdfstring{$\theta$}{theta} angles}
\label{sec:U1_cubed_4d}

The discussion in the preceding section easily generalizes to 4 spacetime dimensions
by changing two of the 1-form symmetries to 2-form symmetries.
A simple free-QFT example has a $U(1)$ gauge field $a$ and two $2\pi$-periodic compact scalars $\varphi_1$ and $\varphi_
2$:
\begin{align}
    \int\!\calD a\calD \varphi_1\calD \varphi_2\ \exp\left( - \frac{1}{2g^2}\int \d a \wedge \star \d a -\sum_{k=1}^{2} \frac{R_k^2}{4\pi} \int \d \varphi_k \wedge \star \d \varphi_k \right)\,.
    \label{eq:4d_cubic_action}
\end{align}
This symmetry of this theory includes 
\begin{equation}
   \U =  U(1)^{[1]}_m\times U(1)^{[2]}_1\times U(1)^{[2]}_2\,.
\end{equation}
as a subgroup.  In the variables of Eq.~\eqref{eq:4d_cubic_action}, $\U$ is a solitonic symmetry.  The $\U$ symmetry $\theta$ angle can thus be written as a Lagrangian $\theta$ angle:
\begin{align}\label{eq:cubic_theta_4d}
   \int\!\calD a\calD \varphi_1\calD \varphi_2\, \exp\Bigg( &- \frac{1}{2g^2} \!\int\! \d a \wedge \star \d a -\sum_{k=1}^{2} \frac{R_k^2}{4\pi} \!\int\! \d \varphi_k \wedge \star \d \varphi_k \\
   &+ \frac{\i\theta}{8\pi^3} \!\int\! \d a \wedge  \d\varphi_1 \wedge \d\varphi_2 \Bigg)\,.
   \nonumber
\end{align}
We should note that everything we said so far will go through if we add arbitrary electrically-charged matter for $a$ and $2\pi$-periodic potentials for $\varphi_k$. 

\begin{figure}[!ht]
    \centering
    \includegraphics[width=0.6\textwidth]{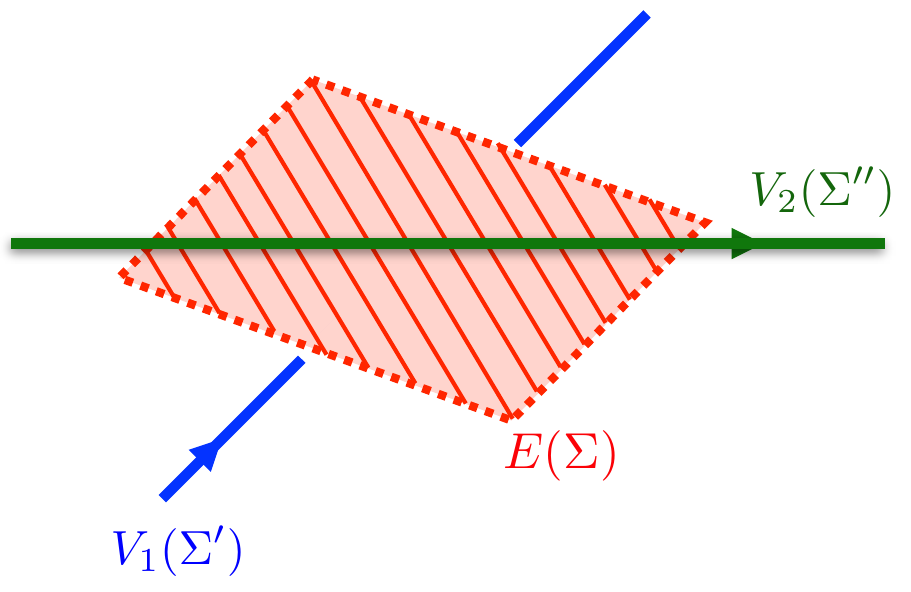}
    \caption{In the presence of a cubic symmetry $\theta$ angle in Eq.~\eqref{eq:cubic_theta_4d}, the physical electric charge $E$ measured on a Gaussian surface $\Sigma$ can be fractional.  The figure shows a (blue) $\varphi_1$ unit vortex with a worldsheet $\Sigma'$, $V_1(\Sigma')$, and a (green) $\varphi_2$ unit vortex with a worldsheet $\Sigma''$, $V_2(\Sigma'')$.  The worldsheets $\Sigma'$ and $\Sigma''$ do not intersect.  In this situation, the physical electric charge measured on the Gaussian surface $\Sigma$ (red) will be fractional, $E(\Sigma) = \theta/(2\pi)$.}
    \label{fig:fractional_electric_charge}
\end{figure}

The cubic $\theta$ angle in Eq.~\eqref{eq:cubic_theta_4d} induces an unusual charge Witten effect, just as in Section~\ref{sec:U1_cubed_3d}.  
For example, the topological operator for the $U(1)_{e}^{[1]}$ symmetry of $a$ is
\begin{align}
    U_{e}(\alpha;\Sigma) = \exp \left[\i \alpha 
    \int \left(- \frac{\i}{g^2} \star \d a - \frac{\theta}{8\pi^3} \d\varphi_1 \wedge  \d\varphi_2 \right)\right] \,.
\end{align}
Because $U_{e}(\alpha;\Sigma)$ always detects integer-valued charges, the physical electric charge $E(\Sigma) =  -\frac{\i}{g^2} \int_{\Sigma} \star \d a$ can be fractional in the presence of vortices when $\theta \neq 0$.  We illustrate a situation where the physical electric charge is fractional in Fig.~\ref{fig:fractional_electric_charge}.

We should note that most of our discussion would have applied just as well if we had used $U(1)^{[1]}_e$ in place of $U(1)^{[1]}_m$.  
The difference would come once the theory is coupled to electrically charged matter with charge $p >1$. 
This would destroy the continuous cubic $\theta$ angle, but a $\Z_p$-valued cubic $\theta$ angle would remain.  
It would be very interesting to realize either the continuous or the discrete cubic $\theta$ angle e.g.~in condensed-matter systems or BSM physics.



\section{Conclusion and outlook}
\label{sec:outlook}

In this paper, we have introduced the notion of symmetry $\theta$ angles for non-anomalous Abelian invertible symmetries.  
The definition of these $\theta$ angles is independent of the choice of path-integral representation.  
While symmetry $\theta$ angles can sometimes be written as more familiar Lagrangian $\theta$ angles, the two concepts are generally quite different. 
A frequent consequence of symmetry $\theta$ angles is a topological Witten effect, which is a reshuffling of twisted sectors with fixed charge under the defining symmetry of the symmetry $\theta$ angle.  
Indeed, we were able to show that all possible topological Witten effects must be associated with symmetry $\theta$ angles.
When there are suitable mixed 't Hooft anomalies, symmetry $\theta$ angles can also lead to the classic charge-fractionalization Witten effect of Ref.~\cite{Witten:1979ey}, as well as its generalizations like those in Ref.~\cite{Hsin:2020nts} and in this paper.  

We believe that we have only scratched the surface of both the concept and the applications of symmetry $\theta$ angles.  
There are many tempting applications and generalizations to explore in the future.  A few of them include:
\begin{itemize}
    \item In this paper, we only discussed symmetry $\theta$ angles for non-anomalous Abelian invertible symmetries. 
    It would be interesting to cover more complicated non-anomalous symmetries, such as non-Abelian, non-invertible, and spacetime symmetries. 
    \item 
    Discrete symmetry $\theta$ angles are RG invariants, and correspond to an extra label that must be specified when proposing an EFT describing a physical system.  It is therefore important to understand how to compute the values of symmetry $\theta$ angles that arise during RG flows to IR EFTs from UV theories with reduced symmetries.
    
    \item
    The existence of discrete symmetry $\theta$ angles means that EFTs have discrete parameters that have not been appreciated in the past.  This can lead to the existence of new phases of matter, at least in the context of gapless systems.
    
    \item We briefly mentioned $1$-form-symmetry $\theta$ angles in $3$-dimensional gauge theories several times in this paper, but have not explored them in detail.  They are associated with the bordism groups 
    \begin{equation}
        \tOmega^{Spin}_{3}\bigl(U(1)\bigr) = \Z_2\,,\qquad
        \tOmega^{Spin}_{3}\bigl(B\Z_2\bigr) = \Z_8\,,\qquad
        \tOmega^{SO}_{3}\bigl(B\Z_n\bigr) = \Z_n\,.
        \label{eq:3d_bordisms}
    \end{equation}
    The $\theta$ angles associated with Eq.~\eqref{eq:3d_bordisms} have a number of interesting features, such as the absence of self-duality and modular group structures (see Section~\ref{sec:counting_params}). We also expect that they produce a more subtle version of the topological Witten effect, seeSection~\ref{sec:beyond-mapping-tori}.

    \item As we very briefly discussed in Section~\ref{sec:U(1)_remark}, for $U(1)$-types symmetries we can generalize $\Theta$ transformations so  that they influence the local physics.
    These  generalized $\Theta$ transformations deserve to be explored much more deeply.
    
    \item Last but not least, we comment on 
    the global structure of the gauge group of the Standard Model, which attracted a surge of interest in recent years; see e.g.~Refs.~\cite{Tong:2017oea,Anber:2021upc,Cordova:2022fhg,Reece:2023iqn,Choi:2023pdp,Cordova:2023her,Cordova:2024ypu,Alonso:2024pmq,Li:2024nuo,Koren:2024xof,Dierigl:2024cxm,Cao:2024lwg,Hsin:2024lya,Wan:2024kaf}.  
    The goal of the recent literature on this subject has been to develop observational strategies to determine which of the $4$ global forms of the SM gauge group
    \begin{align}
        G_{\textrm{SM gauge}} = \frac{SU(3)_c \times SU(2)_L \times U(1)_Y}{\Z_n}\,, \qquad n = 1, 2, 3, \textrm{ or } 6\,.
        \label{eq:SM_gauge}
    \end{align} 
    describes nature.
    However, counting topological manipulations with respect to a $\Z_6^{[1]}$ symmetry in a 4D fermionic QFT (such as the SM) 
    gives the order-$72$ group,
    \begin{align}
        PSL(2,\Z_2)\times PSL(2,\Z_3)\,.
        \label{eq:SM_modular}
    \end{align}
    Some of these $72$ QFTs may be equivalent to each other in the context of the SM. As we briefly mentioned in Footnote~\ref{ft:degeneracy}, this would be related to some symmetries of the SM.  Indeed, there ought to be a $3$-fold degeneracy due to the $(\Z_3)_{B+L}$ symmetry.  Any further degeneracies may be related to the presence of non-invertible symmetries. It would be very interesting to contemplate what experimental probes could tell us which of the many theories described by Eq.~\eqref{eq:SM_modular} describe our universe.
\end{itemize}

\acknowledgments

We are very grateful to Jing-Yuan Chen, Gongjun Choi, Yichul Choi, Anatoly Dymarsky, Isabel Garcia-Garcia, Marius Kongsore, Sungjay Lee, Linhao Li, Theo Jacobson, Mithat \"Unsal, Arkady Vainshtein, and Ken Van Tilburg for helpful discussions.  This
work was supported in part by the Simons Foundation award
number 994302 (A. C., S. C., M. N.), and in part by the National Science
Foundation Graduate Research Fellowship under Grant
No. 1842400 (M. N.).  We are also grateful to the Kavli Institute for Theoretical Physics (KITP) for its hospitality during the ``Lattice and Continuum Approaches to Strongly Coupled QFT'' conference, which was supported in part by the US National Science Foundation under Grant No. NSF PHY-1748958.

\appendix

\section{A heuristic introduction to the classification of invertible theories}
\label{app:topological_data}
\label{app:heuristic}

In this appendix, we give a brief heuristic introduction to the bordism classification of $\calG$-symmetric invertible theories, which may be helpful for understanding Eq~\eqref{eq:classification_discrete}.
The generalization to strictly-topological $\U$-symmetric invertible theories and thus Eq.~\eqref{eq:classification_continuous} is almost trivial.
However, the classification for general non-strictly-topological $\U$-symmetric invertible theories is profoundly different.
Since we do not explore them in any detail in this paper, we shall not review them here.

To begin, let us consider an intuitive but somewhat naive Ansatz for the partition functions of a $\calG$-symmetric invertible theory $\calI$ on a $d$-dimensional oriented spacetime $X$:
\begin{equation}\label{eq:I_naive}
    \calI(X,y)=\exp\left\{\i\int_X\omega(y)\right\}\,,\qquad y\text{ is a $\calG$ gauge field on }X\,,
\end{equation}
for a certain Lagrangian $\omega(y)\in H^d(X,\tfrac{\R}{2\pi\Z})$.
The Lagrangian should be independent of the spacetime $X$.
It is thus helpful to describe the gauge fields and their Lagrangians without specifying the  spacetime manifold $X$.
This can be achieved by using the notion of $\calG$'s \emph{classifying space} $\calB\calG$, an auxiliary topological space with the defining property
\begin{equation}
    \left[X,\calB\calG\right] \,\simeq \, \{\calG \text{ gauge fields on }X\}\,,\qquad\text{for any }X\,.
\end{equation}
where the left-hand side is the set of homotopy classes of continuous maps from $X$ to $\calB \calG$, while the right-hand side is the set of deformation classes of $\calG$ gauge fields on $X$, and $\simeq$ is an isomorphism.  Through the lens of classifying spaces, the Lagrangian is given by
\begin{equation}\label{eq:ordinary_cohomology}
    \omega \in 
    H^d\left(\calB\calG, \tfrac{\R}{2\pi\Z}\right)\,.
\end{equation}
Therefore, based on the intuitive Ansatz~\eqref{eq:I_naive}, $\calG$-symmetric invertible theories appear to be classified by the above cohomology group $H^d\left(\calB\calG, \tfrac{\R}{2\pi\Z}\right)$.

However, this cannot be the correct result, because $H^d\left(\calB\calG, \tfrac{\R}{2\pi\Z}\right)$ completely ignores the ambient spacetime, and does not keep track of the role of spacetime symmetries.
For example, the bosonic spacetime symmetry $SO(d)$ and the fermionic spacetime symmetry $Spin(d)$ are supposed to support quite different invertible theories.
To motivate the resolution of this problem, we notice that Ansatz~\eqref{eq:I_naive} is a \textit{bordism invariant}:
\begin{itemize}
    \item $\calI(X_1\sqcup X_2,y_1+y_2)=\calI(X_1,y_1)\ \calI(X_2,y_2)$,
    \item $\calI(X_1,y_1)=\calI(X_2,y_2)$ when $(X_1,y_1)$ is bordant to $(X_2,y_2)$.
\end{itemize}
Here $\sqcup$ means disjoint union, and saying that $(X_1,y_1)$ is bordant to $(X_2,y_2)$ means $X_1\sqcup-X_2 = \partial W$ for some $(d\!+\!1)$-dimensional compact oriented manifold $W$ to which $y_1$ and $y_2$ can be consistently extended.
We use the minus sign to reverse the orientation.
It is then natural to contemplate generalizing Ansatz~\eqref{eq:I_naive} to general cobordism invariants.

According to the above properties, bordism invariants are 1-dimensional unitary representations of \emph{bordism groups}.
For a topological space $Y$, the collection of bordism classes of oriented or spin $d$-dimensional closed manifolds equipped with a map to $Y$ is denoted by
\begin{equation}\label{eq:Omega(Y)}
    \Omega^{SO}_{d}(Y)\quad\text{or}\quad\Omega^{Spin}_{d}(Y)\,,
\end{equation}
respectively.
They are naturally Abelian groups, with the addition operation given by the disjoint union $\sqcup$ and the inverse given by the orientation reversal.
The groups~\eqref{eq:Omega(Y)} are called $Y$'s oriented and spin bordism groups, respectively.
The two bordism theories satisfy almost all the properties of a homology theory and are thus dubbed generalized homology theories.
The only exceptional property is that while a point's (ordinary) homology groups $H_{d}(\{*\},G)=0$ for all $d\neq 0$, a point's bordism groups
\begin{equation}
    \Omega^{SO}_{d}(\{*\})\quad\text{and}\quad\Omega^{Spin}_{d}(\{*\})
\end{equation}
can be nonzero for any non-negative $d$.
A point's bordism group essentially contains bordism classes without an outward map, and always contributes a direct summand to $Y$'s bordism groups for any $Y$.
One can thus introduce the notion of \emph{reduced} bordism groups%
\footnote{This parallels the definition of the reduced (ordinary) homology and cohomology groups $\widetilde{H}_{\bullet}, \widetilde{H}^{\bullet}$.}
\begin{equation}
    \Omega^{SO}_{d}(Y)=\Omega^{SO}_{d}(\{*\})\oplus\tOmega^{SO}_{d}(Y) \quad\text{and}\quad\Omega^{Spin}_{d}(Y)=\Omega^{Spin}_{d}(\{*\})\oplus\tOmega^{Spin}_{d}(Y)\,.
\end{equation}
The discussion above makes it clear
that, to construct the partition functions of $\calG$-symmetric invertible theories, we need to upgrade the naive Ansatz~\eqref{eq:I_naive} to 1-dimensional unitary representations of the reduced bordism groups 
\begin{equation}
    \tOmega^{SO}_{d}\bigl(\calB\calG) \quad\text{and}\quad\tOmega^{Spin}_{d}\bigl(\calB\calG\bigr)\,,
\end{equation}
in the bosonic and the fermionic cases, respectively.
If we consider more complicated spacetime symmetry structures, $\Omega_{d}$ can be other bordism groups.

Let us now look back at the naive Ansatz~\eqref{eq:I_naive}. 
It should provide a possibly-incomplete possibly-degenerate list of oriented and spin bordism invariants.
The precise relationship is captured by the natural homomorphisms,
\begin{equation}
    \Hom\left(\tOmega^{Spin}_{d}(Y),\tfrac{\R}{2\pi\Z}\right) \ \overset{f^*}{\longleftarrow}\ \Hom\left(\tOmega^{SO}_{d}(Y),\tfrac{\R}{2\pi\Z}\right) \ \overset{g^*}{\longleftarrow}\ \tH^{d}\left(Y, \tfrac{\R}{2\pi\Z}\right)\,,
\end{equation}
which are the $\tfrac{\R}{2\pi\Z}$-duals of the natural homomorphisms
\begin{equation}
    \tOmega^{Spin}_{d}(Y) \ \overset{f}{\longrightarrow}\  \tOmega^{SO}_{d}(Y) \ \overset{g}{\longrightarrow}\ \tH_{d}\left(Y, \Z\right)\,.
\end{equation}
How nicely these homomorphisms behave depends on $Y$'s connectivity.
Let us consider an $n$-connected $Y$, which means all the above groups are trivial for $d\leq n$.
Then it turns out that $g$ and $g^*$ are isomorphisms as long as $d\leq 4\!+\!n$.%
\footnote{
Since the natural map from the Thom spectrum $MSO$ to the Eilenberg-MacLane spectrum $H\Z$ is 4-connected, the natural map from $MSO\wedge Y$ to $H\Z\wedge Y$ is $(5\!+\!n)$-connected.
One can also read this off from the Atiyah-Hirzebruch spectral sequence.}
This means that the intuitive Ansatz~\eqref{eq:I_naive} actually manages to faithfully capture all the bosonic $\calG$-symmetric invertible theories
in low spacetime dimensions, which include the physically interesting cases of $d = 2,3,4$.
However, $f$ and $f^*$ are isomorphisms only for $d\leq 1\!+\!n$.%
\footnote{
Since the natural map from the Thom spectrum $MSpin$ to the Thom spectrum $MSO$ is only 1-connected, the natural map from $MSpin\wedge Y$ to $MSO\wedge Y$ is merely $(2\!+\!n)$-connected.
}
This implies that the intuitive Ansatz~\eqref{eq:I_naive} provides a quite poor approximation to the partition functions of fermionic $\calG$-symmetric invertible theories.
\section{Modular-group structure of self-dual symmetries}
\label{sec:STSTST=1}

This appendix contains a discussion on the appearance of the modular-group relations
\begin{equation}
    S^2=-1\,,\qquad (ST)^3=-1\,,
\end{equation}
for a self-dual non-anomalous finite Abelian invertible symmetry,%
\footnote{
The modular relation exists for self-dual $U(1)$-type symmetries as well but for non-strictly-topological $T$ transformations only.
See Ref.~\cite{Witten:2003ya} for the classic account in the case of $3$-dimensional QFTs with $U(1)^{[0]}$ symmetry.
}
\begin{equation}
    \calG=\widehat{\calG}\,.
\end{equation}
The $-1$ above just denotes an overall sign flip of the background gauge fields.  
This sign flip can be interpreted as a charge-conjugation operation and does not change the physics at all.
The discussion in this appendix is used in Section~\ref{sec:Theta_vs_others}, as well as in the examples of Sections~\ref{sec:electric_theta} and~\ref{sec:4d_YM}. 

There are basically two classes of self-dual symmetries.
The first class are symmetries of the form $\G\!\times\!\widehat{\G}$, which are self-dual by construction.
The less-trivial second class only exist in even spacetime dimension $d$, and is given by $G^{[\frac{d}{2}-1]}$ with a finite group $G$.
Both classes are commonly considered in the literature, with typical examples including
\begin{subequations}
\begin{gather}
    d=2:\qquad \Z_N^{[0]}\,,\quad\Z_N^{[0]}\times\Z_N^{[0]}\\
    d=3:\qquad \Z_N^{[0]}\times\Z_N^{[1]}\,,\\
    d=4:\qquad \Z_N^{[1]}\,,\quad \Z_N^{[0]}\times\Z_N^{[2]}\,.
\end{gather}
\end{subequations}
We will discuss the first class of self-dual symmetries, and then discuss 4-dimensional $\Z_N^{[1]}$ symmetry, a particular example of the second class.
The common feature shared by the two examples will eventually lead us to a uniform understanding on the modular-group structure on all the self-dual symmetries.

\subsection{Example I: $\G\!\times\!\widehat{\G}$ symmetry}

Let us start with the first class, the symmetries of the form $\calG\!\times\!\widehat{\G}$.
It is useful to take the $S$ transformation to have the following form:%
\footnote{
The normalization factor comes from $N(X,\G\!\times\!\widehat{\G})=N(X,\G)\,N(X,\widehat{\G})=|\H(X,\G)|=|\H(X,\widehat{\G})|$.
}
\begin{equation}\label{eq:S^2=-1_first}
    \widehat{\calZ}(\rho,\hat{\rho}) = \frac{1}{|\H(X,\G)|} \sum_{\substack{\sigma\in\H(X,\G)\\\hat{\sigma}\in\H(X,\widehat{\G})}} \!\!\!\!\!\!\calZ(\sigma,\hat{\sigma}) \, \exp\left( - \i\int \rho\cup\hat{\sigma} -\i\int \sigma\cup\hat{\rho} \right),\ \ \ 
    \begin{cases}
        \rho\in\H(X,\G)\\
        \hat{\rho}\in\H(X,\widehat{\G})
    \end{cases}\!\!\!\!\!\!\!\!\,.
\end{equation}
This $S$ transformation is self-dual in the sense that $S^2=-1$.  It differs from our original $S$ transformation in Eq.~\eqref{eq:fore-gauging} by a map $(\rho,\hat{\rho})\to(\rho,\hat{\rho}^{\vee})$ on the background gauge fields.

There is always a particular $\calG\!\times\!\widehat{\G}$-symmetric invertible theory $\calI$ given by
\begin{equation}\label{eq:Q_first}
    \calI(\rho,\hat{\rho}) = \exp\left(\i\int\rho\cup\hat{\rho}\right)\,,\qquad \rho\in\H(X,\G),\ \hat{\rho}\in\H(X,\widehat{\G})\,.
\end{equation}
The theory $\calI$ becomes its own inverse $\calI^{-1}$ under the $S$ transformation~\eqref{eq:S^2=-1_first}.
Hence $\calI$ is simultaneously an invertible theory and a Dijkgraaf-Witten theory.
Let us consider the $T$ transformation $\otimes\calI$.
Performing the $(ST)^3$ transformation to an arbitrary $\G\!\times\!\widehat{\G}$-symmetric theory $\calZ$, we obtain the following chain.  First,
\begin{equation}
\begin{gathered}
    \calZ(\rho,\hat{\rho})\\
    \Big\downarrow T\\
    \calZ(\rho,\hat{\rho})\  \exp\left( \i\int\rho\cup\hat{\rho} \right)\\
    \Big\downarrow S\\
    \frac{1}{|\H(X,\G)|} \sum_{\rho,\hat{\rho}} \calZ(\rho,\hat{\rho}) \, \exp\,\i\!\int \Bigl(\rho\cup\hat{\rho} - \sigma\cup\hat{\rho} -\rho\cup\hat{\sigma} \Bigr)\\
    \Big\downarrow T\\
    \frac{1}{|\H(X,\G)|} \sum_{\rho,\hat{\rho}} \calZ(\rho,\hat{\rho}) \, \exp\,\i\!\int \Bigl(\rho\cup\hat{\rho} - \sigma\cup\hat{\rho} - \rho\cup\hat{\sigma} + \sigma\cup\hat{\sigma} \Bigr)
\end{gathered}
\end{equation}
Applying $S$ again, we find
\begin{equation}
\begin{gathered}
    \begin{split}
        &\: \frac{1}{|\H(X,\G)|^2} \sum_{\rho,\hat{\rho},\sigma,\hat{\sigma}} \calZ(\rho,\hat{\rho}) \, \exp \,\i\!\int \Bigl(\rho\cup\hat{\rho} - \sigma\cup\hat{\rho} - \rho\cup\hat{\sigma} + \sigma\cup\hat{\sigma} -\chi\cup\hat{\sigma} -\sigma\cup\hat{\chi} \Bigr)\\
        = &\: \frac{1}{|\H(X,\G)|} \sum_{\rho,\hat{\rho},\sigma} \calZ(\rho,\hat{\rho}) \, \exp \,\i\!\int \Bigl(\rho\cup\hat{\rho} - \sigma\cup\hat{\rho} -\sigma\cup\hat{\chi} \Bigr)\ \delta(\sigma-\chi-\rho)\\
        = &\: \frac{1}{|\H(X,\G)|} \sum_{\rho,\hat{\rho}} \calZ(\rho,\hat{\rho}) \, \exp \,\i\!\int \Bigl( - \chi\cup\hat{\rho} -\chi\cup\hat{\chi} - \rho\cup\hat{\chi} \Bigr)
    \end{split}
\end{gathered}
\end{equation}
Next, we apply $T$ again to get
\begin{equation}
    \frac{1}{|\H(X,\G)|} \sum_{\rho,\hat{\rho}} \calZ(\rho,\hat{\rho}) \, \exp \,\i\!\int \Bigl( - \chi\cup\hat{\rho} - \rho\cup\hat{\chi} \Bigr)
\end{equation}
and then apply $S$ to get
\begin{align} 
    \calZ(-\rho, -\hat{\rho})\,.
\end{align}
This demonstrates the modular-group relation $(ST)^3=-1$.

\subsection{Example II: 4D \texorpdfstring{$\Z_N^{[1]}$}{ZN[1]} symmetry}
\label{sec:quadratic_refinement}

We now move to a particularly important example of the second class of self-dual symmetries,  $\Z_N^{[1]}$ in $4$ spacetime dimensions. 
First, the $S$ transformation~\eqref{eq:fore-gauging} simplifies to
\begin{equation}
    \widehat{\calZ}(\rho) = \frac{1}{\sqrt{|H^2(X,\Z_N)|}} \sum_{\sigma\in H^2(X,\Z_N)} \!\!\!\!\calZ(\sigma)\,\exp\left(-\i\int\rho\cup\sigma\right)\,,\qquad\rho\in H^2(X,\Z_N)\,,
\end{equation}
and is self-dual in the sense that $S^2=-1$.

Four-dimensional $\Z_N^{[1]}$-symmetric invertible theories are classified by
\begin{subequations}
\begin{gather}
    \Hom\left(\tOmega^{SO}_4(B^2\Z_N),\,\tfrac{\R}{2\pi\Z} \right) = 
    \begin{cases}
        \Z_N\,, & \text{odd }N\\
        \Z_{2N}\,, & \text{even }N
    \end{cases}\,,\\
    \Hom\left(\tOmega^{Spin}_4(B^2\Z_N),\,\tfrac{\R}{2\pi\Z} \right) = \Z_N\,,
\end{gather}
\end{subequations}
where the natural map from the bosonic group to the fermionic group is \textit{surjective} for any $N$.
This surjection implies that the fermionic invertible theories actually do not depend on the choice of spin structures on the spacetime, and can always be extended to bosonic theories defined on general oriented manifolds.
There is only one extension when $N$ is odd, but there are two extensions when $N$ is even.

On any oriented spacetime, a generator of the invertible theories above can be chosen as (see e.g.~\cite{Aharony:2013hda,Gaiotto:2017yup,Kaidi:2021xfk,Duan:2024xbb})
\begin{equation}\label{eq:Pontryagin_square}
    \calI(\rho) \equiv
    \begin{cases}
        \exp\left(2\pi\i\frac{N+1}{2N} \int \rho \cup \rho\right) & \text{odd }N\\
        \exp\left(2\pi\i\frac{N+1}{2N}\int \calP(\rho)\right) & \text{even }N
    \end{cases},
\end{equation}
where $\cup$ is $\Z_N$-valued --- rather than $\tfrac{\R}{2\pi\Z}$-valued as we use everywhere else throughout this paper ---
and the cohomology operation $\calP:H^2(X,\Z_N)\mapsto H^4(X,\Z_{2N})$ is the Pontryagin square that generates $H^4(B^2\Z_N,\Z_{2N}) = \Z_{2N}$.
We have $\calI(\rho)=\calI(\rho)^{N+1}$ on any oriented manifold for odd $N$.
They are distinct for even $N$ but coincide on spin manifold because $\int\calP(\rho)$ is even on spin manifolds.

The invertible theory $\calI$ defined above satisfies a distinguished property,
\begin{subequations}\label{eq:quadratic_refinement_4D}
\begin{gather}
    \calI(n\rho) = \calI(\rho)^{n^2}\,,\qquad\rho\in H^{2}(X,\Z_N)\,,\ n\in\Z\,,\\
    \calI(\rho_1+\rho_2) = \calI(\rho_1)\,\calI(\rho_2)\exp\left(\i\int \rho_1\cup\rho_2\right)\,,\qquad \rho_1,\rho_2\in H^{2}(X,\Z_N)\,.
\end{gather}
\end{subequations}
Namely, the theory $\calI$ is a \textit{quadratic refinement} of the non-degenerate $U(1)$-valued bilinear form $\exp(\i\int\rho_1\cup\rho_2)$.
Because of this fact, the sum
\begin{equation}\label{eq:Gauss-sum}
    \Omega(X) \equiv \frac{1}{\sqrt{|H^2(X,\Z_N)|}}\sum_{\rho\in H^2(X,\Z_N)} \calI(\rho)
\end{equation}
is a typical \textit{Gauss sum}.
It is a pretty non-trivial but well-known lemma that \textit{a Gauss sum is always a phase factor that is always an 8th root of unity}.%
\footnote{
Since the reciprocity formula of Cauchy, Dirichlet, and Kronecker, there have been many results on Gauss sums.
To prove the 8th-root-of-unity property, Milgram's formula~\cite{Milgram:1974soc} is convenient, and is also famous in physics due to its role in 3D Abelian TQFTs; see e.g.~Ref.~\cite{Belov:2005ze}.
For the precise evaluation of $\Phi(X)$, the formula of van der Blij~\cite{van:1959inv} (or a slight generalization due to Turaev~\cite{Turaev:1998rci}) is more useful.
}
More explicitly,
\begin{equation}
    \Omega(X) = \e^{2\pi\i\frac{\Phi(X)}{8}}\,,\qquad \Phi(X)\in\Z\,.
\end{equation}
This immediately implies that $\widehat{\calI}$, the $S$ transformation of $\calI$, is also an invertible theory.
Namely, $\calI$ as well as $\widehat{\calI}$ is an invertible Dijkgraaf-Witten theory.
The quantity $\Phi(X)$ is a gravitational counterterm%
\footnote{
A precise evaluation of $\Phi(X)$ is complicated.
When $\pi_1(X)$ contains no torsion, one can show that
\begin{equation}\label{eq:signature}
    \Phi(X)=(1-N)\,\sigma(X)\,, \mod 8\,.
\end{equation}
Here $\sigma(X)\in\Z$ is the signature of $X$. 
According to the Hirzebruch signature theorem, we can express $\sigma(X)=\frac{1}{3}\int p_1(X)$ in terms of the $p_1$ Pontryagin class.
This gravitational counterterm trivializes on spin manifolds because $\sigma(X)=0 \mod 16$ on spin $X$ according to the Rokhlin theorem.
When $\pi_1(X)$ has $N$-torsion elements, $\Phi(X)$ may also receive contributions from the torsional intersection form.
}
(or say a gravitational $\theta$ angle).

Using above properties, we can further show that $\widehat{\calI}$ is actually $\calI^{-1}$, up to a gravitational counterterm:
\begin{equation}\label{eq:quadraitc-selfdual}
\begin{split}
    \widehat{\calI}(\rho) &= \frac{1}{\sqrt{|H^2(X,\Z_N)|}} \sum_{\sigma} \calI(\sigma)\,\exp\left(-\i\int\rho\cup\sigma\right)\\
    &= \frac{1}{\sqrt{|H^2(X,\Z_N)|}} \sum_{\sigma} \calI(\sigma+\rho)\,\exp\left(-\i\int\rho\cup(\sigma+\rho)\right)\\
    &= \frac{1}{\sqrt{|H^2(X,\Z_N)|}} \sum_{\sigma} \calI(\sigma+\rho)\ \calI^{-2}(\rho)\,\exp\left(-\i\int\rho\cup\sigma\right)\\
    &= \frac{\calI^{-1}(\rho)}{\sqrt{|H^2(X,\Z_N)|}} \sum_{\sigma} \calI(\sigma) \quad=\quad\calI^{-1}(\rho)\ \e^{2\pi\i\frac{\Phi(X)}{8}}\,.
\end{split}
\end{equation}
With the $T$ transformation $\otimes\calI$, we now perform the $(ST)^3$ transformation to an arbitrary 4-dimensional $\Z_N^{[1]}$-symmetric theory $\calZ$.
We first obtain the following chain
\begin{equation}
\begin{gathered}
    \calZ(\rho)\\
    \Big\downarrow T\\
    \calZ(\rho)\ \calI(\rho)\\
    \Big\downarrow S\\
    \frac{1}{\sqrt{|H^2(X,\Z_N)|}} \sum_{\rho} \calZ(\rho)\ \calI(\rho)\,\exp\left(-\i\int\sigma\cup\rho\right)\\
    \Big\downarrow T\\
    \frac{1}{\sqrt{|H^2(X,\Z_N)|}} \sum_{\rho} \calZ(\rho)\ \calI(\rho)\ \calI(\sigma)\,\exp\left(-\i\int\sigma\cup\rho\right)\\
    \Big\downarrow S\\
    \begin{split}
        &\:\frac{1}{|H^2(X,\Z_N)|} \sum_{\rho,\sigma} \calZ(\rho)\ \calI(\rho)\ \calI(\sigma)\,\exp\left(-\i\int\sigma\cup\rho -\i\int\chi\cup\sigma \right)\\
        = &\: \frac{\e^{2\pi\i\frac{\Phi(X)}{8}}}{\sqrt{|H^2(X,\Z_N)|}} \sum_{\rho} \calZ(\rho)\ \calI(\rho)\ \calI^{-1}(\rho+\chi)\\
    \end{split}
\end{gathered}
\end{equation}    
We can then apply a $T$ transformation to the last line above, giving
\begin{equation}
\begin{gathered}  
    \begin{split}
        &\: \frac{\e^{2\pi\i\frac{\Phi(X)}{8}}}{\sqrt{|H^2(X,\Z_N)|}} \sum_{\rho} \calZ(\rho)\ \calI(\rho)\ \calI(\chi)\ \calI^{-1}(\rho+\chi)\\
        =&\: \frac{\e^{2\pi\i\frac{\Phi(X)}{8}}}{\sqrt{|H^2(X,\Z_N)|}} \sum_{\rho} \calZ(\rho)\,\exp\left( -\i\int\rho\cup\chi \right)\,.
    \end{split}
\end{gathered}
\end{equation}
Finally, we apply $S$ and obtain
\begin{equation}
    \calZ(-\rho)\ \e^{2\pi\i\frac{\Phi(X)}{8}}\,.
\end{equation}
Hence we again demonstrated the modular-group relation $(ST)^3=-1$, up to a gravitational counterterm.

\subsection{The general pattern}

Actually, the $\calG\!\times\!\widehat{\G}$-symmetric invertible theory $\calI(\rho,\hat{\rho})$ given by Eq.~\eqref{eq:Q_first} also satisfies a similar property to Eq.~\eqref{eq:quadratic_refinement_4D}.
Namely, it is a quadratic refinement of
\begin{equation}
    \exp\left(\i\int \rho_1\cup\hat{\rho}_2 + \i\int \rho_2\cup\hat{\rho}_1 \right)\,.
\end{equation}
One can explicitly verify that this property is all we need to prove that the theory~\eqref{eq:Q_first} is an invertible Dijkgraaf-Witten theory and that the modular-group relation $(ST)^3=-1$ holds there.

We have now caught the essence of the modular-group structure for general self-dual symmetries.
In general, for a self-dual symmetry $\G$, there is a $\G$-symmetric invertible theory $\calI(\rho)$ that is a \textit{quadratic refinement} of the $U(1)$-valued bilinear form,
\begin{equation}
    \exp\left(\i\int\rho_1\cup\rho_2\right)\,.
\end{equation}
Although it does not happen in our two examples discussed above, in general, the existence of a quadratic refinement may require additional structures on the manifolds, such as a spin structure.%
\footnote{
As a well-known example, in 2-dimensional QFTs with a $\Z_2^{[0]}$ symmetry, $\Hom\left(\tOmega^{SO}_2(B\Z_2),\,\tfrac{\R}{2\pi\Z} \right) = 0$ and $\Hom\left(\tOmega^{Spin}_2(B\Z_2),\,\tfrac{\R}{2\pi\Z} \right) = \Z_2$.
The corresponding fermionic invertible theory can be given by
\begin{equation}
    \calI(\rho,s)=(-1)^{\mathrm{Arf}(s+\rho) + \mathrm{Arf}(s)}\,,\qquad \rho\in H^1(X,\Z_2)
\end{equation}
where $s$ is a spin structure on $X$ and $\mathrm{Arf}(s)$ is the Arf invariant of the spin structure $s$.
}
Then, the phase-factor nature of Gauss sums implies that $\calI(\rho)$ is an invertible Dijkgraaf-Witten theory.
Based on all these, we can derive the modular-group relation 
\begin{equation}
    (ST)^3=-1
\end{equation}
(possibly up to a gravitational counterterm).
This argument provides a uniform derivation of the modular-group structure for an arbitrary self-dual symmetry.%
\footnote{
Actually, this argument even works for $U(1)$-type symmetries with non-strictly-topological $T$ transformations.
For example, Witten's derivation~\cite{Witten:2003ya} of the modular-group relation for 3-dimensional $U(1)^{[0]}$ symmetry is also of this type, as long as one recognizes that the level-1 $U(1)$ Chern-Simons theory is a quadratic refinement of the Beilinson-Deligne cup product we mentioned around Eq.~\eqref{eq:AxB}.
}
\newpage
\bibliographystyle{JHEP}
\bibliography{non_inv}

\end{document}